\documentclass[aps,prx,twocolumn,superscriptaddress,nobalancelastpage]{revtex4-2}

\usepackage{amsmath, amssymb}
\usepackage{amsthm}
\usepackage{ascmac}
\usepackage{bbm}
\usepackage{bm}
\usepackage{comment}
\usepackage{dsfont}
\usepackage{graphicx}
\usepackage{hhline}
\usepackage[hidelinks]{hyperref}
\usepackage{mathcomp}
\usepackage{mathrsfs}
\usepackage{mathtools}
\usepackage{newtxtext}
\usepackage{physics}
\usepackage{setspace}
\usepackage{url}
\usepackage{xcolor}

\newcommand{\calJ}{\mathcal{J}}
\newcommand{\Wsub}{\mathcal{W}_\mathrm{sub}}
\newcommand{\calF}{\mathcal{F}}
\newcommand{\calX}{\mathcal{X}}
\newcommand{\calY}{\mathcal{Y}}

\newcommand{\cyclebasis}{\bm{\mathcal{S}}}
\newcommand{\bbR}{\mathbb{R}}
\newcommand{\bmh}{\bm{h}}

\newcommand{\px}{p_{X}}
\newcommand{\py}{p_{Y}}
\newcommand{\Px}{\bm{p}_{X}}
\newcommand{\Py}{\bm{p}_{Y}}
\newcommand{\Pxy}{\bm{p}}
\newcommand{\pxy}{p}
\newcommand{\Pz}{\bm{p}}
\newcommand{\pz}{p}
\newcommand{\Pa}{\bm{p}_{\alpha}}

\newcommand{\Current}{\bm{J}}
\newcommand{\Force}{\bm{F}}

\newcommand{\force}{\bm{f}}
\newcommand{\Forceth}{\bm{F}^{\mathrm{th}}}
\newcommand{\Forceit}{\bm{F}^{\mathrm{info}}}
\newcommand{\Currenthk}{\bm{J}^{\mathrm{hk}}}
\newcommand{\Forcehk}{\bm{F}^{\mathrm{hk}}}

\renewcommand{\div}{\nabla^{\top}}

\def\innerprod#1#2{\langle#1, #2\rangle}
\def\norm#1{\|#1\|}

\newcommand{\edgeset}{\mathcal{E}}
\newcommand{\start}{\mathrm{s}}
\newcommand{\target}{\mathrm{t}}
\newcommand{\startx}{\mathrm{s}_X}
\newcommand{\targetx}{\mathrm{t}_X}
\newcommand{\starty}{\mathrm{s}_Y}
\newcommand{\targety}{\mathrm{t}_Y}
\newcommand{\starta}{\mathrm{s}_{\alpha}}
\newcommand{\targeta}{\mathrm{t}_{\alpha}}

\newcommand{\EPR}{\dot{\Sigma}}
\newcommand{\epr}{\sigma}
\newcommand{\eprhk}{\sigma^{\mathrm{hk}}}
\newcommand{\eprex}{\sigma^{\mathrm{ex}}}
\newcommand{\EPRhk}{\dot{\Sigma}^{\mathrm{hk}}}
\newcommand{\EPRex}{\dot{\Sigma}^{\mathrm{ex}}}

\newcommand{\diag}{\mathrm{diag}}

\newcommand{\im}{\mathop{\mathrm{im}}}

\newcommand{\iflow}{\dot{I}}
\newcommand{\iflowhk}{\dot{I}^{\mathrm{hk}}}
\newcommand{\iflowex}{\dot{I}^{\mathrm{ex}}}

\newcommand{\alphac}{{\alpha}^{\rm c}}
\newcommand{\potentialit}{\bm{i}}
\newcommand{\energy}{\bm{\epsilon}}
\newcommand{\pseudoenergy}{\bm{\epsilon}^*}
\newcommand{\optimalpotential}{\bm{\phi}^*}

\definecolor{electric_red}{HTML}{B82850}
\definecolor{tsuyukusa}{HTML}{239DDA}
\definecolor{copenhagen_blue}{HTML}{1B4A8B}

\hypersetup{
  colorlinks   = true, %links
  urlcolor     = tsuyukusa, %external hyperlinks
  linkcolor    = electric_red, %internal links
  citecolor   = copenhagen_blue %citations
}

\begin{document}

\title{Geometric decomposition of information flow: New insights into information thermodynamics}
\date{\today}

\author{Yoh Maekawa}
\email{yoh.maekawa@ubi.s.u-tokyo.ac.jp}
\affiliation{Department of Physics, The University of Tokyo, 7-3-1 Hongo, Bunkyo-ku, Tokyo 113-0033, Japan}
\author{Ryuna Nagayama}
\affiliation{Department of Physics, The University of Tokyo, 7-3-1 Hongo, Bunkyo-ku, Tokyo 113-0033, Japan}
\author{Kohei Yoshimura}
\affiliation{Nonequilibrium Quantum Statistical Mechanics RIKEN Hakubi Research Team, Pioneering Research Institute (PRI), RIKEN, 2-1 Hirosawa, Wako, Saitama 351-0198, Japan}
\affiliation{Universal Biology Institute, The University of Tokyo, 7-3-1 Hongo, Bunkyo-ku, Tokyo 113-0033, Japan}
\author{Sosuke Ito}
\email{sosuke.ito@ubi.s.u-tokyo.ac.jp}
\affiliation{Department of Physics, The University of Tokyo, 7-3-1 Hongo, Bunkyo-ku, Tokyo 113-0033, Japan}
\affiliation{Universal Biology Institute, The University of Tokyo, 7-3-1 Hongo, Bunkyo-ku, Tokyo 113-0033, Japan}
\date{\today}

\begin{abstract}
We propose a decomposition of information flow into housekeeping and excess components for autonomous bipartite systems described by Markov jump processes. We introduce this decomposition using the geometric structure of probability currents and the conjugate thermodynamic forces. The housekeeping component arises from cyclic modes caused by violations of detailed balance and maintains the correlations between the two subsystems. In contrast, the excess component arises from conservative forces and alters the mutual information between the two subsystems. With this decomposition, we generalize previous results, such as the second law of information thermodynamics, the cyclic decomposition, and the information-thermodynamic extensions of thermodynamic trade-off relations.
\end{abstract}

\maketitle

\section{Introduction}
Information thermodynamics is the field that studies the conversion of information into energy, and vice versa. 
This framework can be traced back to Maxwell's demon~\cite{maxwell2012theory} and has enabled the thermodynamic treatment of information processing, most notably through Landauer’s principle concerning information erasure~\cite{landauer1991information, esposito2011second} and the Szilard engine~\cite{szilard1929}, which addresses information-energy conversion. The development of stochastic thermodynamics, which describes mesoscopic nonequilibrium systems~\cite{schnakenberg1976network,sekimoto2010stochastic,seifert2012stochastic,seifert2025stochastic}, has significantly advanced the theoretical foundation of information thermodynamics~\cite{parrondo2015thermodynamics, sagawa2010generalized, goold2016role}. 
Information-energy conversion can now be analyzed in a wide variety of systems, including quantum systems~\cite{sagawa2008second, ptaszynski2019thermodynamics, yada2022quantum} and biological processes~\cite{barato2014efficiency, sartori2014thermodynamic, ito2015maxwell}. Several experiments~\cite{toyabe2010experimental,berut2012experimental,jun2014high,koski2014experimental,vidrighin2016photonic,cottet2017observing,ciliberto2017experiments,masuyama2018information,naghiloo2018information,ribezzi2019large,rico2021dissipation,saha2021maximizing,amano2022insights,yada2025experimentally, oikawa2025experimentally} have demonstrated these concepts, such as information erasure and information-energy conversion.

In the context of information thermodynamics for autonomous information processing, information flow quantifies the information exchanged between two subsystems~\cite{ito2013information, hartich2014stochastic, horowitz2014thermodynamics, horowitz2014second, shiraishi2015fluctuation, horowitz2015multipartite, rosinberg2016continuous, ito2016backward, spinney2016transfer, crooks2019marginal, auconi2019information, ito2020unified, wolpert2020uncertainty, nakazato2021geometrical,leighton2024information,tojo2025optimizing, matsumoto2025learning}. When the system is out of equilibrium, the entropy change associated with the subsystems can be negative, apparently violating the second law of thermodynamics. However, taking into account information flow resolves this apparent violation, resulting in a modified inequality known as the second law of information thermodynamics~\cite{parrondo2015thermodynamics}. Information flow has been actively studied especially in the context of Markov jump processes for bipartite systems~\cite{hartich2014stochastic,horowitz2014thermodynamics}, where the two subsystems do not change simultaneously.  It has been analyzed in various contexts, including biological processes such as sensory adaptation~\cite{ito2013information,sartori2014thermodynamic, barato2014efficiency,ito2015maxwell,hartich2016sensory}, membrane transport~\cite{yoshida2022thermodynamic,flatt2023abc}, and molecular motors~\cite{takaki2022information,amano2022insights,leighton2024information}.

Information flow stems from the nonequilibrium nature of a system and is closely related to its entropy production rate (EPR). The EPR is considered a measure of irreversibility, and its nonnegativity is regarded as the second law of thermodynamics~\cite{seifert2025stochastic}. In information thermodynamics, the partial EPR, which is the contribution of the EPR within the subsystem, provides a measure of irreversibility of subsystems, and its nonnegativity is considered the second law of information thermodynamics~\cite{parrondo2015thermodynamics,horowitz2014second}. Information flow is a contribution from the interaction between the subsystems in the partial EPR, and it also becomes non-zero when a subsystem is irreversible.

There are two distinct aspects of the nonequilibrium nature of Markov jump processes: nonconservativeness and nonstationarity. One way to determine whether irreversibility stems from nonconservativeness or nonstationarity is by decomposing the EPR. A geometric decomposition of the EPR into the excess and housekeeping EPRs~\cite{maes2014nonequilibrium,dechant2022geometric1,dechant2022geometric2,yoshimura2023housekeeping,ito2024geometric} is a way to separate the EPR into contributions of nonconservativeness and nonstationarity. Unlike other EPR decompositions such as the Hatano--Sasa decomposition~\cite{hatano2001steady} or the adiabatic-nonadiabatic decomposition~\cite{esposito2010three}, the excess EPR provides tighter bounds on dynamic speeds, such as the speed limits~\cite{nakazato2021geometrical,aurell2012refined, ito2024geometric,yoshimura2023housekeeping} and thermodynamic uncertainty relations (TURs)~\cite{dechant2022geometric2,yoshimura2023housekeeping}. The housekeeping EPR, on the other hand, can be expressed through the cyclic decomposition~\cite{yoshimura2023housekeeping} or the oscillatory mode decomposition~\cite{sekizawa2024decomposing}, meaning it accounts for oscillatory contributions of the EPR due to detailed balance violations. This geometric decomposition of the EPR is universal and can be generalized to various systems, including deterministic chemical reactions, reaction-diffusion systems, open quantum systems, and fluid systems~\cite{nagayama2025geometric,yoshimura2024two,yoshimura2025force}. However, it is unclear whether this universal decomposition applies to the partial EPR in information thermodynamics or to information flow.

In this paper, we show that, similarly to the EPR decomposition, information flow can be decomposed into two nonequilibrium contributions for bipartite systems described by Markov jump processes. The excess information flow, which quantifies nonstationarity, induces temporal changes in the correlation between the two subsystems. By contrast, the housekeeping information flow, which is caused by nonconservativeness, enables information to be transferred between the subsystems without altering their correlations. 
The former has been partially discussed in the context of non-autonomous information processing by a Maxwell's demon via a measurement-feedback scheme~\cite{sagawa2010generalized, parrondo2015thermodynamics}.
We can also find concepts similar to the latter in information thermodynamics for Markov jump processes in the steady state~\cite{horowitz2014thermodynamics, yamamoto2016linear}. Therefore, our new decomposition of information flow appropriately separates the two concepts of information flow in information thermodynamics.

%Furthermore, our results provide a decomposition of the partial EPR into excess and housekeeping components. The non-negativity of these quantities provides two generalizations of the second law of information thermodynamics, offering a novel perspective in information thermodynamics where the contributions of both housekeeping and excess information flows appear to violate the second law of thermodynamics.
The decomposition into housekeeping and excess information flows leads to generalizations of the second law of information thermodynamics. These generalizations offer a fresh perspective in information thermodynamics, particularly in situations where both the housekeeping and excess information flows appear to violate the second law of thermodynamics. These two distinct types of apparent violations suggest novel concepts: Maxwell's demons corresponding to housekeeping and excess dissipation.

Moreover, by considering a geometrical interpretation of the partial excess EPR, we can derive generalizations of TURs and speed limits for subsystems in information thermodynamics. We note that TURs for the partial excess EPR are regarded as generalizations of TURs for the EPR~\cite{barato2015thermodynamic,horowitz2020thermodynamic, otsubo2020estimating} and
for the partial EPR~\cite{otsubo2020estimating,wolpert2020uncertainty, tanogami2023universal}.
These speed limits correspond to the information-thermodynamic speed limits~\cite{nakazato2021geometrical,kamijima2024optimal} based on optimal transport theory~\cite{villani2008optimal} in Langevin systems. These speed limits are obtained by extending the 2-Wasserstein distance~\cite{maas2011gradient, yoshimura2023housekeeping} for Markov jump processes to subsystems. Our speed limits, which are based on the 2-Wasserstein distance, are directly related to the natural decomposition of information flows. This differs from speed limits based on the 1-Wasserstein distance~\cite{dechant2022minimum, van2023thermodynamic, van2023topological,kolchinsky2024generalized, nagayama2025infinite} and its applications to information thermodynamics~\cite{nagase2024thermodynamically,kamijima2025finite,nagase2025thermodynamic}.
In contrast, the partial housekeeping EPR and housekeeping information flow can be decomposed into cycles, as in information thermodynamics in the steady state~\cite{horowitz2014thermodynamics,yamamoto2016linear}. Two information flows in the two subsystems satisfy an antisymmetric relation for a cycle, meaning that they are identical except for their signs. We illustrate these results numerically using a simple four-state model. 

\begin{figure}
    \centering
    \includegraphics[width=\linewidth]{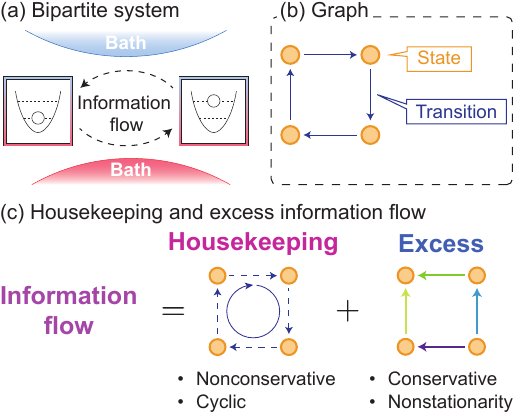}
    \caption{(a) Schematic of a bipartite system. Two subsystems interact while in contact with (potentially multiple) baths. (b) Example of a graph. The nodes represent the system's states, and the edges represent the transitions. (c) Information flow is decomposed into housekeeping and excess components. The housekeeping information flow arises from nonconservative cyclic modes, while the excess information flow arises from conservative nonstationary modes.}
    \label{fig:concept}
\end{figure}

This paper is organized as follows. In Sec.~\ref{sec:review}, we introduce basic concepts and previous results. As a preliminary, we present a geometric approach to nonequilibrium thermodynamics based on Markov jump processes in Secs.~\ref{subsec:mastereq} and~\ref{subsec:EPR}.
In Sec.~\ref{subsec:decomposition}, we review the geometric housekeeping-excess decomposition of the EPR and its consequences.
In Sec.~\ref{sec:geometric_information_thermodynamics}, we introduce a geometric reformulation of information thermodynamics for bipartite systems (see Fig.~\ref{fig:concept}). 
Our formulation allows for a geometric interpretation of the second law of information thermodynamics.
In Sec.~\ref{sec:main_results}, we present our main results.
In Sec.~\ref{subsec:information_flow_decomposition}, we introduce the housekeeping-excess information flow, which is the core idea of this paper. 
In Sec.~\ref{subsec:generalization_second_law}, we demonstrate the thermodynamic interpretation of this decomposition by generalizing the second law of information thermodynamics. 
In Sec.~\ref{subsec:generalization_cyclic_decomposition}, we introduce the cyclic decomposition of the housekeeping information flow and the partial housekeeping EPR. The latter is a dissipation in a subsystem caused by nonconservativeness.
In Sec.~\ref{subsec:TUR}, we derive TURs, which provide tighter bounds than the second law of information thermodynamics regarding the excess information flow.
In Sec.~\ref{subsec:Wasserstein_geometry_subsystem}, we formulate the Wasserstein geometry for subsystems.
To this end, in Sec.~\ref{subsubsec:Wasserstein_distance_marginal}, we first define the $2$-Wasserstein distance for subsystems, which is a conceptually new quantity for discrete systems. We also relate the $2$-Wasserstein distance for subsystems to the notion of housekeeping-excess decomposition of the partial EPR in Sec.~\ref{subsubsec:decomposition_partial_EPR}. 
In Sec.~\ref{subsubsec:speed_limit}, we derive the information-thermodynamic speed limits using the $2$-Wasserstein distance for subsystems.
In Sec.~\ref{sec:example}, we illustrate our results numerically using a four-state model. In Sec.~\ref{sec:discussion}, we conclude the paper with some remarks.

\section{Review of geometric approach to stochastic thermodynamics}
\label{sec:review}
In this section, we briefly review a geometric approach to stochastic thermodynamics. Throughout this paper, the vector space of real column vectors of dimension $d$ is denoted by $\bbR^d$, and the space of real matrices of size $d_1\times d_2$ is denoted by $\bbR^{d_1\times d_2}$. For a matrix $A\in\bbR^{d_1\times d_2}$, its kernel and image are defined as $\ker A = \{\bm{v}\in\bbR^{d_2}\mid A\bm{v} = \bm{0}\}$ and $\im A = \{A\bm{v}\mid \bm{v}\in\bbR^{d_2}\}$, respectively.

\subsection{Master equation}
\label{subsec:mastereq}
We consider a Markov jump process in the system $Z$, and the state of the system $Z$ is represented by $z$. The set of states of $Z$ is denoted by $\mathcal{Z} = \{z_1, \ldots, z_{|\mathcal{Z}|}\}$. 
We assume that $z$ does not contain any odd variables, such as velocity, where the odd variable changes sign in the time-reversal operation. 
The probability that the system $Z$ is in state $z\in\mathcal{Z}$ at time $t$ is denoted by $\pz(z; t)$. 
The system is in contact with several baths, and a bath is denoted by the symbol $\nu$. Each transition between two states is associated with one of these baths. The time evolution of $\pz(z; t)$ is governed by the following master equation:
\begin{align}
    d_t\pz(z; t) = \sum_{z'\in\mathcal{Z}, \nu}R^{(\nu)}(z|z')\pz(z'; t). \label{master_eq}
\end{align}
Here, $R^{(\nu)}(z|z')$ is the transition rate from $z'$ to $z$ associated with the bath $\nu$. For $z\neq z'$, we require $R^{(\nu)}(z|z') \geq 0$, and the diagonal elements satisfy $R^{(\nu)}(z|z) = -\sum_{z': z'\neq z}R^{(\nu)}(z'|z)$. In the following, we can omit the time argument $t$ for the sake of simplicity. For convenience, we define the probability vector $\Pz\in\bbR^{|\mathcal{Z}|}$ by giving its components as $[\Pz]_z = \pz(z)$.

We introduce a graph-theoretic description of the master equation. We write the transition from $z$ to $z'$ associated with the bath $\nu$ as $(\nu, z\to z')$ and interpret it as a directed edge with source $z$ and target $z'$. The set of all such edges is denoted by $\edgeset^{\mathrm{all}} = \{e = (\nu, z\to z')\mid R^{(\nu)}(z'|z)> 0\}$. We impose the condition that for every $e = (\nu, z\to z')\in\edgeset^{\mathrm{all}}$, the reverse edge $-e\coloneqq(\nu, z'\to z)$ is also in $\edgeset^{\mathrm{all}}$. Let $\edgeset = \{e_1,\ldots, e_{|\edgeset|}\}$ be a set of edges
containing only either $e$ or $-e$ for each pair $(e, -e)$. In this way, we define a loopless directed graph $G = (\mathcal{Z}, \edgeset)$. For convenience, we label each edge $e =(\nu (e), \start(e) \to \target(e))$ by its associated bath $\nu(e)$, source state $\start(e)$, and target state $\target(e)$, and write the corresponding transition rate as $R^{(\nu(e))}(\target(e)|\start(e)) = R_e$. The forward flux of edge $e$ is defined as $J_e^+ \coloneqq R_e\pz(\start(e))$, and the reverse flux is $J_e^- \coloneqq R_{-e}\pz(\target(e))$. Therefore, the reverse flux of the edge $e$ is equal to the forward flux of the reverse edge $-e$, i.e., $J_e^-=J_{-e}^+$. The net probability flux of the edge $e$, given by $J_e \coloneqq J_e^+ - J_e^-$, is called the current of $e$.
We define the current vector $\Current\in\bbR^{|\edgeset|}$ by $[\Current]_e = J_e$. If we emphasize the distribution dependence of $\Current$, we may write $\Current(\Pz)$. 
Hereafter, we assume that $p(z)>0$ for any $z\in\mathcal{Z}$, and hence $J^+(\Pz) > 0$ and $J^-(\Pz) > 0$ for allowed transitions.

\begin{figure}
    \centering
   \includegraphics[width=\linewidth]{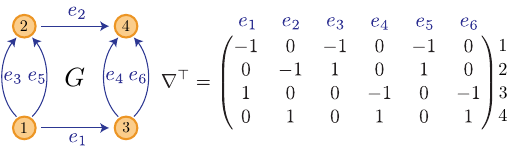}
\caption{A graph and its incidence matrix (transposed, giving the divergence).  The graph $G$ consists of the nodes $\mathcal{Z} = \{1, 2, 3, 4\}$ and the edges $\edgeset = \{e_1, e_2, e_3, e_4, e_5, e_6\}$. Each node represents a state of the system, and the edges indicate reversible transitions between two different states.
We will refer to this graph repeatedly throughout the paper. }
   \label{fig:graph_incidencematrix}
\end{figure}

Next we introduce a matrix $\nabla\in\bbR^{|\edgeset|\times|\mathcal{Z}|}$ by $[\nabla]_{e, z} = \delta_{z, \target(e)} - \delta_{z, \start(e)}$, where $\delta_{x,y}$ is the Kronecker delta.
This matrix can be regarded as the discrete gradient operator, since $[\nabla\bm{\phi}]_e=\phi(\target(e))-\phi(\start(e))$ is equal to the change of $\bm{\phi}$ in the transition corresponding to the edge $e$.
We can then rewrite the master equation~\eqref{master_eq} as
\begin{align}
    d_t\Pz = \div\Current(\Pz), \label{continuity_eq}
\end{align}
where $\div$ is the transpose of $\nabla$, also known as the incidence matrix of the directed graph $G$ in graph theory~\cite{bapat2010graphs} (see also Fig.~\ref{fig:graph_incidencematrix} for an example). Because $\div$ is considered to be the adjoint of the discrete gradient, namely,  the discrete divergence, Eq.~\eqref{master_eq} can be interpreted as a discrete version of the continuity equation, where $\div$ multiplied by the current $\Current$ is equal to $d_t\Pz$. Note that the divergence of the current vanishes in the steady state $\Pz^{\mathrm{st}}$, i.e.,
\begin{align}
    \div\Current(\Pz^{\mathrm{st}}) = \bm{0}. \label{steady_continuity_eq}
\end{align}

For simplicity, we assume that the directed graph $G$ is connected. Physically, this assumption limits our attention to a single physical system. If $G$ had multiple components, each would correspond to an independent system. In this paper, we do not consider such cases.

\subsection{Entropy production rate}
\label{subsec:EPR}
We discuss the local detailed balance and the detailed balance conditions. In the setup of Sec.~\ref{subsec:mastereq}, we assumed the reversibility of the transitions: $R_e> 0 \Rightarrow R_{-e}> 0$ for all $e\in\edgeset^\mathrm{all}$. As a first step to introduce thermodynamics, we impose the local detailed balance condition: 
\begin{align}
    \frac{R_e}{R_{-e}} = e^{s_e}. \label{LDB}
\end{align}
The quantity $s_e \in \mathbb{R}$ represents the entropy change of the bath $\nu(e)$ in the transition corresponding to the edge $e$, with the Boltzmann constant set to unity. Note that $s_e$ satisfies the antisymmetric relation $s_e = -s_{-e}$. 

Next, we introduce the equilibrium distribution, $\Pz^{\mathrm{eq}}$. It is defined as the distribution that satisfies the following detailed balance condition:
\begin{align}
    J_e(\Pz^{\mathrm{eq}}) = R_e[\Pz^{\mathrm{eq}}]_{\start(e)} - R_{-e}[\Pz^{\mathrm{eq}}]_{\target(e)} = 0 \label{DB}
\end{align}
for any $e\in\edgeset$. This condition can be written compactly as $\Current(\Pz^{\mathrm{eq}}) = \bm{0}$, and thus the equilibrium distribution is a stationary distribution. 
Note that $\Pz^{\mathrm{eq}}$ exists if and only if there exists an observable $\energy \in \bbR^{|\mathcal{Z}|}$ with $[\energy]_z = \epsilon_z$ such that $s_e = -\epsilon_{\target(e)} + \epsilon_{\start(e)}$ holds for all $e\in\edgeset$. We refer to a system as being detailed balanced if an equilibrium distribution exists.
In this case, the equilibrium distribution is expressed as $\Pz^{\mathrm{eq}} = e^{-\energy}/Z(\energy)$, where the normalization factor is given by $Z(\energy) = \sum_z e^{-\epsilon_z}$ and we use the notation $[e^{-\energy}]_z = e^{-\epsilon_z}$. By explicitly separating the inverse temperature $\beta > 0$ and the energy level $\bm{E}$ as $\energy = \beta\bm{E}$, the equilibrium distribution takes the form of a canonical distribution $\Pz^{\mathrm{eq}} = e^{-\beta\bm{E}}/Z$.

Based on $s_e$ introduced by the local detailed balance condition [Eq.~\eqref{LDB}], we can define the entropy production rate (EPR) as 
\begin{align}
    \EPR(\Pz)
    = d_tS^{\mathrm{sys}}(\Pz) + \dot{S}^{\mathrm{env}}(\Pz) \label{eq:epr_def}
\end{align}
with the system entropy $S^{\mathrm{sys}}$ and the environmental entropy change rate $\dot{S}^{\mathrm{env}}$ defined as 
\begin{align}
    S^{\mathrm{sys}}(\Pz) 
    = -\sum_z p(z)\ln p(z), \;
    \dot{S}^{\mathrm{env}}(\Pz) = \sum_e s_eJ_e(\Pz).
\end{align}
If we define the thermodynamic force $\Force(\Pz) \in \bbR^{|\edgeset|}$ as 
\begin{align}
    F_e(\Pz)=s_e -[\ln p(\target(e))-\ln p(\start(e))], 
\end{align}
the EPR is rewritten into the insightful form~\cite{schnakenberg1976network}
\begin{align}
    \EPR(\Pz) = \sum_{e\in\edgeset} J_e(\Pz) F_e(\Pz) = \innerprod{\Current(\Pz)}{\Force(\Pz)}. \label{eq:EPR_inner_prod}
\end{align}
Here, $\langle \cdot, \cdot \rangle$ denotes the inner product in $\mathbb{R}^{|\edgeset|}$.
The thermodynamic force is deeply associated with the current because we have 
\begin{align}
    F_e(\Pz) = \ln J_e^+(\Pz) - \ln J_{e}^-(\Pz) \label{eq:fandj}
\end{align}
due to the local detailed balance. 
The EPR is often regarded as a quantitative measure of the irreversibility of thermodynamic processes in a system and its environment. 
This is formalized by the second law of thermodynamics, which states that the EPR is non-negative: $\EPR \geq 0$. 
The expression in Eq.~\eqref{eq:fandj} proves this inequality because it shows that $J_e(\Pz)$ and $F_e(\Pz)$ always have the same sign. 

The thermodynamic force is divided into a thermal part and an informational part as 
\begin{align}
    \Force(\Pz) = \Forceth + \Forceit(\Pz), \label{force_th_it}
\end{align}
where $[\Forceth]_e = \ln (R_{e}/R_{-e}) = s_e$ and $[\Forceit(\Pz)]_e = \ln[\pz(\start(e))/\pz(\target(e))]$. 
Note that these terms correspond to $\dot{S}^{\mathrm{env}}$ and $d_tS^{\mathrm{sys}}$ in the EPR, respectively. 
The informational component quantifies the Shannon entropy change, which becomes clearer if we define $\bmh\in\mathbb{R}^{|\mathcal{Z}|}$ as $[\bmh]_z=-\ln p(z)$; then, it is rewritten as $\Forceit(\Pz)=\nabla\bmh$. 
On the other hand, the thermal part is given as a gradient only if the system is detailed balanced. 
When $s_e = -\epsilon_{\target(e)} + \epsilon_{\start(e)}$ holds for all $e\in\edgeset$, the thermal force reads $\Forceth = -\nabla\energy$, and we obtain 
\begin{align}
    \Force(\Pz) = -\nabla(\energy - \bmh).
\end{align}
As discussed in the next section, we define a thermodynamic force as \textit{conservative} if it is the gradient of a potential (i.e., a vector in $\mathbb{R}^{|\mathcal{Z}|}$); the above discussion shows that the thermodynamic force is conservative if and only if the system is detailed balanced. 

We now introduce the geometric expression for the EPR. 
First, we define the edgewise Onsager coefficient $l_e$ by the ratio of $J_e$ to $F_e$ as
\begin{align}
    l_e(\Pz) \coloneqq \frac{J_e^+(\Pz) - J_e^-(\Pz)}{\ln J_e^+(\Pz) - \ln J_e^-(\Pz)} = \frac{J_e(\Pz)}{F_e(\Pz)} > 0.
\end{align}
If $J^+_e(\Pz) = J^-_e(\Pz) > 0$ holds, we let $l_e(\Pz) \coloneqq (J^+_e(\Pz) + J^-_e(\Pz))/2$ to ensure the continuity.
The relation $J_e = l_e F_e$ can be interpreted as a microscopic analog of the Onsager relation in linear response theory~\cite{onsager1931reciprocal1, onsager1931reciprocal2}, and $l_e$ is a discrete analog of the diffusion coefficient in Langevin systems~\cite{van2021geometrical}. 
Introducing the diagonal matrix of the edgewise Onsager coefficients $L(\Pz) = \diag(l_1(\Pz), l_2(\Pz), \dots ,l_{|\edgeset|}(\Pz) ) \in \bbR^{|\edgeset|\times|\edgeset|}$, we obtain the compact expression $\Current(\Pz) = L(\Pz)\Force(\Pz)$. 
We call a pair of vectors $(\Current, \Force)$ a conjugate pair if they are connected by $L(\Pz)$ as $\Current=L(\Pz)\Force$. 
Since $L(\Pz)$ is positive definite, we can define an inner product and norm with the metric $L(\Pz)$ for vectors $\bm{v}, \bm{v}' \in \bbR^{|\edgeset|}$ as $\innerprod{\bm{v}}{\bm{v}'}_{L(\Pz)} = \bm{v}^{\top}{L(\Pz)}\bm{v}'$ and $\norm{\bm{v}}_{L(\Pz)} = \sqrt{\innerprod{\bm{v}}{\bm{v}}_{L(\Pz)}}$, respectively. The EPR is thus expressed as the squared norm of the thermodynamic force in the vector space of forces, and is therefore non-negative, 
\begin{align}
    \EPR(\Pz) = \norm{\Force(\Pz)}^2_{L(\Pz)} \geq 0. \label{geometric_EPR}
\end{align}
The inequality on the right-hand side is nothing more than the second law of thermodynamics.
In the following sections, we sometimes make $\Pz$ dependence implicit for simplicity when clear from the context.

\subsection{Geometric housekeeping-excess decomposition of EPR}
\label{subsec:decomposition}
We briefly review the geometric housekeeping-excess decomposition of the EPR. The content of this section follows Ref.~\cite{yoshimura2023housekeeping}. Important consequences of this decomposition are summarized in Secs.~\ref{subsubsec:cyclic_decomposition} and \ref{subsubsec:Wasserstein_geometry}. 

We begin by introducing the conservativeness of thermodynamic forces, which plays a central role in geometric decomposition. 
A force $\Force$ is  called a conservative force if there exists a potential $\bm{\phi} \in \bbR^{|\mathcal{Z}|}$ such that $\Force = -\nabla\bm{\phi}$. 
Furthermore, we write the set of all conservative forces as $\im\nabla$
and call this set the conservative subspace (or the image of $\nabla$).
If and only if the system is detailed balanced, the force is always expressed as $\Force(\Pz) = -\nabla(\energy - \bmh)$, and thus is conservative. 
In other words, the important point is that the nonconservativeness characterizes the violation of the detailed balance.

Although the thermodynamic force $\Force$ is not necessarily conservative, any dynamics can be reproduced by a conservative force. Let us explain this fact concretely; for any $\Pz$, the following linear equation for $\Current'$ has a unique solution:
\begin{align}
        \div\Current(\Pz) = \div\Current',
    \label{Onsager_equation}
\end{align}
under the condition
\begin{align}
    \Current' = L(\Pz)\Force'\;\text{and}\;\Force'\in\im\nabla. \label{constraints_Onsager_equation}
\end{align}
We write the solution as $\Current^*$ and the conjugate force as $\Force^* = L^{-1}\Current^*$. Let us explain the physical meaning of the solution. We can rewrite the continuity equation~\eqref{continuity_eq} as
\begin{align}
    d_t\Pz = \div\Current^*, \label{modified_continuity_eq}
\end{align}
so we can interpret $\Current^*$ as the current that effectively drives the dynamics at each moment. 
Furthermore, considering that $\Current^*$ is conjugate to $\Force^*$, we obtain the picture that any time evolution $d_t\Pz$ can be interpreted as the dynamics driven by a virtual conservative force $\Force^*$. Note that $\Current^*$ satisfying only Eq.~\eqref{Onsager_equation} may not be uniquely determined because $\ker\div$ can contain non-trivial elements. The addition of the condition~\eqref{constraints_Onsager_equation} to Eq.~\eqref{Onsager_equation} establishes the existence and uniqueness of the solution, as explained in the next paragraph.

We now interpret the solution $\Current^*$ from a geometric perspective by focusing on the conjugate force $\Force^*$. Formally decomposing the force as $\Force = \Force^* + (\Force - \Force^*)$, we find that $\Force^* \in \im\nabla$ and $\Force - \Force^* \in \ker\div L$, since $\div L(\Force - \Force^*) = \div\Current - \div\Current^* = \bm{0}$ holds. Note that $\div L$ is the adjoint of $\nabla$ with respect to the inner product $\innerprod{\cdot}{\cdot}_L$. Thus, we find that the decomposition $\Force = \Force^* + (\Force - \Force^*)$ is the orthogonal decomposition of $\Force$ into $\im\nabla$ and $\ker\div L$, with respective components $\Force^*$ and $\Force - \Force^*$. Therefore, $\Force^*$ is given by the orthogonal projection of $\Force$ onto $\im\nabla$ with respect to the inner product $\innerprod{\cdot}{\cdot}_L$. We can express $\Force^*$ through the following two equivalent minimization problems:
\begin{align}
    \Force^*(\Pz) &= \underset{\Force':\Force(\Pz) - \Force'\in \ker\div L(\Pz)}{\arg\min}\norm{\Force'}_{L(\Pz)}^2 \label{projected_force_def1}\\
    &= \underset{\Force'\in \im\nabla}{\arg\min}\norm{\Force(\Pz) - \Force'}^2_{L(\Pz)}. \label{projected_force_def2}
\end{align}
The fact that the projection is uniquely determined justifies the existence and uniqueness of the solution to Eq.~\eqref{Onsager_equation} with the constraints~\eqref{constraints_Onsager_equation}.

Let us now introduce the geometric housekeeping-excess decomposition of the EPR. Take a potential $\optimalpotential \in \bbR^{|\mathcal{Z}|}$ such that $\Force^* = -\nabla\optimalpotential$. Note that $\optimalpotential$ is not unique; 
$\Force^*$ is invariant under the transformation $\optimalpotential \to \optimalpotential + r(1, 1, \ldots, 1)^\top$ for any scalar $r$, and there is no other freedom~\footnote{Since $\ker\nabla = \{r(1, 1, \ldots, 1)^\top\mid r\in\bbR\}$, this fact can be shown as follows. For a potential $\bm{\psi}\in\bbR^{|\mathcal{Z}|}$ satisfying $\nabla\bm{\psi} = \bm{0}$, $[\bm{\psi}]_{\start(e)} - [\bm{\psi}]_{\target(e)} = 0$ holds for all $e\in\edgeset$. Since $G$ is connected there exists a path connecting any two states $(z, z')\in\mathcal{Z}^2$, so we obtain $[\bm{\psi}]_{z_1} = [\bm{\psi}]_{z_2} = \cdots = [\bm{\psi}]_{z_{|\mathcal{Z}|}}$.}. 
If the system is detailed balanced, we have $\Force = -\nabla(\energy - \bmh)$, hence we can set $\optimalpotential = \energy - \bmh \eqqcolon \bm{\phi}^{\mathrm{eq}}$. As a property of the projection, $\Force^*$ and $\Force - \Force^*$ are always orthogonal with respect to $\innerprod{\cdot}{\cdot}_L$, or we can directly show  $\innerprod{\Force^*}{\Force - \Force^*}_L = \innerprod{-\nabla\optimalpotential}{\Current - \Current^*} = -(\optimalpotential)^{\top}\div(\Current - \Current^*) = 0$. Using this orthogonality, we can decompose the EPR into two non-negative terms:
\begin{gather}
    \EPR(\Pz) = \EPRhk(\Pz) + \EPRex(\Pz), \label{hk_ex_decomposition}\\
    \EPRhk(\Pz) \coloneqq \norm{\Force(\Pz) - \Force^*(\Pz)}^2_{L(\Pz)}, \label{eq:housekeeping_EPR}\\
    \EPRex(\Pz) \coloneqq \norm{\Force^*(\Pz)}^2_{L(\Pz)}.
\end{gather}
Here, $\EPRhk$ and $\EPRex$ are referred to as the housekeeping EPR and the excess EPR, respectively~\cite{yoshimura2023housekeeping}. 

Let us briefly characterize these two types of EPRs through the following two situations. 
The first one is the steady state, where $\div\Current(\Pz^\mathrm{st}) = \div L(\Pz^\mathrm{st})\Force(\Pz^\mathrm{st}) = \bm{0}$, i.e., $\Force(\Pz^\mathrm{st})\in\ker\div L(\Pz^\mathrm{st})$ holds. Focusing on the minimization problem~\eqref{projected_force_def1}, we find that $\Force^*(\Pz) = \bm{0}$ is the minimizer. Therefore, the following two equalities hold:
\begin{align}
    \EPRhk(\Pz^{\mathrm{st}}) = \EPR(\Pz^{\mathrm{st}}),~\EPRex(\Pz^{\mathrm{st}}) = 0. \label{decomposition_in_steady_state}
\end{align}
Second, assume the system is detailed balanced; i.e., $\Force(\Pz)\in\im\nabla$ holds. Focusing on the second minimization~\eqref{projected_force_def2}, we then find that $\Force^* =\Force(\Pz)\in\im\nabla$ is the minimizer, which implies
\begin{align}
    \EPRhk(\Pz) = 0,~\EPRex(\Pz) = \EPR(\Pz).
\end{align}

Therefore, we can interpret the housekeeping EPR and the excess EPR as follows.
The housekeeping EPR, which arises from the nonconservative component of the thermodynamic force ($\Force - \Force^*$), quantifies the dissipation required to maintain a steady state out of equilibrium.
On the other hand, the excess EPR, which is a contribution from the conservative component of the thermodynamic force ($\Force^*$), represents the dissipation due to the system's time evolution.

The physical meaning of the excess EPR can also be clarified by rewriting it as follows:
\begin{gather}
    \EPRex(\Pz) = d_tS^\mathrm{sys}(\Pz) + \dot{S}^\mathrm{ex,\,env}(\Pz), \label{excessepr}
\end{gather}
which provides the excess EPR as the sum of the Shannon entropy's time derivative and the excess entropy change rate in the environment $\dot{S}^\mathrm{ex,\,env}(\Pz)$ as in Eq.~\eqref{eq:epr_def}. 
Specifically, $\dot{S}^\mathrm{ex,\,env}(\Pz)$ is defined as
\begin{align}
    \dot{S}^\mathrm{ex,\,env}(\Pz)
    \coloneqq -\innerprod{\Current^*(\Pz)}{\nabla\pseudoenergy(\Pz)}, \label{eq:ex_alt}
\end{align}
with the pseudo-energy $\pseudoenergy$, which is defined as a potential satisfying the relation
\begin{align}
    \pseudoenergy(\Pz) \coloneqq \optimalpotential(\Pz) + \bmh(\Pz). \label{pseudo_energy_level}
\end{align}
Note that this relation is analogous to $\energy = \bm{\phi}^{\mathrm{eq}} + \bmh$ in detailed balanced systems. 
Because $\Force^* = -\nabla\bm{\phi}^* = \nabla(\bmh - \pseudoenergy)$, we obtain Eq.~\eqref{excessepr} as 
\begin{align}
    d_tS^{\mathrm{sys}} + \dot{S}^{\mathrm{ex,\,env}}
    &=(\nabla^{\top} \Current^*)^{\top} \bmh - \innerprod{\Current^*}{\nabla\pseudoenergy} \notag\\
    &=\innerprod{\Current^*}{\nabla (\bmh -\pseudoenergy)}  \notag\\
    &=\innerprod{\Current^*}{\Force^*}=\EPRex.
\end{align}

%We discuss when these quantities vanish based on the above expressions.
%Since the excess EPR is also given by $\EPRex(\Pz)= -(d_t \Pz)^{\top}\bm{\phi}^*$, it vanishes when the system is in the steady state, $d_t \Pz =\boldsymbol{0}$. Therefore, the excess EPR quantifies dissipation caused by nonstationarity.
%On the other hand, the housekeeping EPR vanishes when $\Force^{\rm nc} (\Pz)=\Force^{\mathrm{th}} - \Force^{\mathrm{c}}(\Pz)=\boldsymbol{0}$. 
%If the system is detailed balanced, then $\Forceth$ becomes conservative and the condition $\Force^{\rm  nc}(\Pz) =\boldsymbol{0}$ satisfies. 
%Therefore, the housekeeping EPR quantifies dissipation caused by nonconservativeness.

Moreover, we can find an alternative expression for the excess entropy change rate in terms of the excess heat flow. If the system is detailed balanced, we obtain $\dot{S}^{\mathrm{ex,\,env}}(\Pz) = \dot{S}^{\mathrm{env}}(\Pz)$. The pseudo-energy level $\energy^*$ is also equal to $\energy$, and it could be the product of the inverse temperature $\beta$ and energy level $\bm{E}$, i.e., $\energy^* =\energy= \beta\bm{E}$. The heat flow $\dot{Q}$ is introduced as $\dot{Q} \coloneqq \innerprod{\Current(\Pz)}{\nabla\bm{E}} = (d_t\Pz )^{\top}\bm{E}$
and we can obtain the relation $\dot{S}^{\mathrm{ex,\,env}} =\dot{S}^{\mathrm{env}} = -\beta\dot{Q}$. Analogously to the case of the detailed balance condition, we may define the excess heat flow $\dot{Q}^{\rm ex}$ as $\dot{Q}^{\rm ex} \coloneqq \beta^{-1} \innerprod{\Current^*}{\nabla\energy^*} = (d_t\Pz )^{\top} \beta^{-1} \energy^*$ if the system is attached to a single bath with inverse temperature $\beta$. 
From this definition, we obtain $\dot{S}^{\mathrm{ex,\,env}} = -\beta\dot{Q}^{\mathrm{ex}}$. When multiple baths are present, the excess heat flow becomes ambiguous, and only the excess entropy change rate in the environment $\dot{S}^\mathrm{ex,\,env}$ may be well defined in general. 

We note that the geometric housekeeping-excess decomposition is different from the traditional Hatano--Sasa (adiabatic-nonadiabatic) decomposition~\cite{hatano2001steady, esposito2010three, esposito2007entropy, ge2010physical}. The former requires only information about the time evolution at that instant, while the latter needs details of the steady state.
The former is more closely connected to the speed of the time evolution, which is advantageous for deriving thermodynamic trade-off relations.

\begin{figure}
    \centering
   \includegraphics[width=\linewidth]{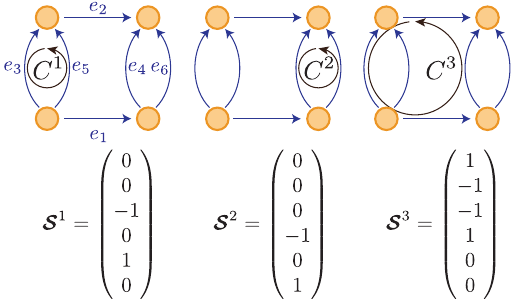}
 \caption{ An example of cycles and their associated vectors in the graph $G$ presented in Fig.~\ref{fig:graph_incidencematrix}. The graph $G$ has three fundamental cycles $C^1 = \{-e_3, e_5\}, C^2 = \{-e_4, e_6\}$ and $C^3 = \{e_1, -e_2, -e_3, e_4\}$, whose corresponding vectors are denoted by $\cyclebasis^1, \cyclebasis^2$, and $\cyclebasis^3$, respectively. Any cycle on $G$ can be expressed as a superposition of $C^1$, $C^2$, and $C^3$, and the kernel of $\div$ is given by $\ker\div = \{r_1\cyclebasis^1 + r_2\cyclebasis^2 + r_3\cyclebasis^3 \mid r_1, r_2, r_3\in\bbR\}$.
}
   \label{fig:cycles}
\end{figure}

\subsubsection{Cyclic decomposition}
\label{subsubsec:cyclic_decomposition}
In this section, we review an expression for the housekeeping EPR~\cite{yoshimura2023housekeeping} that generalizes a well-known formula for the steady-state EPR provided by Schnakenberg in Ref.~\cite{schnakenberg1976network}.
To this end, we define a cycle $C$ in $G$ as a nonempty subset of $\edgeset^\mathrm{all}$ satisfying the following two properties:  (1) $C$ is a closed path in $G$. When we express $C$ explicitly as $C = \{e_{i(1)}, \ldots, e_{i(|C|)}\}$, $\target(e_{i(|C|)}) = \start(e_{i(1)})$ and $\target(e_{i(k)}) = \start(e_{i(k + 1)})$ hold for $k = 1, \ldots, |C| - 1$. 
(2) $C$ does not self-intersect; that is, there is at most one $e\in C$ satisfying $\start(e) = z$ for any $z\in\mathcal{Z}$, and there is at most one $e \in C$ satisfying $\target(e) = z$ for any $z\in\mathcal{Z}$.

We can deal with cycles concisely in a linear-algebraic manner. 
We define a vector $\cyclebasis(C)\in\mathbb{R}^{|\edgeset|}$ corresponding to a cycle $C$ by 
\begin{align}
    [\cyclebasis(C)]_e = 
    \begin{cases}
        1 & e\in C,\\
        -1 & -e\in C,\\
        0 & \text{otherwise.}
    \end{cases}
\end{align}
From a direct computation, we can find $\div\cyclebasis(C)=\bm{0}$ for any cycle $C$. 
Thus, every vector $\cyclebasis(C)$ generated from a cycle belongs to $\ker\div$. 
Furthermore, it is known that by appropriately choosing cycles, we can create a basis of $\ker\div$ from such vectors. 
We label cycles whose corresponding vectors form a basis by a label $\mu\in\mathcal{M}=\{1,2,\dots,\dim\ker\div\}$ so that a basis is given by $\{\cyclebasis(C^\mu)\}_{\mu\in\mathcal{M}}$, and refer to such cycles as fundamental cycles~\cite{bapat2010graphs}. 
For simplicity, we write $\cyclebasis(C^\mu)$ as $\cyclebasis^\mu$ and call the set $\{\cyclebasis^\mu\}_{\mu\in\mathcal{M}}$ the cycle basis of $G$
(see also Fig.~\ref{fig:cycles} for an example).

In his seminal work~\cite{schnakenberg1976network}, Schnakenberg gave an expression for the steady-state EPR in terms of cycles. 
Note that since $\div\Current(\Pz^{\mathrm{st}}) = \bm{0}$, the steady-state current $\Current(\Pz^{\mathrm{st}})$ is in ${\rm ker}\div$, and can be expanded in terms of the cycle basis as
\begin{align}
    \Current(\Pz^{\mathrm{st}}) = \sum_{\mu\in\mathcal{M}}\calJ(\Pz^{\mathrm{st}}; C^\mu)\cyclebasis^\mu. \label{steady_current_cyclic_decomp}
\end{align}
Here, the expansion coefficients $\calJ(\Pz^\mathrm{st}; C^\mu)$ represent the cyclic current that flows along the cycle $C^\mu$. Thus, the steady-state current can be interpreted as a superposition of such cyclic currents. 
We note that the set of the coefficients $\{\calJ(\Pz^\mathrm{st}; C^\mu)\}_{\mu\in\mathcal{M}}$ is uniquely determined by $\Current(\Pz^{\mathrm{st}})$ because $\mathrm{Span}\{\cyclebasis^\mu\}_{\mu\in\mathcal{M}} = \ker\div$. 
The cycle affinity $\calF(\Pz^{\mathrm{st}}; C^\mu)$ conjugate to $\calJ(\Pz^{\mathrm{st}}; C^\mu)$ is defined as 
\begin{align}
    \calF(\Pz^{\mathrm{st}}; C^\mu) \coloneqq \sum_{e\in\edgeset}[\cyclebasis^\mu]_e[\Force(\Pz^{\mathrm{st}})]_e = \innerprod{\cyclebasis^\mu}{\Force(\Pz^{\mathrm{st}})}.
\end{align}
The cyclic currents and conjugate affinities lead to the decomposition of the steady-state EPR into contributions from individual cycles~\cite{schnakenberg1976network}:
\begin{align}
    \EPR(\Pz^{\mathrm{st}}) &= \bigg\langle\sum_{\mu\in\mathcal{M}}\calJ(\Pz^\mathrm{st}; C^\mu)\cyclebasis^\mu,
    \Force(\Pz^\mathrm{st})\bigg\rangle \notag\\
    &= \sum_{\mu\in\mathcal{M}}\calJ(\Pz^{\mathrm{st}}; C^\mu)\innerprod{\cyclebasis^\mu}{\Force(\Pz^\mathrm{st})} \notag\\
    &=\sum_{\mu\in\mathcal{M}}\calJ(\Pz^{\mathrm{st}}; C^\mu)\calF(\Pz^{\mathrm{st}}; C^\mu).
    \label{EPR_cyclic_decomposition}
\end{align}

Ref.~\cite{yoshimura2023housekeeping} generalized this formula to nonstationary situations, giving a similar expression for the housekeeping EPR. For convenience, we write the housekeeping current and the housekeeping force as follows:
\begin{gather}
    \Currenthk(\Pz) = \Current(\Pz) - \Current^*(\Pz),\;\Forcehk(\Pz) = \Force(\Pz) - \Force^*(\Pz).
\end{gather}
Since $\div\Currenthk (\Pz) = \bm{0}$ and $\Currenthk (\Pz) \in {\rm ker} \div$, a decomposition similar to Eq.~\eqref{steady_current_cyclic_decomp} is obtained for the housekeeping current: 
\begin{align}
    \Currenthk(\Pz) = \sum_{\mu\in\mathcal{M}}\calJ^{\mathrm{hk}}(\Pz; C^\mu)\cyclebasis^\mu
    \label{housekeeping_cyclic_decomp}
\end{align}
with coefficients $\{\calJ^\mathrm{hk}(\Pz; C^\mu)\}_{\mu\in\mathcal{M}}$. The cycle affinity $\calF^{\mathrm{hk}}(\Pz; C^\mu)$ conjugate to $\calJ^{\mathrm{hk}}(\Pz; C^\mu)$ is defined as
\begin{align}
    \calF^{\mathrm{hk}}(\Pz; C^\mu) \coloneqq \sum_{e\in\edgeset}[\cyclebasis^\mu]_e[\Forcehk(\Pz)]_e = \innerprod{\cyclebasis^\mu}{\Forcehk(\Pz)}. \label{housekeeping_cycle_affinity}    
\end{align}
Therefore, $\EPRhk$ can always be decomposed into contributions from each cycle, even in a nonstationary regime~\cite{yoshimura2023housekeeping}:
\begin{align}
    \EPRhk(\Pz) = \sum_{\mu\in\mathcal{M}}\calJ^{\mathrm{hk}}(\Pz; C^\mu)\calF^{\mathrm{hk}}(\Pz; C^\mu). \label{EPRhk_cyclic_decomposition}
\end{align}
This expression is a generalization of Eq.~\eqref{EPR_cyclic_decomposition} in Schnakenberg network theory. 

\subsubsection{Wasserstein geometry}
\label{subsubsec:Wasserstein_geometry}
This section briefly reviews the connection between the Wasserstein geometry and the geometric housekeeping-excess decomposition. 

The Wasserstein distance was originally introduced to quantify the cost to transport one distribution to another in a continuous space~\cite{villani2008optimal}.
Extensions compatible with dynamics with discrete degrees of freedom have been accomplished in recent works~\cite{maas2011gradient,yoshimura2023housekeeping}. 
The extension, which we simply call the $2$-Wasserstein distance, is defined by the following minimization problem: for two probability distributions $\Pz^{(0)}$ and $\Pz^{(1)}$, we define 
\begin{align}
    \mathcal{W}(\Pz^{(0)}, \Pz^{(1)}) = \inf_{(\Pz(t), \Force'(t))_{0\leq t \leq 1}}\left(\int_0^1 \norm{\Force'(t)}^2_{L(\Pz(t))}dt\right)^{1/2}, \label{Wasserstein_distance}
\end{align}
subject to 
\begin{align}
    d_t\Pz(t) = \div L(\Pz(t))\Force'(t),~\Pz(0) = \Pz^{(0)},~\Pz(1) = \Pz^{(1)}.
    \label{continuity_condition}
\end{align}
If there is no pair of $\Pz(t)$ and $\Force'(t)$ that satisfies Eq.~\eqref{continuity_condition} for a given $(\Pz^{(0)},\Pz^{(1)})$, we may let $\mathcal{W}(\Pz^{(0)}, \Pz^{(1)}) =\infty$.
We can equivalently express this $2$-Wasserstein distance with the integration interval $[0, \tau]$ by performing the change of variables $t \to t /\tau$ for arbitrary positive values of $\tau$. The minimization problem is rewritten as
\begin{align}
    \mathcal{W}(\Pz^{(0)}, \Pz^{(1)}) = \inf_{(\Pz(t), \Force'(t))_{0\leq t \leq \tau}}\left(\tau\int_0^\tau \norm{\Force'(t)}^2_{L(\Pz(t))}dt\right)^{1/2},
\end{align}
subject to
\begin{align}
    d_t\Pz(t) = \div L(\Pz(t))\Force'(t),~\Pz(0) = \Pz^{(0)},~\Pz(\tau) = \Pz^{(1)}.
\end{align}

It should be pointed out that the $2$-Wasserstein distance defined by Eqs.~\eqref{Wasserstein_distance} and~\eqref{continuity_condition} vanishes under the large rate limit. We clarify this fact in the following way.
First, let us rewrite the minimization problem [Eqs.~\eqref{Wasserstein_distance} and~\eqref{continuity_condition}] in terms of the current:
\begin{align}
    \mathcal{W}(\Pz^{(0)}, \Pz^{(1)}) = \inf_{(\Pz(t), \Current'(t))_{0 \leq t \leq 1}}\left(\int_0^1 \norm{\Current'(t)}^2_{L(\Pz(t))^{-1}}dt\right)^{1/2}, 
\end{align}
subject to 
\begin{align}
    d_t\Pz(t) = \div\Current'(t),~\Pz(0) = \Pz^{(0)},~\Pz(1) = \Pz^{(1)}.
\end{align}
Here, we introduced a norm as $\norm{\Current'(t)}_{L(\Pz(t))^{-1}} \coloneqq \sqrt{(\Current'(t))^\top L(\Pz(t))^{-1}\Current'(t)}$, which is a norm because $L(\Pz(t))^{-1}$ is positive definite.
For simplicity, let us consider the situation where the rates that define $\mathcal{W}$ are parameterized by a dimensionless parameter $\lambda$ as $R_e=\lambda r_e$ with reference rates $r_e$ independent of $\lambda$ (note that $\mathcal{W}$ is defined using a matrix-valued function $L$ that refers to some rates $R_e$). 
Then, if we write the Onsager coefficient matrix as $L_\lambda$ when $R_e=\lambda r_e$, we can obtain the simple relation $L_\lambda(\Pz(t))^{-1}=\lambda^{-1} L_1(\Pz(t))^{-1}$, which leads to 
\begin{align}
    \mathcal{W}_\lambda(\Pz^{(0)},\Pz^{(1)}) = \lambda^{-1/2}\mathcal{W}_1(\Pz^{(0)},\Pz^{(1)}), 
\end{align}
where $\mathcal{W}_\lambda(\Pz^{(0)},\Pz^{(1)})$ indicates the 2-Wasserstein distance at the given value of $\lambda$. 
Therefore, if the rates are simultaneously increased, the 2-Wasserstein distance decreases and approaches zero. 
% Then, consider a case where a representative timescale $T>0$ exists and the transition rates scale as $\min_{e\in\edgeset}R_e = O(1/T)$.
% In this case, since we have $\max_{e\in\edgeset}l^{-1}_e = O(T)$, the limit $T\to 0$ yields $\norm{\Current'}^2_{L^{-1}} = 0$ for any vector $\Current'$ independent of $T$, which in turn gives $\mathcal{W}(\cdot, \cdot) = 0$.

The connection between the $2$-Wasserstein distance and the geometric housekeeping-excess decomposition is discussed in terms of the instantaneous speed in the space of the $2$-Wasserstein distance. The instantaneous speed is defined as 
\begin{align}
    v(\Pz(t)) \coloneqq \lim_{\Delta t\to + 0}\frac{\mathcal{W}(\Pz(t), \Pz(t + \Delta t))}{\Delta t}.
\end{align}
Ref.~\cite{yoshimura2023housekeeping} shows that the square of the instantaneous speed is equivalent to the excess EPR, i.e., $v(\Pz(t))^2 = \EPRex(\Pz(t))$. Therefore, the instantaneous speed $v(\Pz(t))$ is always well-defined. The relation $v(\Pz(t))^2 = \EPRex(\Pz(t))$ is useful to obtain the thermodynamic speed limit~\cite{yoshimura2023housekeeping, van2021geometrical},
\begin{align}
    \frac{\mathcal{W}(\Pz(0), \Pz(\tau))^2}{\tau} \leq \int_0^\tau\EPRex(\Pz(t))dt. \label{TSL}
\end{align}
This inequality indicates that the lower bound on the integrated excess EPR (i.e., the excess entropy production) comes from squaring the distance between the initial and final distributions and dividing by the time interval. We can interpret the speed limit as follows: The speed required to change the system's state is limited by dissipation due to nonstationarity.

\section{Geometric approach in information thermodynamics}
\label{sec:geometric_information_thermodynamics}
In this section, we offer a geometric interpretation of the conventional theory of information thermodynamics for bipartite systems. This reformulation is more than just a mathematical rewrite. It refines the conventional formulation and establishes a foundation for a deeper understanding of information thermodynamics, which will be illustrated by the main results.

Let us introduce the setup and notation. We restrict our focus to a situation where the system $Z$ consists of two subsystems $X$ and $Y$ so that we can discuss information theory. The states of $X$ and $Y$ are denoted by $x$ and $y$, respectively, so we write the state of $Z$ as $z = (x, y)$. Let $\calX = \{x_1, \ldots, x_{|\calX|}\}$ and $\calY = \{y_1, \ldots, y_{|\calY|}\}$ be the sets of states of $X$ and $Y$, respectively. 
Then the set of states of $Z$ is given by $\mathcal{Z} = \calX\times\calY$. The joint probability of $(x, y)$ is denoted by $\pxy(x, y)$, and we define a vector $\Pxy\in \bbR^{|\calX\times\calY|}$ by $[\Pxy]_{z=(x, y)} = \pxy(x, y)$. 
The marginal distributions are given by $\px(x) = \sum_{y\in \calY}\pxy(x, y)$ and $\py(y) = \sum_{x\in \calX}\pxy(x, y)$. Here, we define the marginalization matrices $\Pi_X\in \bbR^{|\calX|\times|\calX\times\calY|}$ and $\Pi_Y\in \bbR^{|\calY|\times|\calX\times\calY|}$ as $[\Pi_X]_{x', z=(x, y)} = \delta_{x', x}$ and $[\Pi_Y]_{y', z=(x, y)} = \delta_{y', y}$, respectively. Using these matrices, we can express the marginal distributions as follows:
\begin{align}
    \Px = \Pi_X\Pxy,\quad \Py = \Pi_Y\Pxy,
\end{align}
where $[\Px]_x = \px(x)$ and $[\Py]_y = \py(y)$. We use the index $\alpha \in \{ X, Y \}$ to denote either $X$ or $Y$. For example, the above equations can be  written as $\Pxy_{\alpha} = \Pi_{\alpha}\Pxy$.
We also write the other subsystem as $\alphac \in \{ X, Y \}$ (thus, $X^{\rm c} =Y$ and $Y^{\rm c} =X$). We call $\alphac$ the complementary system of $\alpha$. 

\subsection{Bipartite systems}
\label{subsec:bipartite}
We discuss bipartite systems~\cite{barato2013information, hartich2014stochastic, horowitz2014thermodynamics}, which are the simplest setup for studying information thermodynamics based on Markov jump processes.  Bipartite systems are introduced through the following condition on transition rates:
\begin{align}
    R^{(\nu)}((x, y)|(x', y')) = 0
\end{align}
if $x\neq x'$ and $y\neq y'$. That is,  simultaneous transitions in subsystems $X$ and $Y$ are prohibited in the dynamics of bipartite systems.
In other words, jumps can induce a change in either $\calX$ or $\calY$. Thus, when the bipartite system is designated by a directed graph $G = (\calX\times\calY, \edgeset)$, the set of edges $\edgeset$ can be partitioned into two sets $\edgeset_X$ and $\edgeset_Y$ as
\begin{align}
    \begin{split}
        \edgeset_X = \left\{\left(\nu, 
        (
            x, y) 
        \to
        (
            x', y') \right)
        \in\edgeset\mid x \neq x', y = y'\right\},\\
        \edgeset_Y = \left\{
        \left(\nu, 
        (
            x,y )
        \to
        (
            x', y') \right)
        \in\edgeset\mid x = x', y \neq y'\right\},
    \end{split} \label{bipartite_def1}
\end{align}
which satisfy $\edgeset= \edgeset_X\cup\edgeset_Y $ and $\edgeset_X\cap\edgeset_Y = \emptyset$. 
We also split the set of all the edges $\edgeset^{\mathrm{all}}$ into $\edgeset^\mathrm{all}_\alpha = \bigcup_{e\in\edgeset_\alpha}\{e, -e\}$ with $\alpha \in \{ X, Y \}$ (i.e., $\edgeset^\mathrm{all}_X\cup\edgeset^\mathrm{all}_Y = \edgeset^\mathrm{all}$ and $\edgeset^\mathrm{all}_X\cap\edgeset^\mathrm{all}_Y = \emptyset$).

For the sake of simplicity, we label edges by an index $i \in \{1, \dots, |\edgeset_X| + |\edgeset_{Y}| \}$ as $e_i \in \edgeset$ such that $\edgeset_X = \{e_1, \ldots, e_{|\edgeset_X|}\}$ and $\edgeset_Y = \{e_{|\edgeset_X| + 1}, \ldots, e_{|\edgeset_X| + |\edgeset_{Y}|}\}$ without loss of generality. 
We also assume that the $i$th row of a matrix (possibly a vector) with $|\edgeset|$ rows corresponds to $e_i$.
Then, it is split vertically, where the upper components correspond to $\edgeset_X$ and the lower to $\edgeset_Y$ (examples follow soon). 
Similarly, assuming the correspondence in matrices with $|\edgeset|$ columns, we partition them horizontally, where a left component corresponds to $\edgeset_X$ and a right component corresponds to $\edgeset_Y$.

\begin{figure}
    \centering
    \includegraphics[width=\linewidth]{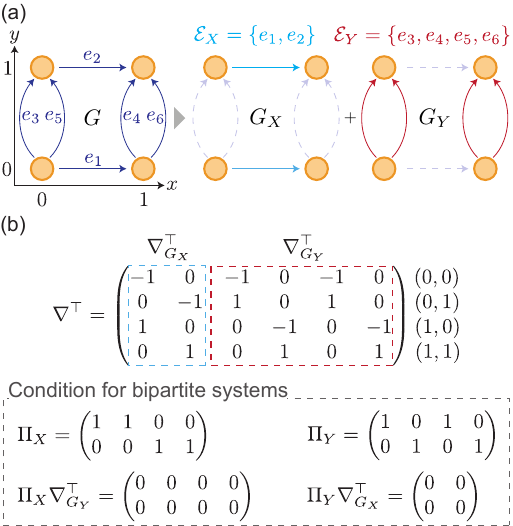}
    \caption{(a) An example of a graph representing a bipartite system, which is structurally the same as the graph $G$ presented in Fig.~\ref{fig:graph_incidencematrix}. 
    Let $\calX = \{0, 1\}$ and $\calY = \{0, 1\}$. 
    The original states $\{1,2,3,4\}$ are identified with $\{(0,0),(0,1),(1,0),(1,1)\}$. 
    In this setting, the edges are split into $\edgeset_X = \{e_1, e_2\}$ and $\edgeset_Y = \{e_3, e_4, e_5, e_6\}$, and the original graph $G$ is decomposed into the subgraphs $G_X = (\calX\times\calY, \edgeset_X)$ and $G_Y = (\calX\times\calY, \edgeset_Y)$. (b) Corresponding to the decomposition of the graph, the incidence matrix $\div$ is partitioned into $\div_{G_{X}}$ and $\div_{G_{Y}}$. We also show the marginalization matrix $\Pi_{\alpha}$, and the condition for bipartite systems $\Pi_{\alpha}\div_{G_{\alpha^{\rm c}}} = O$ for $\alpha \in \{X, Y\}$.}
    \label{fig:decomposed_graphs}
\end{figure}

We can find a special form of the discrete continuity equation [Eq.~\eqref{continuity_eq}] in bipartite systems. Note that the decomposition $\edgeset = \edgeset_X\cup\edgeset_Y$ leads to the two graphs $G_X = (\calX\times\calY, \edgeset_X)$ and $G_Y = (\calX\times\calY, \edgeset_Y)$ (see also Fig.~\ref{fig:decomposed_graphs}(a) for an example). It further provides us with the corresponding incidence matrices $\div_{G_X}$ and $\div_{G_Y}$ to $G_X$ and $G_Y$, respectively. The incidence matrix of the original graph $G$ can be reconstructed as
\begin{align}
    \div = (\div_{G_X}~\div_{G_Y}),
    \label{divergence_decomposition}
\end{align}
because $[\div]_{z, e}$ is equal to $[\div_{G_\alpha}]_{z, e'}$ when the edge corresponding to $e' \in \edgeset_{\alpha}$ is the same as the edge corresponding to the index $e \in \edgeset$ (see also Fig.~\ref{fig:decomposed_graphs}(b)). 

The current $\Current(\Pxy)$ is also decomposed into $\Current_X(\Pxy)$ and $\Current_Y(\Pxy)$ 
where the vector $\Current_X(\Pxy) = (J_{e_1}(\Pxy), \ldots, J_{e_{|\edgeset_X|}}(\Pxy))^\top$ is the vector of currents associated with $\edgeset_X$, and the vector $\Current_Y(\Pxy) = (J_{e_{|\edgeset_X|+1}}(\Pxy), \ldots, J_{e_{|\edgeset_X|+|\edgeset_Y|}}(\Pxy))^\top$ is the vector of currents associated with $\edgeset_Y$ (the ``vertical partition''). 
In particular, we use a symbol $\oplus$ to express the recomposition of partitioned vectors; for example, we express the current $\Current(\Pxy)$ as $\Current(\Pxy)=\Current_X (\Pxy)\oplus \Current_Y(\Pxy)$. Generally, the symbol $\oplus$ represents the operation of creating a column vector $(v_{e_1}, \dots, v_{e_{|\edgeset_X| + |\edgeset_Y|}})^{\top} \in \mathbb{R}^{|\edgeset_X| + |\edgeset_Y|}$ from the column vectors $(v_{e_1}, \dots, v_{e_{|\edgeset_X|}})^{\top} \in \mathbb{R}^{|\edgeset_X|}$ and $(v_{e_{|\edgeset_X|+1}}, \dots, v_{e_{|\edgeset_X|+|\edgeset_Y|}})^{\top} \in \mathbb{R}^{|\edgeset_Y|}$. 

With the partition of $\div$ and $\Current(\Pxy)$, the time evolution of the joint probability distribution $\Pxy$ governed by Eq.~\eqref{continuity_eq} can be rewritten as
\begin{align}
    d_t\Pxy &= \div \Current(\Pxy) 
    =(\div_{G_X}~\div_{G_Y})
    \begin{pmatrix}
        \Current_X(\Pxy)\\
        \Current_Y(\Pxy)
    \end{pmatrix}\notag\\
    &= \div_{G_X}\Current_X(\Pxy) + \div_{G_Y}\Current_Y(\Pxy).\label{continuity_eq_XY}
\end{align}
This equation can be regarded as a discrete version of the two-dimensional continuity equation.

The marginal distributions also obey a characteristic equation of motion. For an edge $e = (\nu(e), (x, y) \to (x', y'))$, we introduce the notations $\startx(e) = x$, $\targetx(e) = x'$, $\starty(e) = y$, and $\targety(e) = y'$. Using these notations, the $(x, e)$-component of $\Pi_X\div_{G_Y}$ for $e\in\edgeset_Y$ is calculated as
\begin{align}
    &[\Pi_X\div_{G_Y}]_{x, e}
    = \sum_{(x', y) \in\calX\times\calY}[\Pi_X]_{x, (x', y)}[\div_{G_Y}]_{(x', y), e} \notag\\
    &= \sum_{(x', y)\in\calX\times\calY}\delta_{x, x'}(\delta_{(x', y), \target(e)} - \delta_{(x', y), \start(e)}) \notag\\
    &= \sum_{(x', y)\in\calX\times\calY}\delta_{x, x'}(\delta_{x', \targetx(e)}\delta_{y, \targety(e)} - \delta_{x', \startx(e)}\delta_{y, \starty(e)}) \notag\\
    &=\sum_{y\in\calY}(\delta_{x, \targetx(e)}\delta_{y, \targety(e)} - \delta_{x, \startx(e)}\delta_{y, \starty(e)}) \notag\\
    &= \delta_{x, \targetx(e)} - \delta_{x, \startx(e)}. \label{cross_conservative}
\end{align}
For the third equality, we used the identities $\delta_{(x', y), \start(e)} = \delta_{x', \startx(e)}\delta_{y, \starty(e)}$ and $\delta_{(x', y),\target(e)} = \delta_{x', \targetx(e)}\delta_{y, \targety(e)}$. Because $\targetx(e) = \startx(e)$ for $e\in\edgeset_Y$, we obtain $\delta_{x, \targetx(e)} = \delta_{x, \startx(e)}$ and $[\Pi_X\div_{G_Y}]_{x, e}=0$. Therefore, we find $\Pi_X\div_{G_Y} = O$, where $O$ indicates the zero matrix. Due to symmetry, we also have $\Pi_Y\div_{G_X} = O$ and thus $\Pi_{\alpha} \div_{G_{\alpha^{\rm c}}} = O$ for $\alpha \in \{X, Y \}$. Hence, the time evolution of the marginal distribution $\Pa$ for $\alpha \in \{X, Y \}$ is given by
\begin{align}
    d_t\Pa 
    &= \Pi_{\alpha}\div_{G_{\alpha}}\Current_{\alpha}(\Pz) + \Pi_{\alpha}\div_{G_{\alpha^{\rm c}}}\Current_{\alpha^{\rm c}}(\Pz)\notag\\
    &= \Pi_{\alpha}\div_{G_{\alpha}}\Current_{\alpha}(\Pz). 
    \label{partial_continuity_eq}
\end{align}
This equation can also be considered a discrete version of the continuity equation (see also Appendix~\ref{app:graph_projection} for a more precise explanation).

Let us see what happens in the steady state of a bipartite system. Because the current satisfies $d_t\Pxy^{\rm st} = \div_{G_X}\Current_X (\Pxy^{\rm st})+ \div_{G_Y}\Current_Y (\Pxy^{\rm st})= \bm{0}$, we have
\begin{align}
    \div_{G_X}\Current_X(\Pxy^{\mathrm{st}}) = -\div_{G_Y}\Current_Y(\Pxy^{\mathrm{st}}). \label{condition_ss1}
\end{align}
This equation shows that, in the steady state, the divergence of the $X$-component of the current is canceled by that of the $Y$-component globally. 
By using $\Pi_{\alpha} \div_{G_{\alpha^{\rm c}}} = O$ in Eq.~\eqref{condition_ss1}, we also obtain
\begin{align}
\Pi_{\alpha}\div_{G_{\alpha}}\Current_{\alpha}(\Pxy^{\mathrm{st}}) = \bm{0},\label{condition_ss2}
\end{align}
for $\alpha \in \{X, Y \}$.  
It shows that the inflow and outflow of current are balanced in each subsystem.

\subsection{Information flow}
\label{subsec:information_flow}

In this section, we introduce a thermodynamic measure of continuous information flow for bipartite systems called \textit{information flow}~\cite{parrondo2015thermodynamics}. First, we define the Kullback--Leibler divergence between probability distributions $\Pz = (p_1,\ldots, p_{|\mathcal{Z}|})^\top$ and $\bm{q} = (q_1,\ldots, q_{|\mathcal{Z}|})^\top$ as $D_{\mathrm{KL}}(\Pz\|\bm{q}) = \sum_{z}p_z\ln (p_z/q_z)$. 
Using the Kullback–Leibler divergence, we can introduce mutual information, which is a measure of dependence between two random variables. Mutual information between the random variable for $X$ at time $s$, denoted by $\hat{X}_s$, and the random variable for $Y$ at time $t$, denoted by $\hat{Y}_t$, is defined as
\begin{align}
    I(\hat{X}_s; \hat{Y}_t) = D_{\mathrm{KL}}(\bm{P}(s, t)\| \Px(s)\otimes\Py(t)).
\end{align}
Here, $[\bm{P}(s, t)]_{(x, y)} \coloneqq P(x; s, y; t)$ is the joint probability of the state $x$ at time $s$ in the system $X$ and the state $y$ at time $t$ in the system $Y$, and $[\Px(s)\otimes\Py(t)]_{(x, y)} \coloneqq p_X(x; s)p_Y(y; t)$ is given by the marginal distributions $p_X(x; s) = \sum_y P(x; s, y; t)$ and $p_Y(y; t) = \sum_x P(x; s, y; t)$. Let the distribution $\Pz (t)$ be $\Pz (t) =\bm{P}(t, t)$.
Finally, information flow from system $Y$ to $X$ and that from $X$ to $Y$ are defined as
\begin{align}
    \begin{split}
        \iflow_X(\Pz(t)) = \lim_{\Delta t \to +0}\frac{I(\hat{X}_{t+\Delta t }; \hat{Y}_t) - I(\hat{X}_t; \hat{Y}_t)}{\Delta t },\\
        \iflow_Y(\Pz(t)) = \lim_{\Delta t \to +0}\frac{I(\hat{X}_t; \hat{Y}_{t+\Delta t }) - I(\hat{X}_t; \hat{Y}_t)}{\Delta t }, 
    \end{split} \label{information_flow}
\end{align}
respectively~\cite{allahverdyan2009thermodynamic, horowitz2014thermodynamics}.
Intuitively, information flow is expressed in terms of the partial derivative of mutual information. The quantity $\iflow_{\alpha}$ measures the rate of change of mutual information due to transitions in the subsystem $\alpha \in \{ X, Y\}$. 
Alternatively, it can be interpreted as the rate at which the subsystem $\alpha$ learns about the other system $\alphac$, and is thus also referred to as the learning rate~\cite{hartich2014stochastic} or merely the change in mutual information between the two subsystems~\cite{sagawa2012fluctuation, still2012thermodynamics, ito2013information}.

We next introduce the explicit form of information flow for bipartite systems. We define the vectors of stochastic Shannon entropy $\bmh(\Pxy) \in \mathbb{R}^{|\calX\times\calY|}$, $\bmh_X(\Pxy) \in \mathbb{R}^{|\calX|}$ and $\bmh_Y(\Pxy) \in \mathbb{R}^{|\calY|}$ as
$[\bmh(\Pxy)]_{z=(x, y)} = -\ln \pxy(x, y)$, $[\bmh_{X}(\Pxy)]_x = -\ln p_{X}(x)$, and $[\bmh_{Y}(\Pxy)]_y = -\ln p_{Y}(y)$, respectively. Using the stochastic Shannon entropy, the vector of stochastic mutual information $\bm{i}(\Pz)\in\bbR^{|\mathcal{Z}|}$ is defined as
\begin{align}
    \bm{i}(\Pz) = -\bmh(\Pxy) + \Pi_X^{\top}\bmh_X(\Pxy) + \Pi_Y^{\top}\bmh_Y(\Pxy). \label{information_potential}
\end{align}
The $(x, y)$-component of $\bm{i}(\Pxy(t))$ is given by
\begin{align}
    [\bm{i}(\Pxy(t))]_{(x, y)} = \ln\frac{p(x, y; t)}{p_X(x; t)p_Y(y; t)}, \label{information_potential_component}
\end{align}
and hence its expectation value gives the mutual information when $t=s$:
\begin{align}
    (\Pz(t))^{\top}\bm{i} (\Pz(t))&=  \sum_{x, y}\pxy(x, y;t)\ln\frac{p(x, y; t)}{p_X(x; t)p_Y(y; t)} \nonumber \\
    &= I(\hat{X}_t; \hat{Y}_t).
\end{align}
Using $\bm{i}(\Pxy(t))$, we can express information flow as the inner product of the current and the gradient of stochastic mutual information, with respect to the subsystem $\alpha \in \{X, Y \}$ (see Appendix \ref{app:information_flow_potential} for the derivation): 
\begin{align}
    \iflow_{\alpha}(\Pz(t)) = \innerprod{\Current_{\alpha}(\Pz(t))}{\nabla_{G_{\alpha}}\potentialit(\Pz(t))}. \label{information_flow_potential}
\end{align}
An expression corresponding to Eq.~\eqref{information_flow_potential} in continuous systems has also been found in two-dimensional overdamped Langevin systems ~\cite{allahverdyan2009thermodynamic,
 horowitz2014second, rosinberg2016continuous, nakazato2021geometrical} where two-dimensional overdamped Langevin systems with independent noises can be regarded as bipartite systems.

The condition for bipartite systems enables us to split the time derivative of the mutual information into the sum of information flows:
\begin{align}
    d_tI(\hat{X}_t; \hat{Y}_t) = \iflow_X(\Pz(t)) + \iflow_Y(\Pz(t)). \label{iflow_derivative}
\end{align}
To derive Eq.~\eqref{iflow_derivative} from Eq.~\eqref{information_flow_potential}, we focus on the following identity
\begin{align}
    &\Pxy^{\top} d_t\potentialit(\Pxy) \notag\\
    &= \sum_{x, y}p(x, y)d_t(\ln p(x, y) - \ln p_X(x) - \ln p_Y(y)) \notag\\
    &= \sum_{x, y}d_t p(x, y) - \sum_{x} d_t p_X(x)- \sum_{y} d_t p_Y(y)\notag\\
    &= 0,
\end{align}
where we used $d_t\sum_{x, y}p(x, y) = 0$ and $d_t\sum_{a\in\mathcal{A}}p_\alpha(a) = 0$ for $(\alpha, \mathcal{A}) \in \{(X, \calX), (Y, \calY) \}$. Using this identity, we can obtain Eq.~\eqref{iflow_derivative} as follows:
\begin{align}
    &d_tI(\hat{X}_t; \hat{Y}_t) \notag \\
    &= (d_t\Pxy(t))^{\top}\potentialit(\Pxy(t)) +(\Pxy(t))^{\top} d_t\potentialit(\Pxy(t))\notag\\
    &= [\div_{G_X}\Current_X(\Pxy(t)) + \div_{G_Y}\Current_Y(\Pxy(t))]^{\top}\potentialit(\Pxy(t))\notag\\
    &= \innerprod{\Current_X(\Pxy(t))}{\nabla_{G_X}\potentialit(\Pxy(t))} + \innerprod{\Current_Y(\Pxy(t))}{\nabla_{G_Y}\potentialit(\Pxy(t))}\notag\\
    &= \iflow_X(\Pxy(t)) + \iflow_Y(\Pxy(t)).
    \label{calculationmi}
\end{align}

We discuss information flow in the steady state. In this case, $\left. d_t I(\hat{X}_t;\hat{Y}_t)\right|_{\Pz(t)=\Pz^{\rm st}} = 0$ holds because the steady state condition $d_t \Pz^{\rm st} =0$ leads to $\left. d_t I(\hat{X}_t;\hat{Y}_t)\right|_{\Pz(t)=\Pz^{\rm st}} =(d_t\Pxy^{\rm st})^{\top}\potentialit(\Pxy^{\rm st}) = 0$.
Therefore, Eq.~\eqref{iflow_derivative} implies that information flow in the steady state satisfies the relation
\begin{align}
    \iflow_X(\Pz^{\mathrm{st}}) = -\iflow_Y(\Pz^{\mathrm{st}}). \label{antisymmetry_information_flow}
\end{align}
This relation means that information flow can be non-zero in the steady state, where the time derivative of mutual information is zero. Therefore, information flow between $X$ and $Y$ can still exist in the steady state.

\subsection{The second law of information thermodynamics}
\label{subsec:second_law}
This section discusses the second law of information thermodynamics for the subsystems, which is a generalization of the second law of thermodynamics. 

First, we introduce the entropy change rate associated with a subsystem $\alpha \in \{X, Y \}$, which we write as $\epr_{\alpha}$. 
It is given by the sum of the entropy change rate in the subsystem $\alpha$, that is $d_tS^{\mathrm{sys}}_{\alpha}$, and the entropy change rate in the environment due to transitions in $\alpha$, that is $\dot{S}^{\mathrm{env}}_{\alpha}$, as 
\begin{align}
    \epr_{\alpha} = d_tS^{\mathrm{sys}}_{\alpha} + \dot{S}_{\alpha}^{\mathrm{env}}, 
\end{align}
where $S^{\mathrm{sys}}_{\alpha}$ and $\dot{S}^{\mathrm{env}}_{\alpha}$ are defined as 
\begin{gather}
    S^{\mathrm{sys}}_{\alpha} = -\sum_{a\in\mathcal{A}}{p}_{\alpha}(a)\ln p_{\alpha}(a),\\
    \dot{S}^{\mathrm{env}}_{\alpha} = \sum_{e\in\edgeset_{\alpha}}s_e(J^+_e-J^-_e),
\end{gather}
respectively for $(\alpha, \mathcal{A})\in\{(X, \mathcal{X}), (Y, \mathcal{Y})\}$. If the complementary system $\alphac$ does not exist, then $\sigma_{\alpha}$ corresponds to the EPR and is non-negative. However, due to the influence of the complementary system $\alphac$, $\sigma_{\alpha}$ can generally take negative values. This negativity appears to contradict the second law of thermodynamics, in which case the complementary system acts as Maxwell's demon. Nevertheless, the total system's EPR $\dot{\Sigma}(\Pz)$ remains non-negative, and the second law of thermodynamics is not violated.

Similarly to the expression for the EPR given by the inner product between the current and force vectors [Eq.~\eqref{eq:EPR_inner_prod}], we can find an expression for the entropy change rate.
We can clarify this fact by introducing the concept of apparent thermodynamic force for subsystem $\alpha$ as $\tilde{\Force}_\alpha \coloneqq \Forceth_\alpha + \nabla_{G_\alpha}\Pi^\top_\alpha\bmh_\alpha$, which yields the expression,
\begin{align}
    \epr_\alpha = \innerprod{\Current_\alpha}{\tilde{\Force}_\alpha}. \label{eq:ecr_inner_prod}
\end{align}
It can be shown as follows. By using Eq.~\eqref{partial_continuity_eq}, we obtain $d_tS^\mathrm{sys}_\alpha = (d_t\Pa)^{\top}\bmh_\alpha = (\Pi_\alpha\div_{G_\alpha}\Current_{\alpha})^{\top}\bmh_\alpha = \innerprod{\Current_{\alpha}}{\nabla_{G_\alpha}\Pi^\top_\alpha\bmh_\alpha}$ and $\dot{S}^\mathrm{env}_\alpha = \innerprod{\Current_{\alpha}}{\Forceth_{\alpha}}$, so we arrive at $d_tS^{\mathrm{sys}}_{\alpha} + \dot{S}^{\mathrm{env}}_\alpha = \innerprod{\Current_\alpha}{\nabla_{G_\alpha}\Pi^\top_\alpha\bmh_\alpha + \Forceth_{\alpha}} = \innerprod{\Current_\alpha}{\tilde{\Force}_\alpha}$.
We can intuitively say that $\tilde{\Force}_\alpha$ functions as the thermodynamic force for a subsystem $\alpha$.

It is worth noting that $\nabla_{G_\alpha}\Pi^\top_\alpha \eqqcolon \nabla_{g_\alpha}$ can be interpreted as the gradient operator on a graph $g_\alpha$ that reflects the structure of transitions in a subsystem $\alpha$.
The graph $g_\alpha$ is constructed using the \textit{graph projection method}, the details of which are explained in Appendix~\ref{app:graph_projection}.
This method provides a unified understanding of our framework.
For example, we find that $\nabla_{G_\alpha}\Pi^\top_\alpha\bmh_\alpha = \nabla_{g_\alpha}\bmh_\alpha$ in the apparent thermodynamic force is given by the gradient of the stochastic Shannon entropy for the marginal distribution $\Pa$.

Accounting for information flow $\iflow_{\alpha}$ from $\alphac$ to $\alpha$, the thermodynamic irreversibility of the subsystem $\alpha \in \{ X, Y\}$ can be properly quantified. Similar to the second law of thermodynamics, the following inequalities can be proved for the subsystems:
\begin{align}
    \epr_X\geq \iflow_X\;\text{and}\;\epr_Y\geq\iflow_Y.\label{information_second_law}
\end{align}
This type of inequality in a subsystem is generally referred to as the second law of information thermodynamics~\cite{parrondo2015thermodynamics}. Since $\iflow_{\alpha}$ can take negative values, $\epr_{\alpha}$ can also be negative. In this case, we can interpret that the complementary system acts as Maxwell's demon, and information flow compensates for the negative entropy changes in the subsystem $\alpha$ and its environment. Based on the second law of information thermodynamics, we can define the partial EPR $\EPR_{\alpha}$, which quantifies the thermodynamic irreversibility associated with subsystem $\alpha$, as
\begin{align}
    \EPR_{\alpha} = \epr_{\alpha} - \iflow_{\alpha} \geq 0. \label{partial_EPR}
\end{align}
Its non-negativity is due to the second law of information thermodynamics [Eq.~\eqref{information_second_law}]. 

We briefly outline the derivation of Eq.~\eqref{information_second_law}. First, we decompose the thermodynamic force $\Force (\Pz)$ as follows,
\begin{gather}
    \Force (\Pz)= \Force_X (\Pz) \oplus\Force_Y (\Pz),\\
    \Forceth = \Forceth_X \oplus\Forceth_Y,~
    \Forceit (\Pz) = \Forceit_X (\Pz)\oplus\Forceit_Y (\Pz),
\end{gather}
where $\Force_{\alpha}(\Pz)$ means the force corresponding to $\mathcal{E}_{\alpha}$ in the system $\alpha \in \{X, Y\}$ and $\Forceth_{\alpha}  = \Force_{\alpha} (\Pz) -\Forceit_{\alpha}(\Pz)$ with $\Forceit_{\alpha} (\Pz) = \nabla_{G_{\alpha}}\bmh$. 
%Using the notation and Eq.~\eqref{partial_continuity_eq}, we obtain $d_tS^\mathrm{sys}_\alpha = (d_t\Pa)^{\top}\bmh_\alpha = (\Pi_\alpha\div_{G_\alpha}\Current_{\alpha})^{\top}\bmh_\alpha = \innerprod{\Current_{\alpha}}{\nabla_{G_\alpha}\Pi^\top_\alpha\bmh_\alpha}$ and $\dot{S}^\mathrm{env}_\alpha = \innerprod{\Current_{\alpha}}{\Forceth_{\alpha}}$. 
We then use Eqs.~\eqref{information_flow_potential} and~\eqref{eq:ecr_inner_prod} to show the non-negativity of the EPR as follows:
\begin{align}
    \begin{split}
        \epr_{\alpha} - \iflow_\alpha &= \innerprod{\Current_\alpha}{\nabla_{G_\alpha}\Pi^\top_\alpha\bmh_\alpha + \Forceth_\alpha}\\
        &\quad- \innerprod{\Current_\alpha}{\nabla_{G_\alpha}(-\bmh + \Pi^\top_X\bmh_X + \Pi^\top_Y\bmh_Y)}
    \end{split} \notag\\
    &=\innerprod{\Current_\alpha}{\Forceth_\alpha + \nabla_{G_\alpha}\bmh} = \Force_\alpha^\top L_\alpha\Force_\alpha\geq 0, \label{nonnegativity_partial_EPR}
\end{align}
where we define the positive-definite diagonal matrix $L_{\alpha}\in\bbR^{|\edgeset_{\alpha}|\times|\edgeset_{\alpha}|}$ as $[L_{\alpha}]_{e', e} = l_e \delta_{e, e'}$ and used $\nabla_{G_\alpha}\Pi^\top_{\alphac}=O$, $\Force_\alpha=\Forceth_\alpha + \nabla_{G_\alpha}\bmh$ and the relation $\Current_\alpha = L_\alpha\Force_\alpha$. Similarly to Eq.~\eqref{geometric_EPR}, we can also express the partial EPR geometrically. Here, we define the inner product and norm with respect to $L_{\alpha}$ as follows: $\innerprod{\bm{v}}{\bm{v}'}_{L_{\alpha}} = \bm{v}^{\top}L_{\alpha}\bm{v}'$ and $\norm{\bm{v}}_{L_{\alpha}} = \sqrt{\innerprod{\bm{v}}{\bm{v}}_{L_\alpha}}$ for $\bm{v}, \bm{v}'\in\bbR^{|\edgeset_\alpha|}$. Using this norm, we obtain
\begin{align}
    \EPR_{\alpha} = \innerprod{\Current_{\alpha}}{\Force_{\alpha}}
    = \norm{\Force_{\alpha}}^2_{L_{\alpha}} \geq 0. \label{geometric_partial_EPR}
\end{align}

From Eq.~\eqref{geometric_partial_EPR}, we find that $\EPR_\alpha = 0$ is achieved if and only if $\Force_{\alpha} = \bm{0}$, and thus the equality condition of the second law of information thermodynamics~\eqref{information_second_law} is also given by $\Force_{\alpha} = \bm{0}$. When this equality holds, the relation $\Current_\alpha = L_\alpha\Force_\alpha = \bm{0}$ provides $\epr_{\alpha} =0$ and $\iflow_\alpha=0$. Thus, the equality condition is only achieved when information flow does not exist. We note that this fact does not prevent finite information transfer with zero partial entropy production. This situation can be viewed as an optimal information-thermodynamic engine, such as the Szilard engine~\cite{szilard1929, sagawa2014thermodynamic}. If we consider the situation where $\iflow_\alpha \propto \tau^{-1}$ and $\EPR_\alpha \propto \tau^{-2}$ for a time interval $\tau$, then its time integral $\int_0^{\tau} dt\iflow_\alpha$ could be finite but the partial entropy production $\int_0^{\tau}
 dt \EPR_\alpha$ can become zero in the limit $\tau \to \infty$. 

\section{Main results}
\label{sec:main_results}
%In this section, we present the main results, which are the decomposition of information flow and its consequences. 

In this section, we present our main result, a geometric decomposition of information flow, along with several consequences that follow from it. 
These results connect the two techniques presented in previous sections, that is, the geometric housekeeping-excess decomposition and the information thermodynamic framework. 
Although both have provided important insights into thermodynamic dissipation, they have so far been developed separately. 
In this section, we fill this gap by exploiting the geometric viewpoint introduced in Sec.~\ref{subsec:bipartite}. 

\subsection{Geometric housekeeping-excess decomposition of information flow}
\label{subsec:information_flow_decomposition}
We will decompose information flow into housekeeping and excess parts based on the decomposition of currents. 
When the system is bipartite, currents that appear in the EPR decomposition are partitioned into $X$ and $Y$ components. 
Recall that $\Current^*(\Pz)$ is the current that reproduces the dynamics as in Eq.~\eqref{Onsager_equation} and is generated by a conservative force as in Eq.~\eqref{constraints_Onsager_equation}, whereas the remainder $\Currenthk(\Pxy)=\Current(\Pxy)-\Current^*(\Pxy)$ consists of cyclic contributions. 
Then, they are decomposed as $\Current^*(\Pxy) = \Current^*_X(\Pxy)\oplus\Current^*_Y(\Pxy)$ and $\Currenthk(\Pxy) = \Currenthk_X(\Pxy)\oplus\Currenthk_Y(\Pxy)$ in bipartite systems. 
Given these decompositions, the continuity equation~\eqref{continuity_eq_XY} is rewritten as 
\begin{align}
    d_t\Pxy = \div_{G_X}\Current_X^*(\Pxy) + \div_{G_Y}\Current_Y^*(\Pxy). \label{excess_continuity_eq}
\end{align}
Because of the condition $\Pi_\alpha\div_{G_{\alphac}}=O$, the time evolution of the marginal probability distribution in Eq.~\eqref{partial_continuity_eq} can also be rewritten as
\begin{align}
    d_t\Pa = \Pi_\alpha\div_{G_\alpha}\Current^*_\alpha(\Pxy),\label{modified_partial_continuity_eq}
\end{align}
for $\alpha \in \{X, Y \}$.
Since $\nabla^{\top}\Currenthk(\Pxy) = \boldsymbol{0}$, the decomposition $\Currenthk(\Pxy) = \Currenthk_X(\Pxy)\oplus\Currenthk_Y(\Pxy)$ leads to the antisymmetric relation
\begin{align}
    \div_{G_X}\Currenthk_X(\Pxy) = -\div_{G_Y}\Currenthk_Y(\Pxy) \label{current_antisymmetry}.
\end{align}

Then, housekeeping and excess information flow are defined by 
\begin{align}
    \iflowhk_{\alpha}(\Pz) &= \innerprod{\Currenthk_{\alpha}(\Pz)}{\nabla_{G_{\alpha}}\potentialit(\Pz)}, \label{iflowhk}\\
    \iflowex_{\alpha}(\Pz) &= \innerprod{\Current^*_{\alpha}(\Pz)}{\nabla_{G_{\alpha}}\potentialit(\Pz)}, \label{iflowex}
\end{align}
for $\alpha \in \{X, Y \}$. Due to the relation $\Current_{\alpha}(\Pz) = \Currenthk_{\alpha}(\Pz)+\Current^*_{\alpha}(\Pz)$ and the explicit formula~\eqref{information_flow_potential}, we find that they actually decompose information flow: 
\begin{align}
\iflow_{\alpha}(\Pxy) = \iflowhk_{\alpha}(\Pxy) + \iflowex_{\alpha}(\Pxy).
\end{align}

\begin{comment}
Besides, from Eq.~\eqref{current_antisymmetry} we find that the housekeeping information flow satisfies the antisymmetric relation
\begin{align}
    \iflowhk_X(\Pz) = -\iflowhk_Y(\Pz), \label{iflowhk_antisymmetry}
\end{align}
reproducing the relation in Eq.~\eqref{antisymmetry_information_flow} fulfilled by the total information flow in the steady state. 
Combining this property with Eq.~\eqref{iflow_derivative} further implies that the time derivative of the mutual information is solely determined by the excess information flow (compare with Eq.~\eqref{iflow_derivative}) 
\begin{align}
    d_tI(\hat{X}_t;\hat{Y}_t) = \iflowex_X(\Pz(t)) + \iflowex_Y(\Pz(t)). \label{information_flow_derivative_excess}
\end{align}
\end{comment}

We briefly demonstrate how $\iflowhk_\alpha(\Pxy)$ and $\iflowex_\alpha(\Pxy)$ can be interpreted as ``housekeeping'' and ``excess'', respectively, by considering the following two scenarios.

When the system is in the steady state, since we have  $\Current^*_{\alpha}(\Pz^{\mathrm{st}}) = \bm{0}$, we can show that
\begin{align}
    \iflow_{\alpha}(\Pz^{\mathrm{st}}) = \iflowhk_{\alpha}(\Pz^{\mathrm{st}}),~\iflowex_{\alpha}(\Pz^{\mathrm{st}}) = 0, \label{eq:housekeeping_stationary} %housekeeping information flow for the stationary distribution
\end{align}
for $\alpha \in \{X, Y \}$; i.e., the housekeeping information flow coincides with the total information flow in the steady state. 
If the system is detailed balanced, we generally have $\Current^*(\Pz) = \Current(\Pz)$ and we obtain 
\begin{align}
    \iflowhk_{\alpha}(\Pz) = 0,~\iflow_{\alpha}(\Pz) = \iflowex_{\alpha}(\Pz),\label{conservative_information_flow}
\end{align}
for $\alpha \in \{X, Y \}$ and any $\Pz$, which means that the housekeeping information flow is absent if the system is detailed balanced. 

The physical relevance of the housekeeping and excess information flow can also be demonstrated by the fact that they generalize the properties of information flow.
From Eq.~\eqref{current_antisymmetry},
we find that the housekeeping information flow satisfies the antisymmetric relation for any distribution $\Pxy$:
\begin{align}
    \iflowhk_X(\Pxy) = -\iflowhk_Y(\Pxy).
    \label{iflowhk_antisymmetry}
\end{align}
According to Eq.~\eqref{eq:housekeeping_stationary}, the antisymmetric relation of steady-state information flow [Eq.~\eqref{antisymmetry_information_flow}] is reproduced from this equality if $\Pxy$ is chosen as $\Pxy^\mathrm{st}$. 
Combining this property with Eq.~\eqref{iflow_derivative} further implies that the time derivative of the mutual information is solely determined by the excess information flow: 
\begin{align}
    d_tI(\hat{X}_t;\hat{Y}_t) = \iflowex_X(\Pz(t)) + \iflowex_Y(\Pz(t)). \label{information_flow_derivative_excess}
\end{align}

Let us conclude this section with some additional comments on housekeeping and excess information.
Originally, information flow $\iflow_\alpha (\Pz)$ in Eq.~\eqref{information_flow} represents the change in mutual information caused by the time evolution in subsystem $\alpha \in \{X, Y\}$. From Eq.~\eqref{conservative_information_flow}, we see that it consists only of the excess contribution if the detailed balance condition is satisfied; if not, $\iflow_\alpha (\Pz)$ potentially consists of two distinct modes. 
Specifically, $\iflowhk_\alpha$ arises when $\Currenthk (\Pz)$ does not vanish.
As already explained, $\Currenthk (\Pz)$ represents the cyclic contribution in the whole dynamics, given as a sum of cyclic currents; therefore, we can interpret $\iflowhk_\alpha$ as quantifying the ``cyclic'' mode in the information flow.
Since the cyclic current leaves the total distribution unchanged, the housekeeping information flow does not change the mutual information [Eq.~\eqref{information_flow_derivative_excess}]. 
Nevertheless, $\iflowhk_X$ and $\iflowhk_Y$ have non-zero values, which are balanced as in Eq.~\eqref{iflowhk_antisymmetry}. 
Reproducing the antisymmetric relation for the total information flow in the steady state, this relation reveals that the housekeeping information flow provides the information flow required to maintain the nonequilibrium steady state. 

On the other hand, $\iflowex_\alpha(\Pz)$ can be interpreted as the conservative mode, since it arises from $\Current^* (\Pz)$, which is conjugate to a conservative force. Because $\Current^* (\Pz)$ gives the time evolution of $\Pz$ in the continuity equation [Eq.~\eqref{modified_continuity_eq}], this mode substantially changes the mutual information, and thus Eq.~\eqref{information_flow_derivative_excess} holds. In the steady state, the mutual information does not change over time. In fact, $\iflowex_\alpha$ also vanishes in the steady state.

\subsection{Generalization of the second law of information thermodynamics}
\label{subsec:generalization_second_law}
In this section, we derive a generalization of the second law of information thermodynamics. We show that the second law of information thermodynamics~\eqref{information_second_law} can be expressed as two inequalities, one corresponding to the housekeeping contribution and the other to the excess contribution.
 
%First, we discuss how to define the housekeeping and excess entropy change rates associated with a subsystem $\alpha \in \{X, Y\}$.
%For this purpose, we refer to the expressions of the housekeeping and excess EPRs given in Eqs.~\eqref{excessepr} and \eqref{eq:hk_alt}.
%In Eq.~\eqref{excessepr}, the excess EPR is defined as the sum of the time derivative of the Shannon entropy $d_tS^{\mathrm{sys}}(\Pxy)$ and the excess environmental entropy change $\dot{S}^{\mathrm{ex,env}}(\Pxy)$. 

First, we discuss how to define the excess entropy change rate of a subsystem $\alpha \in \{X, Y\}$.
In Eq.~\eqref{excessepr}, the excess EPR is characterized as the sum of the time derivative of the Shannon entropy $d_tS^{\mathrm{sys}}(\Pxy)$ and the excess environmental entropy change $\dot{S}^{\mathrm{ex,env}}(\Pxy)$, generalizing the definition of the total EPR [Eq.~\eqref{eq:epr_def}].

These formulas motivate us to define the excess entropy change rate associated with $\alpha$, $\sigma_\alpha^\mathrm{ex}(\Pxy)$, by replacing the Shannon entropy of $\Pxy$, the optimal current $\Current^*(\Pxy)$, and the gradient $\nabla$ with the Shannon entropy of $\Pa$, the subsystem component of the current $\Current_\alpha^*(\Pxy)$, and that of the gradient $\nabla_{G_\alpha}$, respectively. Thus, we define $\sigma_\alpha^\mathrm{ex}(\Pxy)$ as  
\begin{align}
    \epr_{\alpha}^{\mathrm{ex}}(\Pxy) &= d_tS^\mathrm{sys}_\alpha(\Pxy) + \dot{S}^\mathrm{ex,\,env}_\alpha(\Pxy) \label{excess_ECR},
\end{align}
where $\dot{S}^\mathrm{ex,\,env}_\alpha(\Pxy)$ is defined as 
\begin{align}
    \dot{S}^\mathrm{ex,\,env}_\alpha(\Pxy) = -\innerprod{\Current^*_\alpha(\Pxy)}{\nabla_{G_\alpha}\pseudoenergy(\Pxy)}.  
\end{align}
Using the definition of $d_tS^\mathrm{sys}_\alpha(\Pxy)$, $\epr_{\alpha}^{\mathrm{ex}}(\Pxy)$ is calculated as $\epr_{\alpha}^{\mathrm{ex}}(\Pxy) = \innerprod{\Current^*_\alpha(\Pxy)}{\nabla_{G_{\alpha}}(\Pi^\top_\alpha\bmh_\alpha-\pseudoenergy)}$.

Then, we define the housekeeping entropy change rate of subsystem $\alpha\in\{X, Y\}$, denoted by $\eprhk_\alpha$, by focusing on the definition of housekeeping EPR: $\EPRhk = \innerprod{\Current - \Current^*}{\Force - \Force^*}$ [Eq.~\eqref{eq:housekeeping_EPR}].
To this end, we decompose the projected force as $\Force^* = \Force^*_X\oplus\Force^*_Y$, and 
% regard $\tilde{\Force}_\alpha - \Force^*_\alpha$ as an apparent nonconservative component of the thermodynamic force. 
use the apparent thermodynamic force $\tilde{\Force}_\alpha$ instead of $\Force_\alpha$ as in Eq.~\eqref{eq:ecr_inner_prod}. 
Then we define $\eprhk_\alpha$ as
\begin{align}
    \eprhk_\alpha \coloneqq \innerprod{\Current_\alpha - \Current^*_\alpha}{\tilde{\Force}_\alpha - \Force^*_\alpha},
\end{align}
which can be interpreted as an apparent housekeeping dissipation of subsystem $\alpha$.

Let us examine some basic properties of $\eprex_\alpha(\Pxy)$ and $\eprhk_\alpha(\Pxy)$. 
If $\alpha$ represents the entire system and $\alphac$ does not exist, $\eprex_\alpha(\Pxy)$ and $\eprhk_\alpha(\Pxy)$ reproduce $\EPRex(\Pxy) (\geq 0)$ and $\EPRhk(\Pxy)  (\geq 0)$, respectively, because $\edgeset_{\alpha}=\edgeset$ and $\Pa=\Pxy$. 
However, in the presence of $\alphac$, the positivity of excess and housekeeping entropy change rates $\epr_{\alpha}^{\mathrm{ex}}(\Pxy)$ and $\epr_{\alpha}^{\mathrm{hk}}(\Pxy)$ is no longer guaranteed. In fact, they can generally be negative, as demonstrated by numerical calculations in Sec.~\ref{sec:example}. 
If the system is detailed balanced, we have $\Current_\alpha(\Pxy) = \Current^*_\alpha(\Pxy)$ and $-\nabla_{G_\alpha}\pseudoenergy = -\nabla_{G_\alpha}\energy = \Forceth_\alpha$, so that 
\begin{align}
    \eprhk_\alpha(\Pxy) = 0,\;\eprex_\alpha(\Pxy) = \epr_\alpha(\Pxy). \label{eq:ecr_detailed_balanced}
\end{align}
When the system is in the steady state $\Pxy^\mathrm{st}$, we obtain
\begin{align}
    \eprhk_\alpha(\Pxy^\mathrm{st}) = \epr_\alpha(\Pxy^\mathrm{st}),\;\eprex_\alpha(\Pxy^\mathrm{st}) = 0 \label{eq:ecr_steady_state}
\end{align}
because $\Current^*_\alpha(\Pxy^{\mathrm{st}})=\bm{0}$ and $\Force^*_\alpha(\Pxy^{\mathrm{st}})= \bm{0}$ are satisfied. 

We note that $\epr_\alpha(\Pxy)\neq \eprhk_\alpha(\Pxy)+\eprex_\alpha(\Pxy)$ in general. The following calculation confirms this fact:
\begin{align}
    &\eprhk_\alpha + \eprex_\alpha \notag\\
    &= \innerprod{\Current_\alpha - \Current^*_\alpha}{\tilde{\Force}_\alpha - \Force^*_\alpha} + \innerprod{\Current^*_\alpha}{\nabla_{G_{\alpha}}(\Pi^\top_\alpha\bmh_\alpha-\pseudoenergy)} \notag\\
    &= \innerprod{\Current_\alpha - \Current^*_\alpha}{\Forceth_\alpha - \Force^*_\alpha}
    + (d_t\Pa)^\top\bmh_\alpha - \innerprod{\Current^*_\alpha}{\nabla_{G_\alpha}\pseudoenergy}\notag\\  
    \begin{split}
        &= (d_t\Pa)^\top\bmh_\alpha + \innerprod{\Current_\alpha}{\Forceth_\alpha} \\
        &\quad - \innerprod{\Current_\alpha}{\Force^*_\alpha} +\innerprod{\Current^*_\alpha}{\Force^*_\alpha-\Forceth_\alpha - \nabla_{G_\alpha}\pseudoenergy}
    \end{split} \notag\\
    &= \epr_\alpha - \innerprod{\Current^*_\alpha}{\Forcehk_\alpha + \Forceth_\alpha + \nabla_{G_\alpha}\pseudoenergy} \notag\\
    &= \epr_\alpha - 2\innerprod{\Current^*_\alpha}{\Forcehk_\alpha},
\end{align}
where we used $\tilde{\Force}_\alpha = \Forceth_\alpha + \nabla_{G_\alpha}\Pi^\top_\alpha\bmh_\alpha$, $\Pi_\alpha\nabla_{G_\alpha}^{\top} (\Current_\alpha - \Current^*_\alpha)=0$, $d_t\Pa = \Pi_\alpha\div_{G_\alpha}\Current^*_\alpha$, $\innerprod{\Current_\alpha}{\Force^*_\alpha}= \innerprod{\Current_\alpha^*}{\Force_\alpha}$, $\epr_\alpha = (d_t\Pa)^\top\bmh_\alpha + \innerprod{\Current_\alpha}{\Forceth_\alpha}$, and $\Forceth_\alpha + \nabla_{G_\alpha}\pseudoenergy = \Force_\alpha + \nabla_{G_\alpha}(\pseudoenergy - \bmh) = \Forcehk_\alpha$. 
Due to the presence of the system $\alphac$, $\innerprod{\Current^*_\alpha}{\Forcehk_\alpha}$ cannot vanish in general, whereas the sum with $\innerprod{\Current^*_{\alphac}}{\Forcehk_{\alphac}}$ is always zero as 
\begin{align}
    \innerprod{\Current^*_\alpha}{\Forcehk_\alpha}
    + \innerprod{\Current^*_{\alphac}}{\Forcehk_{\alphac}}
    =\innerprod{\Current^*}{\Forcehk}
    =0. 
\end{align}

Given $\eprhk_\alpha$ and $\eprex_\alpha$ with $\alpha\in\{X, Y\}$, we can generalize the second law of information thermodynamics. Since both $\eprhk_\alpha$ and $\eprex_\alpha$ can take negative values, they do not quantify irreversibility of thermodynamic processes as $\EPRhk$ and $\EPRex$ do. On the other hand, combining $\eprhk_\alpha$ with $\iflowhk_\alpha$ and $\eprex_\alpha$ with $\iflowex_\alpha$ results in the following two inequalities:
\begin{gather}
    \eprhk_{\alpha}(\Pxy) \geq \iflowhk_{\alpha}(\Pxy), \label{information_second_law_hk}\\ 
    \eprex_{\alpha}(\Pxy) \geq \iflowex_{\alpha}(\Pxy) \label{information_second_law_ex}.
\end{gather}
These inequalities can be interpreted as generalizations of the second law of information thermodynamics [Eq.~\eqref{information_second_law}].

These two generalizations of the second law of information thermodynamics offer a fresh perspective on existing discussions in information thermodynamics. 
 According to the conventional second law of information thermodynamics [Eq.~\eqref{information_second_law}], the entropy change of a subsystem is lower bounded by information flow. The existence of Maxwell's demon is implied by a negative entropy change of the subsystem, which necessarily leads to the negativity of information flow.
Our results reveal that information flow has two components: housekeeping and excess information flows.
These components impose different bounds [Eqs.~\eqref{information_second_law_hk} and~\eqref{information_second_law_ex}].
Therefore, we can discuss the existence of Maxwell's demon with regard to housekeeping and excess dissipation separately, based on the negativity of the respective quantities $\eprhk_{\alpha}(\Pxy)$ and $\eprex_{\alpha}(\Pxy)$. 
We will refer to the former as the \textit{housekeeping demon} and the latter as the \textit{excess demon}.

Let us briefly consider the existence of the housekeeping and excess demons.
Interestingly, due to the antisymmetric relation $\dot{I}^{\rm hk}_X = - \dot{I}^{\rm hk}_Y$, only one of $\dot{I}^{\rm hk}_X$ and $\dot{I}^{\rm hk}_Y$ has a negative sign. The inequality [Eq.~\eqref{information_second_law_hk}] implies that, if $\eprhk_{\alpha}(\Pxy)$ is negative, $\dot{I}^{\rm hk}_{\alpha}$ is negative and $\dot{I}^{\rm hk}_{\alpha^{\rm c}}$ and $\eprhk_{\alpha^{\rm c}}(\Pxy)$ is positive. 
Therefore, the housekeeping demon can exist in only of the two subsystems, $X$ or $Y$.
In Sec.~\ref{sec:example}, we provide a numerical illustration of several cases in two examples: (i) no demons, $\sigma^{\rm hk}_{\alpha}>0$ and $\sigma^{\rm ex}_{\alpha}>0$, (ii) housekeeping demon only, $\sigma^{\rm hk}_{\alpha}<0$ and $\sigma^{\rm ex}_{\alpha}>0$, (iii) excess demon only, 
$\sigma^{\rm hk}_{\alpha}>0$ and $\sigma^{\rm ex}_{\alpha}<0$, and (iv) both demons, 
$\sigma^{\rm hk}_{\alpha}<0$ and $\sigma^{\rm ex}_{\alpha}<0$.

We provide the proofs for Eqs.~\eqref{information_second_law_hk} and~\eqref{information_second_law_ex}. First, recalling the definitions of $\potentialit$ [Eq.~\eqref{information_potential}] and $\iflowhk_{\alpha}$ [Eq.~\eqref{iflowhk}], we can perform the following calculation:
\begin{align}
    \eprhk_{\alpha} - \iflowhk_{\alpha} &= \innerprod{\Currenthk_{\alpha}}{\Forceth_{\alpha} - \Force^*_\alpha + \nabla_{g_\alpha}\bmh_\alpha} - \innerprod{\Currenthk_\alpha}{\nabla_{G_\alpha}\potentialit}\notag\\
    &= \innerprod{\Currenthk_\alpha}{\Forceth_\alpha - \Force^*_\alpha + \nabla_{G_\alpha}\bmh- \nabla_{G_\alpha}\Pi^\top_{\alphac}\bmh_{\alphac}} \notag\\
    &= \innerprod{\Currenthk_\alpha}{\Forcehk_\alpha},
\end{align}
where we used $\Force_\alpha = \Forceth_\alpha + \nabla_{G_\alpha}\bmh$, $\Forcehk_\alpha = \Force_\alpha - \Force^*_\alpha$, and $\nabla_{G_\alpha}\Pi^\top_{\alphac} = O$. From the linear relation $\Currenthk_\alpha = L_\alpha\Forcehk_\alpha$, we obtain $\eprhk_\alpha - \iflowhk_\alpha =(\Forcehk_\alpha)^{\top} L_\alpha \Forcehk_\alpha =\norm{\Forcehk_\alpha}^2_{L_\alpha} \geq 0$, which is the desired bound [Eq.~\eqref{information_second_law_hk}]. Next, using the definitions of $\potentialit$ [Eq.~\eqref{information_potential}] and $\iflowex_{\alpha}$ [Eq.~\eqref{iflowex}], we obtain
\begin{align}
    \eprex_{\alpha} - \iflowex_{\alpha} &= \innerprod{\Current^*_{\alpha}}{\nabla_{G_{\alpha}}(\Pi^\top_\alpha\bmh_\alpha-\pseudoenergy -\potentialit)} \notag\\
    &= \innerprod{\Current^*_{\alpha}}{\nabla_{G_{\alpha}}(\bmh-\pseudoenergy ) - \nabla_{G_\alpha}\Pi^\top_{\alphac}\bmh_{\alphac}} \notag\\
    &= \innerprod{\Current^*_{\alpha}}{\Force^*_{\alpha}} = \norm{\Force^*_{\alpha}}_{L_\alpha}^2 \geq 0,
\end{align}
where we used $\Force^*_\alpha = \nabla_{G_{\alpha}}(\bmh-\pseudoenergy )$, $\nabla_{G_\alpha}\Pi^\top_{\alphac} = O$ and $\Current^*_{\alpha}=L_{\alpha} \Force^*_{\alpha}$.
This inequality is equivalent to the desired bound [Eq.~\eqref{information_second_law_ex}].

Based on the inequalities [Eqs.~\eqref{information_second_law_hk} and~\eqref{information_second_law_ex}], we define the partial housekeeping EPR $\EPRhk_{\alpha}(\Pxy)$ and the partial excess EPR $\EPRex_{\alpha}(\Pxy)$ as
\begin{gather}
    \EPRhk_{\alpha}(\Pxy) \coloneqq \eprhk_{\alpha}(\Pxy) - \iflowhk_{\alpha}(\Pxy) = \norm{\Forcehk_{\alpha}(\Pxy)}_{L_\alpha(\Pxy)}^2, \label{partial_housekeeping_EPR}\\
    \EPRex_{\alpha}(\Pxy) \coloneqq \eprex_{\alpha}(\Pxy) - \iflowex_{\alpha}(\Pxy) = \norm{\Force^*_{\alpha}(\Pxy)}_{L_\alpha(\Pxy)}^2, \label{partial_excess_EPR}
\end{gather}
respectively. Since these quantities are non-negative, the quantities $\EPRhk_{\alpha}(\Pxy)$ and $\EPRex_{\alpha}(\Pxy)$ can be considered measures of dissipation. 

We note that the partial housekeeping EPR and the partial excess EPR satisfy the properties corresponding to Eqs.~\eqref{eq:ecr_detailed_balanced} and~\eqref{eq:ecr_steady_state}. If the system is detailed balanced, where $\Force = \Force^*$ holds, we have
\begin{align}
    \EPRhk_\alpha(\Pxy) = 0,\;\EPRex_\alpha(\Pxy) = \EPR_\alpha(\Pxy),
\end{align}
for any $\Pxy$. On the other hand, if the system is in a steady state, which is characterized by $\Force^*(\Pxy^\mathrm{st}) = \bm{0}$, we can show
\begin{align}
    \EPRhk_\alpha(\Pxy^\mathrm{st}) = \EPR_\alpha(\Pxy^\mathrm{st}),\;\EPRex_\alpha(\Pxy^\mathrm{st}) = 0.
\end{align}

%Here, $\EPRhk_\alpha$ quantifies the irreversibility due to nonconservative external driving forces that prevent the subsystem $\alpha$ from reaching equilibrium, and $\EPRex_\alpha$ quantifies the irreversibility arising from the nonstationarity in the subsystem $\alpha$.

The partial housekeeping EPR and partial excess EPR can be interpreted in the following way:
$\EPRhk_\alpha$ quantifies the irreversibility due to nonconservative external driving forces that prevent the subsystem $\alpha$ from reaching equilibrium, and $\EPRex_\alpha$ quantifies the irreversibility arising from the nonstationarity in the subsystem $\alpha$.

Because $[L_{\alpha}]_{e,e'} =l_e \delta_{e,e'}$ and $L(\Pz) = \diag(l_{e_1}(\Pz), l_{e_2}(\Pz), \dots ,l_{e_{|\edgeset|}}(\Pz))$, we obtain 
\begin{align}
    \EPRhk_{X}(\Pxy) + \EPRhk_{Y}(\Pxy)&= \norm{\Forcehk(\Pxy)}_{L(\Pxy)}^2=   \EPRhk(\Pxy), \label{additivity_partial_EPRhk}\\
    \EPRex_{X}(\Pxy) + \EPRex_{Y}(\Pxy) &=\norm{\Force^*(\Pxy)}_{L(\Pxy)}^2=   \EPRex(\Pxy). \label{additivity_partial_EPRex}
\end{align}
Therefore, the housekeeping and excess EPRs in the total system are decomposed into the partial components in the subsystems. 
In other words, adding two inequalities [Eq.~\eqref{information_second_law_hk} (Eq.~\eqref{information_second_law_ex}) for $\alpha=X$ and $\alpha=Y$]  $\eprhk_{X}(\Pxy) -\iflowhk_{X}(\Pxy)\geq 0$ ($\eprex_{X}(\Pxy) -\iflowex_{X}(\Pxy) \geq 0$) and $\eprhk_{Y}(\Pxy) -\iflowhk_{Y}(\Pxy)\geq 0$ ($\eprex_{Y}(\Pxy) -\iflowex_{Y}(\Pxy)\geq 0$) on both sides gives the non-negativity of the housekeeping (excess) EPR $\dot{\Sigma}^{\rm hk}(\Pxy) \geq 0$ ($\dot{\Sigma}^{\rm ex}(\Pxy) \geq 0$). This is analogous to how the second law of thermodynamics $\dot{\Sigma}(\Pxy) \geq 0$ can be derived from the two second laws of information thermodynamics  $\sigma_{X}(\Pxy) - \dot{I}_{X}(\Pxy) \geq 0$ and $\sigma_{Y} (\Pxy)- \dot{I}_{Y}(\Pxy) \geq 0$.

On the other hand, the partial EPRs are not reconstructed from the partial housekeeping and excess EPRs; i.e., $\EPRhk_{\alpha}(\Pxy) + \EPRex_{\alpha}(\Pxy) \neq \EPR_{\alpha}(\Pxy)$ for $\alpha \in \{ X, Y \}$ in general because $\sigma^{\rm hk}_{\alpha}(\Pxy)+\sigma^{\rm ex}_{\alpha}(\Pxy) \neq \sigma_{\alpha}(\Pxy)$ while $\iflow_{\alpha}(\Pxy) = \iflowhk_{\alpha}(\Pxy) + \iflowex_{\alpha}(\Pxy)$. 
Instead, we can provide a geometric decomposition of the partial EPRs based on the concept of local conservativeness, which will be discussed in Sec.~\ref{subsec:Wasserstein_geometry_subsystem}.

\subsection{Generalization of cyclic decomposition}
\label{subsec:generalization_cyclic_decomposition}
Here, we present the cyclic decomposition of the partial housekeeping EPR and the housekeeping information flow. 
This result generalizes the cyclic decomposition of the housekeeping EPR [Eq.~\eqref{EPRhk_cyclic_decomposition}] for subsystems. 
Moreover, it extends other information-thermodynamic results to nonstationary situations; the partial EPR~\cite{yamamoto2016linear} and the information flow~\cite{horowitz2014thermodynamics} are known to be expressed in the form of the cyclic decomposition in the steady state, as is the total EPR. We will provide a generalization of these results to non-steady states, with a focus on the housekeeping term introduced in preceding sections.

As a preliminary step to the cyclic decomposition, we categorize the cycles $\{C^\mu\}_{\mu\in\mathcal{M}}$
as either local or global. 
A cycle $C^\mu = \{e_{i(1)}, \ldots, e_{i(|C^\mu|)}\}$ ($\mu\in\mathcal{M}$) is defined to be $\alpha$-local ($\alpha \in \{X, Y \}$) if $C^\mu\subset\edgeset^\mathrm{all}_{\alpha}$ holds. Due to the definition, the edges in $C^\mu$ satisfy
\begin{align}
    \start_{\alphac}(e_{i(k)}) =\start_{\alphac}(e_{i(k+1)})
\end{align}
for any $k\in\{1,2,\dots,|C^\mu|\}$, in addition to the cyclic condition $\target_\alpha(e_{i(k)})=\start_\alpha(e_{i(k+1)})$.   Here, we impose the boundary condition $e_{i(|C^\mu|+1)} = e_{i(1)}$. 
That is, the state in $\alphac$ never changes in an $\alpha$-local cycle. 
We write the set of indices of $\alpha$-local cycles as $\mathcal{M}^\alpha\subset\mathcal{M}$ ($\alpha \in \{X, Y \}$). 
Global cycles are defined as cycles that are neither $X$-local nor $Y$-local, whose set of indices is denoted by $\mathcal{M}^{XY}$; thus, $\mathcal{M}=\mathcal{M}^X\cup\mathcal{M}^Y\cup\mathcal{M}^{XY}$. 

To demonstrate the cyclic decomposition, let us define the partial housekeeping cycle affinity of $C^\mu$ as follows:
\begin{align} 
    \calF^\mathrm{hk}_\alpha(\Pz; C^\mu) \coloneqq \sum_{e\in\edgeset_\alpha}[\cyclebasis^\mu]_e[\Forcehk (\Pxy)]_e,\label{partial_cycle_affinity}
\end{align}
for $\alpha\in\{X, Y\}$. 
The quantity $\calF^\mathrm{hk}_\alpha(\Pxy;C^\mu)$ resembles the housekeeping cycle affinity~\eqref{housekeeping_cycle_affinity}, but is not identical because the sum range is restricted by $\edgeset_\alpha$, resulting in a different value.
They coincide only if the cycle is $\alpha$-local; i.e., we have $\calF^\mathrm{hk}_\alpha(\Pxy;C^\mu) = \calF^\mathrm{hk}(\Pxy;C^\mu)$ only if $\mu\in\mathcal{M}^\alpha$. 
In the other cases, $\mu\in\mathcal{M}^{\alphac}$ or $\mu\in\mathcal{M}^{XY}$, 
we find $\calF^\mathrm{hk}_\alpha(\Pxy;C^\mu)\neq \calF^\mathrm{hk}(\Pxy;C^\mu)$ in general. In particular, $\calF^\mathrm{hk}_\alpha(\Pxy;C^\mu) = 0$ for $\mu\in\mathcal{M}^{\alphac}$ because then $[\cyclebasis^\mu]_e = 0$ for all $e\in\edgeset_\alpha$.

Using the partial housekeeping cycle affinity $\calF^\mathrm{hk}_\alpha(\Pz; C^\mu)$ and the housekeeping cyclic current $\calJ^\mathrm{hk}(\Pz; C^\mu)$ [Eq.~\eqref{housekeeping_cyclic_decomp}], we can decompose the partial housekeeping EPR of the subsystem $\alpha\in\{X, Y\}$ into the contributions from individual cycles:
\begin{align}
    \EPRhk_\alpha(\Pz) = \sum_{\mu\in\mathcal{M}^\alpha\cup\mathcal{M}^{XY}}\calJ^\mathrm{hk}(\Pz; C^\mu)\calF^\mathrm{hk}_\alpha(\Pxy;C^\mu). \label{partial_EPRhk_cyclic_decomposition}
\end{align}
This expression is a generalization of Eq.~\eqref{EPRhk_cyclic_decomposition} for $\EPRhk_\alpha$, and a generalization of the expression for the partial EPR in the steady state discussed in Ref.~\cite{yamamoto2016linear}. 

\begin{comment}
It is noteworthy that Eq.~\eqref{partial_EPRhk_cyclic_decomposition} can be considered as a cyclic decomposition for the projected graph $g_\alpha$. Precisely, $\calJ^\mathrm{hk}(\Pz;C^\mu)$ and $\calF^\mathrm{hk}_\alpha(\Pz;C^\mu)$ can be interpreted as the cyclic current and cycle affinity associated with a cycle in $g_\alpha$. The cycles in $g_\alpha$ are called \textit{projected cycles}. A concrete explanation of the concept of a projected cycle is provided in Appendix~\ref{app:projected_cycles}, and the proof of Eq.~\eqref{partial_EPRhk_cyclic_decomposition} is provided in Appendix~\ref{app:proof_cyclic_decomposition_1}.
\end{comment}

Then, we discuss the cyclic decomposition of the housekeeping information flow. To this end, we formally extend the concept of \textit{information affinity}, which was originally introduced by Horowitz and Esposito for the steady state~\cite{horowitz2014thermodynamics}, to include the non-steady state. More specifically, the information affinity along cycle $C^\mu$ is defined as follows: 
\begin{align}
    \calF^\mathrm{info}_\alpha(\Pz; C^\mu) = \sum_{e\in\edgeset_\alpha}[\cyclebasis^\mu]_e([\potentialit(\Pz)]_{\target(e)} - [\potentialit(\Pz)]_{\start(e)})
\end{align}
for $\alpha\in\{X, Y\}$.
Note that $\calF^\mathrm{info}_\alpha(\Pz^\mathrm{st}; C^\mu)$ coincides with the conventional definition.
We can interpret $\calF^\mathrm{info}_\alpha(\Pz; C^\mu)$ as describing the change in mutual information caused by the changes specified by the edges in $C^\mu\cap\edgeset^\mathrm{all}_\alpha$.
Although we formally defined $\calF^\mathrm{info}_\alpha(\Pz; C^\mu)$ for all the cycles in $G$, this quantity is zero for any local cycle: $\calF^\mathrm{info}_\alpha(\Pz; C^\mu) = 0$ holds for $\mu\in\mathcal{M}^X\cup\mathcal{M}^Y$ (see Appendix~\ref{app:proof_cyclic_decomposition_2}). Therefore, information affinity can only be non-zero within a global cycle in which information is exchanged between two subsystems.

The next step is to use the information affinity to derive the cyclic decomposition of the housekeeping information flow. Here, let us simply state the result (see also Appendix~\ref{app:proof_cyclic_decomposition_2} for the proof):
\begin{align}
    \iflowhk_{\alpha}(\Pz) = \sum_{\mu\in \mathcal{M}^{XY}}\calJ^{\mathrm{hk}}(\Pz; C^\mu)\calF^\mathrm{info}_\alpha(\Pz; C^\mu). \label{information_flow_cyclic_decomposition}
\end{align}
Therefore, $\iflowhk_\alpha(\Pz)$ is expressed solely by the contributions from global cycles $\{C^\mu\}_{\mu\in\mathcal{M}^{XY}}$. 
For the steady state, the representation of information flow by information affinity and its decomposition into global cycles was discovered by Horowitz and Esposito~\cite{horowitz2014thermodynamics}, and further analyzed in detail by Yamamoto et al.~\cite{yamamoto2016linear}. Therefore, Eq.~\eqref{information_flow_cyclic_decomposition} is regarded as a generalization of that result for the housekeeping information flow in a non-steady state.

To further characterize Eq.~\eqref{information_flow_cyclic_decomposition}, we define the housekeeping information flow along the global cycle $C^\mu$ ($\mu\in \mathcal{M}^{XY}$) as $\dot{\mathcal{I}}^{\mathrm{hk}}_{\alpha}(\Pz; C^\mu) = \calJ^{\mathrm{hk}}(\Pz; C^\mu)\calF^\mathrm{info}_\alpha(\Pz; C^\mu)$, which leads to the decomposition $\iflowhk_{\alpha}(\Pz) =\sum_{\mu\in \mathcal{M}^{XY}}\dot{\mathcal{I}}^{\mathrm{hk}}_{\alpha}(\Pz; C^\mu)$. We point out that $\dot{\mathcal{I}}^{\mathrm{hk}}_{\alpha}(\Pz; C^\mu)$ satisfies the antisymmetric relation at the level of individual cycles:
\begin{align}
    \dot{\mathcal{I}}^{\mathrm{hk}}_X(\Pz; C^\mu) = -\dot{\mathcal{I}}^{\mathrm{hk}}_Y(\Pz; C^\mu).
\end{align}
The antisymmetric relation can be verified as follows: 
\begin{align}
    &\dot{\mathcal{I}}^{\mathrm{hk}}_X(\Pz; C^\mu) + \dot{\mathcal{I}}^{\mathrm{hk}}_Y(\Pz; C^\mu) \notag\\
    &= \calJ^\mathrm{hk}(\Pz; C^\mu)(\calF^\mathrm{info}_X(\Pz; C^\mu) + \calF^\mathrm{info}_Y(\Pz; C^\mu)) = 0,
\end{align}
because 
\begin{align}
    &\calF^\mathrm{info}_X(\Pz; C^\mu) + \calF^\mathrm{info}_Y(\Pz; C^\mu) \notag\\
    &= \sum_{e\in\edgeset}[\cyclebasis^\mu]_e([\potentialit(\Pz)]_{\target(e)} - [\potentialit(\Pz)]_{\start(e)}) \notag\\
    &= \sum_{e\in\edgeset}[\cyclebasis^\mu]_e\sum_{(x, y)\in\calX\times\calY}[\nabla]_{e, (x, y)}[\potentialit(\Pz)]_{(x, y)}\notag\\
    &= (\div\cyclebasis^\mu)^\top\potentialit(\Pz) = 0,
\end{align}
where the last equality follows from $\div\cyclebasis^\mu = \bm{0}$.

\subsection{Short-time thermodynamic uncertainty relation}
\label{subsec:TUR}
We discuss a short-time thermodynamic uncertainty relation (TUR) for the partial excess EPR, which is a generalization of the short-time TUR for the excess EPR in Ref.~\cite{yoshimura2023housekeeping}.
We use the geometric representation of the partial excess EPR [Eq.~\eqref{partial_excess_EPR}] to obtain a TUR. 

To this end, we introduce the time-independent observable $\bm{\Psi}_\alpha\in\im\Pi^\top_\alpha$ with $\alpha\in\{X, Y\}$. Note that $[\bm{\Psi}_\alpha]_{(x, y)}$ depends only on the state of the subsystem $\alpha$. This fact can be confirmed as follows. By introducing $\bm{\psi}_\alpha\in\bbR^{|\mathcal{A}|}$ ($(\alpha, \mathcal{A})\in\{(X, \calX), (Y, \calY)\}$) such that $\bm{\Psi}_\alpha = \Pi^\top_\alpha\bm{\psi}_\alpha$, we have $[\bm{\Psi}_X]_{(x, y)} = [\bm{\psi}_X]_x$ and $[\bm{\Psi}_Y]_{(x, y)} = [\bm{\psi}_Y]_y$. 

To obtain a TUR, we consider the following Cauchy--Schwarz inequality with respect to the inner product $\innerprod{\cdot}{\cdot}_{L_\alpha(\Pxy)}$:
\begin{align}
    (\innerprod{\Force^*_\alpha(\Pxy)}{\nabla_{G_\alpha}\bm{\Psi}_\alpha}_{L_\alpha(\Pxy)})^2 \leq \norm{\Force^*_\alpha(\Pxy)}^2_{L_\alpha(\Pxy)}\norm{\nabla_{G_\alpha}\bm{\Psi}_\alpha}^2_{L_\alpha(\Pxy)}
    \label{cauchy-schawarz}
\end{align}
for any time-independent  observable $\bm{\Psi}_\alpha \in\im\Pi^\top_\alpha$. Because $d_t\Pa = \Pi_\alpha\div_{G_\alpha}\Current^*_\alpha (\Pxy)= \div_{g_\alpha}\Current^*_\alpha (\Pxy)$ holds, we can rewrite $\innerprod{\Force^*_\alpha(\Pxy)}{\nabla_{G_\alpha}\bm{\Psi}_\alpha}_{L_\alpha(\Pxy)}$ as
\begin{align}
    \innerprod{\Force^*_\alpha(\Pxy)}{\nabla_{G_\alpha}\bm{\Psi}_\alpha}_{L_\alpha(\Pxy)} &= \innerprod{\Current^*_\alpha (\Pxy)}{\nabla_{g_\alpha}\bm{\psi}_\alpha}\notag\\
    &= (d_t\Pa)^\top\bm{\psi}_\alpha \notag\\
    &= d_t(\Pa^\top\bm{\psi}_\alpha) \notag \\
    &= d_t(\Pxy^\top\bm{\Psi}_\alpha). \label{expected_value_change_rate}
\end{align}
From the geometric representation $\EPRex_\alpha = \norm{\Force^*_\alpha}^2_{L_\alpha}$ and Eqs.~\eqref{cauchy-schawarz} and \eqref{expected_value_change_rate}, we obtain the following TUR for the partial excess EPR:
\begin{align}
    \frac{(d_t(\Pxy^\top\bm{\Psi}_\alpha))^2}{\norm{\nabla_{G_\alpha}\bm{\Psi}_\alpha}_{L_\alpha(\Pxy)}^2}\leq \EPRex_\alpha(\Pxy). \label{TUR}
\end{align}
This inequality implies that $\EPRex_\alpha$ is bounded by the time derivative of the expected value $\Pxy^\top\bm{\Psi}_\alpha$, which quantifies the speed of the subsystem $\alpha \in\{X, Y\}$ measured by the observable $\bm{\Psi}_\alpha$.

%Additionally, Eq.~\eqref{TUR} may be applicable for inferring $\EPRex_\alpha$ if $L_{\alpha} (\Pxy)$ is known. Although we cannot know $L_{\alpha}(\Pxy)$, we can still evaluate the denominator $\norm{\nabla_{G_\alpha}\bm{\Psi}_\alpha}_{L_\alpha(\Pxy)}^2$ by considering its upper bound given by the diffusivity of $\bm{\Psi}_\alpha$. 

We can estimate the denominator of the lower bound in the TUR [Eq.~\eqref{TUR}] by considering its upper bound, which is the diffusivity of $\bm{\Psi}_\alpha$.

To clarify this notion, we introduce a stochastic change in the value of an arbitrary time-independent observable $\bm{\Phi}\in\bbR^{|\calX\times\calY|}$ over a sufficiently small $\Delta t (>0)$ as $\delta\bm{\Phi} \coloneqq [\bm{\Phi}]_{(x (t+\Delta t), y (t+\Delta t))} - [\bm{\Phi}]_{({x}(t), {y}(t))}$, where $x(t)$ and $y(t)$ denote the states of $X$ and $Y$ at time $t$, respectively. The diffusivity of $\bm{\Phi}$ is defined as
\begin{align}
    \mathcal{D}(\bm{\Phi}) \coloneqq \lim_{\Delta t\to +0}\frac{\mathrm{Var}[\delta\bm{\Phi}]}{\Delta t},
\end{align}
where $\mathrm{Var}[\delta\bm{\Phi}]$ is the variance of $\delta\bm{\Phi}$.
The upper bound is obtained as
\begin{align}
    \frac{\mathcal{D}(\bm{\Psi}_\alpha)}{2} \geq \norm{\nabla_{G_\alpha}\bm{\Psi}_\alpha}^2_{L_\alpha}. \label{diffusivity_inequality}
\end{align}
The proof of Eq.~\eqref{diffusivity_inequality} is provided in Appendix~\ref{app:diffusivity_inequality}.
The equality holds if and only if $J_e^+ = J_e^-$ for any $e\in\edgeset^\mathrm{all}_\alpha$ such that $[\nabla_{G_\alpha}\bm{\Psi}_\alpha]_e \neq 0$. Thus, we expect the values on both sides of Eq.~\eqref{diffusivity_inequality} to be close to each other near equilibrium.

Finally, combining Eqs.~\eqref{expected_value_change_rate} and~\eqref{diffusivity_inequality} with Eq.~\eqref{TUR} yields the following looser TUR:
\begin{align}
    \frac{2(d_t(\Pxy^\top\bm{\Psi}_\alpha))^2}{\mathcal{D}(\bm{\Psi}_{\alpha})} \leq \EPRex_{\alpha}. \label{TUR_2}
\end{align}
%Although this inequality is looser than Eq.~\eqref{TUR}, it may be more useful for inferring the partial excess EPR. 
%Calculating the lower bound only requires knowledge of how the expected value changes and the diffusivity of the focusing observable. Both of them are experimentally accessible. 

Although this inequality is looser than Eq.~\eqref{TUR}, it has a clearer physical meaning.
That is, calculating the lower bound only requires knowledge of how the expected value changes and the diffusivity of the observable of interest, and both are experimentally accessible. We also note that this result is an information-thermodynamic extension of the TUR for excess EPR in Ref.~\cite{yoshimura2023housekeeping}.

%This result is an information-thermodynamic extension of the TUR for excess EPR in Ref.~\cite{yoshimura2023housekeeping}.

\subsection{Wasserstein geometry for subsystems}
\label{subsec:Wasserstein_geometry_subsystem}
We present a formulation of Wasserstein geometry for subsystems in Sec.~\ref{subsubsec:Wasserstein_distance_marginal}, and demonstrate two consequences derived from it: the geometric housekeeping-excess decomposition of partial EPR and a thermodynamic speed limit, which are discussed in Secs.~\ref{subsubsec:decomposition_partial_EPR} and~\ref{subsubsec:speed_limit}, respectively.

\subsubsection{Wasserstein distance for marginal distributions}
\label{subsubsec:Wasserstein_distance_marginal}
Here, we propose a formalization of the $2$-Wasserstein distance for marginal distributions. As we reviewed in Sec.~\ref{subsubsec:Wasserstein_geometry}, the $2$-Wasserstein distance for joint distributions is defined by the minimization problem~\eqref{Wasserstein_distance} under the constraints~\eqref{continuity_condition}. Inspired by this formulation, we now define the $2$-Wasserstein distance between two marginal distributions, $\Pa^{(0)}$ and $\Pa^{(1)}$ with $\alpha\in\{X, Y\}$, as follows:
\begin{align}
    &\Wsub(\Pa^{(0)}, \Pa^{(1)}) \nonumber\\
 &= \inf_{(\Pxy(t), \force_{\alpha}(t))_{0\leq t\leq \tau} }\left(\tau\int_0^{\tau}\norm{\force_{\alpha}(t)}_{L_\alpha(\Pxy(t))}^2dt\right)^{\frac{1}{2}}, \label{subsystem_Wasserstein_distance}
\end{align}
subject to
\begin{gather}
    \Pi_{\alpha}\Pxy(t) = \Pa(t), \label{cond1_projection}\\
    d_t\Pa (t)= \div_{g_{\alpha}}L_\alpha(\Pxy(t))\force_{\alpha}(t),\label{cond2_dynamics_conservation}\\
    \Pa(0) = \Pa^{(0)},~\Pa(\tau) = \Pa^{(1)}.\label{cond3_initial_final_distributions}
\end{gather}
Here, the infimum is taken over $\boldsymbol{p}(t)$ that remains strictly positive joint probability distributions.
We refer to $\Wsub(\cdot, \cdot)$ as the $2$-Wasserstein distance for the subsystem. Note that we can choose the time $\tau>0$ arbitrarily~\footnote{For any $\tau'>0$, applying the variable transformation $t\to t(\tau/\tau')$ allows us to change the integration interval from $[0, \tau]$ to $[0, \tau’]$.}. If there is no pair $(\Pxy(t), \force_\alpha(t))$ satisfying the constraints~\eqref{cond1_projection}--\eqref{cond3_initial_final_distributions}, we may let $\Wsub(\cdot, \cdot) = \infty$. We will later confirm that $\Wsub(\cdot, \cdot)$ is the distance between the marginal distributions.

Let us discuss a property of the minimizer of Eq.~\eqref{subsystem_Wasserstein_distance}. It is known that the minimizer of the $2$-Wasserstein distance is given by a conservative force~\cite{yoshimura2023housekeeping,maas2011gradient}. We can find that the same property holds for the case of marginal distributions as well. To explain this notion, we introduce the concept of local conservativeness, which is conservativeness on the projected graph $g_\alpha$. The local conservative subspace with respect to the subsystem $\alpha$ is defined as $\im\nabla_{g_\alpha}$. The elements of $\im\nabla_{g_\alpha}$ are called local conservative forces. 
Then, we can demonstrate that the optimal force for the minimization problem in Eq.~\eqref{subsystem_Wasserstein_distance}, denoted by $\{\force^*_{\alpha}(t)\}_{t\in[0, \tau]}$, is given by a local conservative force. 
In other words, there is a potential $\{\bm{\varphi}^*_\alpha(t)\in\bbR^{|\mathcal{A}|}\}_{t\in[0, \tau]}$ ($(\alpha, \mathcal{A})\in\{(X, \calX), (Y, \calY)\}$) such that $\force^*_{\alpha}(t) =-\nabla_{g_{\alpha}}\bm{\varphi}^*_\alpha(t)$ (see Appendix~\ref{app:conservativeness} for the proof).
This fact enables us to rewrite the $2$-Wasserstein distance for the subsystem $\alpha$ as follows:
\begin{align}
    &\Wsub(\Pa^{(0)}, \Pa^{(1)}) \notag\\
    &= \inf_{(\Pxy(t),\bm{\varphi}_{\alpha}(t))_{0\leq t \leq \tau}}\left(\tau\int_0^{\tau}\norm{\nabla_{g_{\alpha}}\bm{\varphi}_{\alpha}(t)}_{L_\alpha(\Pxy(t))}^2dt\right)^{\frac{1}{2}}, \label{subsystem_Wasserstein_distance_potential}
\end{align}
subject to Eqs.~\eqref{cond1_projection} and~\eqref{cond3_initial_final_distributions}, and
\begin{align}
    d_t\Pa(t) = -\div_{g_\alpha}L_\alpha(\Pxy(t))\nabla_{g_{\alpha}}\bm{\varphi}_{\alpha}(t). 
    \label{cond2_potential}
\end{align} 
Here, the infimum is also taken over $\boldsymbol{p}(t)$ that remains strictly positive joint probability distributions.

We emphasize that this definition of the $2$-Wasserstein distance for the subsystem is not simply equivalent to the existing definitions~\cite{maas2011gradient, yoshimura2023housekeeping} for the marginal distribution, which may be defined via an optimization problem regarding the marginal distribution. This is because the Onsager coefficient matrix $L_\alpha(\Pxy(t))$ is introduced only for the joint distribution $\Pxy(t)$, and is not a function of the marginal distribution $\Pa(t)$. This definition optimizes the joint distribution $\Pxy(t)$, rather than just the marginal distribution $\Pa(t)$ of the subsystem $\alpha \in \{X, Y \}$. Therefore, this quantity should be considered new. Since it is unclear whether it satisfies the axioms of a metric, it must be proved. 

Assuming the existence of minimizers, we show that $\Wsub(\cdot, \cdot)$ satisfies the axioms of a metric. We note that the same conclusion may hold without this assumption, by using paths whose costs are arbitrarily close to the infimum. By definition, we confirm that $\Wsub(\cdot, \cdot)$ satisfies the axioms of non-negativity and non-degeneracy, i.e., $\Wsub(\cdot,\cdot) \geq 0$ and $\Wsub(\Pa^{(0)}, \Pa^{(1)}) = 0$ if and only if $\Pa^{(0)} = \Pa^{(1)}$. The $2$-Wasserstein distance $\Wsub(\cdot, \cdot)$ also satisfies the axiom of symmetry $\Wsub(\Pa^{(0)}, \Pa^{(1)}) = \Wsub(\Pa^{(1)}, \Pa^{(0)})$. This symmetry is verified in Appendix~\ref{app:symmetry_subsystem_Wasserstein_distance}. In Appendix~\ref{app:triangle_inequality}, we also show that the axiom of the triangle inequality
\begin{align}
    \Wsub(\Pa^{(0)}, \Pa^{(2)})\leq \Wsub(\Pa^{(0)}, \Pa^{(1)}) + \Wsub(\Pa^{(1)}, \Pa^{(2)}), \label{triangle_inequality}
\end{align}
holds for any $\Pa^{(0)}, \Pa^{(1)}$ and $\Pa^{(2)}$ satisfying $\Wsub(\Pa^{(0)}, \Pa^{(2)})<\infty$.

Using $\Wsub(\cdot,\cdot)$, we introduce the speed of a given trajectory. We consider the trajectory $\{\Pxy(t)\}_{t\in[0, \tau]}$ and its projection $\{\Pa(t)\}_{t\in[0, \tau]} = \{\Pi_{\alpha}\Pxy(t)\}_{t\in[0, \tau]}$ for $\alpha\in\{X, Y\}$. 
The speed of the trajectory $\{\Pa(t)\}_{t\in[0, \tau]}$, measured with the Wasserstein distance, is defined as
\begin{align}
    v_{\alpha}(\Pa(t)) &= \lim_{\Delta t\to +0}\frac{\Wsub(\Pa(t), \Pa(t + \Delta t))}{\Delta t}.
    \label{speed_optimization_form}
\end{align}
We can rewrite $v_{\alpha}(\Pa(t))$ as the following minimization problem:
\begin{align}
    v_{\alpha}(\Pa(t))=\inf_{\Pxy, \force_{\alpha}}\left(\norm{\force_{\alpha}}^2_{L_\alpha(\Pxy)}\right)^{\frac{1}{2}},
    \label{optimizationvalpha}
\end{align}
with the conditions $\Pi_\alpha\Pxy  = \Pa(t)$ and $d_t\Pa = \div_{g_{\alpha}}L_\alpha(\Pxy)\force_{\alpha}$ on $\Pxy$ and $\force_{\alpha}$. Here, the infimum is also taken over $\boldsymbol{p}$ that remains a strictly positive joint probability distribution. This form of $v_{\alpha}(\Pa(t))$ is proved as follows. By substituting $\Pa^{(0)} = \Pa(t)$, $\Pa^{(1)} = \Pa(t + \Delta t)$, and $\tau = \Delta t$ for sufficiently small $\Delta t$ in Eq.~\eqref{subsystem_Wasserstein_distance}, we obtain
\begin{align}
    &\Wsub(\Pa(t), \Pa(t+\Delta t))\\
    &\quad= \Delta t\inf_{\Pxy, \force_{\alpha}} \left(\norm{\force_{\alpha}}^2_{L_\alpha(\Pxy)} \right)^{\frac{1}{2}}+ o(\Delta t).\label{subsystem_Wasserstein_distance_infinitesimal}
\end{align}
The same constraints imposed on Eq.~\eqref{subsystem_Wasserstein_distance} are imposed on the minimization. Dividing both sides by $\Delta t$ and taking the limit $\Delta t\to0$ yields Eq.~\eqref{optimizationvalpha}.

We discuss the geometric properties of  the minimizer of Eq.~\eqref{subsystem_Wasserstein_distance}, which we write by $(\Pxy^*, \force^*_\alpha)$. In terms of the $2$-Wasserstein geometry for the subsystem, the trajectory $\{\Pa^*(t)\}_{t\in[0, \tau]} = \{\Pi_\alpha\Pxy^*(t)\}_{t\in[0, \tau]}$ is considered as a geodesic, i.e.,
\begin{align}
     \norm{\force^*_\alpha (t)}_{L_\alpha(\Pxy^*(t))} = \frac{\Wsub(\Pa^{(0)}, \Pa^{(1)})}{\tau} ={\rm const.},
     \label{equationgeodesic}
\end{align}
holds (see Appendix~\ref{app:geodesic}). Moreover, we can also show 
\begin{align}
    v_\alpha(\Pa^*(t)) = \norm{\force^*_\alpha (t)}_{L_\alpha(\Pxy^*(t))}.
    \label{equalityforvalpha}
\end{align}
This property is analogous to the property that is discussed for joint probability distributions in Refs.~\cite{erbar2012ricci, yoshimura2023housekeeping}. To prove Eq.~\eqref{equalityforvalpha}, we will examine the following inequality, which is a consequence of the triangle inequality for the $2$-Wasserstein distance for the subsystem:  $\Wsub(\Pa^{(0)}, \Pa^{(1)}) \leq \int_0^{\tau}  v_\alpha(\Pa^*(t))  dt$. Because $\Pxy^*(t)$ and  $\force^*_\alpha(t)$ satisfy the constraints of the optimization problem [Eq.~\eqref{optimizationvalpha}], we can obtain $ v_\alpha(\Pa^*(t)) \leq \norm{\force^*_\alpha (t)}_{L_\alpha(\Pxy^*(t))}$, which yields the inequalities $\Wsub(\Pa^{(0)}, \Pa^{(1)}) \leq \int_0^{\tau}  v_\alpha(\Pa^*(t))  dt \leq \int_0^{\tau}  \norm{\force^*_\alpha (t)}_{L_\alpha(\Pxy^*(t))}  dt$. For these inequalities to satisfy Eq.~\eqref{equationgeodesic} with equality, Eq.~\eqref{equalityforvalpha} must hold based on the conditions for equality.

\subsubsection{Geometric decomposition of partial EPR}
\label{subsubsec:decomposition_partial_EPR}

We discuss the geometric decomposition of the partial EPR $\dot{\Sigma}_{\alpha}$ for $\alpha \in \{X, Y\}$. 
We consider a solution to the master equation $\{\Pxy(t)\}_{t\in[0, \tau]}$ (i.e., $d_t\Pxy = \div\Current(\Pxy) = \div L(\Pxy)\Force(\Pxy)$) and the corresponding marginal distribution $\{\Pa(t) = \Pi_\alpha\Pxy(t)\}_{t\in[0, \tau]}$. To perform the geometric decomposition, we focus on the fact that the following linear equation for $\force_{\alpha}'$ has a unique solution:
\begin{align}
    \nabla_{g_{\alpha}}^{\top}\bm{J}_{\alpha}(\Pxy)=\nabla_{g_{\alpha}}^{\top}L_{\alpha}(\Pxy)\force_{\alpha}',
\end{align}
under the condition $\force_{\alpha}'\in\im\nabla_{g_{\alpha}}$. 
We write the solution to this linear equation as $\force'^{\ast}_\alpha=-\nabla_{g_{\alpha}}\bm{\psi}'^{*}_{\alpha}$. 
Then we find $\innerprod{\force'^*_\alpha}{\Force_\alpha -\force'^*_\alpha}_{L_\alpha(\Pxy)} = -\innerprod{\nabla_{g_\alpha}\bm{\psi}'^*_\alpha}{\Force_\alpha + \nabla_{g_\alpha}\bm{\psi}'^*_\alpha}_{L_\alpha(\Pxy)} = -(\bm{\psi}'^*_\alpha)^\top\div_{g_\alpha}L_\alpha(\Pxy) (\Force_\alpha + \nabla_{g_\alpha}\bm{\psi}'^*_\alpha) = 0$ because $\div_{g_\alpha}L_\alpha(\Pxy)\Force_\alpha = \div_{g_\alpha}\bm{J}_{\alpha}=-\div_{g_{\alpha}}L_{\alpha}(\Pxy)\nabla_{g_{\alpha}}\bm{\psi}'^*_{\alpha}$. Thus, $\force'^*_\alpha$ is given by the projection of $\Force_\alpha$ onto the local conservative subspace $\im\nabla_{g_\alpha}$, with respect to the inner product $\innerprod{\cdot}{\cdot}_{L_\alpha}$. Inspired by the analogy between this projection and the projection of $\Force$ discussed in Sec.~\ref{subsec:decomposition}, we define the local housekeeping EPR and the local excess EPR of the subsystem $\alpha\in\{X, Y\}$ as $\EPR^\mathrm{hk,\;loc}_\alpha(\Pxy) \coloneqq \norm{\Force_\alpha - \force'^*_\alpha}^2_{L_{\alpha}(\Pxy)}$ and $\EPR^\mathrm{ex,\;loc}_\alpha(\Pxy) \coloneqq \norm{\force'^*_\alpha}^2_{L_{\alpha}}(\Pxy)$, respectively. Then, the orthogonality leads to the following decomposition:
\begin{align}
    \EPR_\alpha(\Pxy) = \EPR^\mathrm{hk,\;loc}_\alpha(\Pxy) + \EPR^\mathrm{ex,\;loc}_\alpha(\Pxy).
\end{align}
This equality is verified as follows: $\EPR_\alpha(\Pxy)=\norm{\Force_\alpha}^2_{L_\alpha(\Pxy)}=\norm{(\Force_\alpha - \force'^*_\alpha) + \force'^*_\alpha}^2_{L_\alpha(\Pxy)}=\norm{\Force_\alpha - \force'^*_\alpha}^2_{L_\alpha(\Pxy)}+\norm{\force'^*_\alpha}^2_{L_\alpha(\Pxy)}=\EPR^\mathrm{hk,\;loc}_\alpha(\Pxy) + \EPR^\mathrm{ex,\;loc}_\alpha(\Pxy)$ because $\innerprod{\force'^*_\alpha}{\Force_\alpha -\force'^*_\alpha}_{L_\alpha(\Pxy)} = 0$. 

We briefly discuss the connection between the local excess EPR $\EPR^\mathrm{ex,\;loc}_\alpha(\Pxy)$ and the $2$-Wasserstein distance for the subsystem. Specifically, the speed $v_\alpha$ provides a lower bound of $\EPR^\mathrm{ex,\;loc}_\alpha(\Pxy)$ as
\begin{align}
    [v_{\alpha}(\Pa)]^2\leq\EPR^\mathrm{ex,\;loc}_\alpha(\Pxy).
    \label{ineq_localexcess_speed}
\end{align}
We note that while this result corresponds to an equality relationship between the excess EPR and the 2-Wasserstein distance in the total system~\cite{yoshimura2023housekeeping}, this becomes an inequality when focusing on the subsystem.
This inequality is shown as follows. The pair $\{\Pxy(t),\force^*_\alpha\}$ is a candidate of the minimization problem [Eq.~\eqref{optimizationvalpha}] because the pair satisfies $\Pi_\alpha\Pxy(t)  = \Pa(t)$ and $d_t\Pa(t) = \div_{g_{\alpha}}L_\alpha(\Pxy(t))\force'^{\ast}_{\alpha}$. Thus, we obtain $v_{\alpha}(\Pa(t))\leq\|\force'^{\ast}_{\alpha}\|_{L_{\alpha}(\Pxy(t))}=\sqrt{\EPR^\mathrm{ex,\;loc}_\alpha(\Pxy(t))}$, whose square yields Eq.~\eqref{ineq_localexcess_speed}. This inequality arises because $\boldsymbol{p}(t)$ is not necessarily equivalent to the optimal distribution for Eq.~\eqref{optimizationvalpha}. 

The local excess and housekeeping EPRs are different from the partial housekeeping and excess EPRs. For example, we find $\EPR^\mathrm{hk,\;loc}_X (\Pxy)+ \EPR^\mathrm{hk,\;loc}_Y(\Pxy) \neq \EPRhk(\Pxy)$ and  $\EPR^\mathrm{ex,\;loc}_X (\Pxy)+ \EPR^\mathrm{ex,\;loc}_Y (\Pxy)\neq \EPRex(\Pxy)$ in general, while Eqs.~\eqref{additivity_partial_EPRhk} and~\eqref{additivity_partial_EPRex} hold. This difference arises from the order in which the EPRs are decomposed geometrically. The partial housekeeping and excess EPRs $\EPRhk_\alpha(\Pxy)$ and $\EPRex_\alpha(\Pxy)$ are obtained by decomposing $\EPRhk(\Pxy)$ and $\EPRex(\Pxy)$ into contributions from the subsystem $\alpha$, where the housekeeping and excess EPRs $\EPRhk(\Pxy)$ and $\EPRex(\Pxy)$ are obtained by decomposing the total system's EPR $\EPR(\Pxy)$ based on the conservativeness described by $\nabla$. 
On the other hand, $\EPR^\mathrm{hk,\;loc}_\alpha(\Pxy)$ and $\EPR^\mathrm{ex,\;loc}_\alpha(\Pxy)$ are obtained by decomposing the partial EPR $\EPR_\alpha(\Pxy)$ based on the local conservativeness described by $\nabla_{g_\alpha}$. 

Although the local and partial excess EPRs, $\EPR^\mathrm{ex,\;loc}_\alpha(\Pxy)$ and $\EPRex_{\alpha}(\Pxy)$, are fundamentally different quantities, they are linked by the following inequality (see Appendix~\ref{app:EPRex_inequality} for the proof):
\begin{align}
    \EPR^\mathrm{ex,\;loc}_\alpha (\Pxy)\leq \EPRex_{\alpha} (\Pxy)\label{EPRex_inequality}.
\end{align}
We can intuitively understand the inequality in the following way. $\EPR^\mathrm{ex,\;loc}_\alpha(\Pxy)$ is the minimum dissipation required to change (at least) the subsystem $\alpha$, while $\EPRex_\alpha(\Pxy)$ is the contribution of subsystem $\alpha$ to the minimum dissipation required to change the total system. Therefore, $\EPR^\mathrm{ex,\;loc}_\alpha(\Pxy)$ is never greater than $\EPRex_\alpha$. We can also obtain the hierarchy $[v_{\alpha} (\Pxy_{\alpha})]^2\leq \EPR^\mathrm{ex,\;loc}_\alpha (\Pxy)\leq \EPRex_{\alpha}(\Pxy)$ by combining Eqs.~\eqref{ineq_localexcess_speed} and ~\eqref{EPRex_inequality}.

We conclude this section by introducing TURs for the local excess EPR. 
The derivation of the TURs follows the same logic used in Sec.~\ref{subsec:TUR}. Here, we consider the same observable $\bm{\Psi}_\alpha = \Pi^\top_\alpha\bm{\psi}_\alpha$ used in Sec.~\ref{subsec:TUR}. We first introduce the TUR corresponding to Eq.~\eqref{TUR}, which is given by the following inequality:
\begin{align}
    \frac{(d_t(\Pxy^\top\bm{\Psi}_\alpha))^2}{\norm{\nabla_{G_\alpha}\bm{\Psi}_\alpha}_{L_\alpha(\Pxy)}^2}=\frac{(d_t(\Pa^\top\bm{\psi}_\alpha))^2}{\norm{\nabla_{g_\alpha}\bm{\psi}_\alpha}^2_{L_\alpha(\Pxy)}} \leq \EPR^\mathrm{ex,\;loc}_\alpha (\Pxy),\label{eq:TUR_local_excess}
\end{align}
for any time-independent observable $\bm{\psi}_\alpha\in\bbR^{|\mathcal{A}|}$ with $(\alpha, \mathcal{A})\in\{(X, \calX), (Y, \calY)\}$. The inequality [Eq.~\eqref{eq:TUR_local_excess}] can be shown by considering the following Cauchy--Schwarz inequality: $(\innerprod{\force'^*_\alpha}{\nabla_{g_\alpha}\bm{\psi}_\alpha}_{L_\alpha(\Pxy)})^2 \leq \norm{\force'^*_\alpha}^2_{L_\alpha(\Pxy)}\norm{\nabla_{g_\alpha}\bm{\psi}_\alpha}^2_{L_\alpha(\Pxy)}$.
The left-hand side can be rewritten as $(\innerprod{\force'^*_\alpha}{\nabla_{g_\alpha}\bm{\psi}_\alpha}_{L_\alpha(\Pxy)})^2 = (d_t(\Pa^\top\bm{\psi}_\alpha))^2$. 
On the other hand, $\norm{\force'^*_\alpha}^2_{L_\alpha(\Pxy)} = \EPR^\mathrm{ex,\;loc}_\alpha(\Pxy)$ by definition, so we have Eq.~\eqref{eq:TUR_local_excess}. 
Note that the equality is always achievable by choosing $\bm{\psi}_\alpha = \bm{\psi}'^*_\alpha$ (recall $\force'^*_\alpha = -\nabla_{g_\alpha}\bm{\psi}'^*_\alpha$).

Then, by combining Eqs.~\eqref{diffusivity_inequality} and~\eqref{eq:TUR_local_excess}, we obtain a looser TUR:
\begin{align}
    \frac{2(d_t(\Pxy^\top\bm{\Psi}_\alpha))^2}{\mathcal{D}(\bm{\Psi}_{\alpha})} =\frac{2(d_t(\Pa^\top\bm{\psi}_\alpha))^2}{\mathcal{D}(\Pi^\top_\alpha\bm{\psi}_\alpha)} \leq \EPR^\mathrm{ex,\;loc}_\alpha(\Pxy), \label{eq:looser_TUR_local}
\end{align}
corresponding to Eq.~\eqref{TUR_2}.
Although this inequality is looser than Eq.~\eqref{eq:TUR_local_excess}, it may be more useful for inferring the local excess EPR. 
Calculating the lower bound only requires knowledge of how the expected value changes and the diffusivity of the observable of interest. Both are experimentally accessible. 

 We discuss the tightness of inequalities between TURs using the partial excess EPR [Eqs.~\eqref{TUR} and \eqref{TUR_2}] and the local excess EPR [Eqs.~\eqref{eq:TUR_local_excess} and \eqref{eq:looser_TUR_local}]. Since the lower bounds on the left-hand sides are the same, and since there is an inequality between the partial EPR and the local EPR [Eq.~\eqref{EPRex_inequality}], the following hierarchy of inequalities 
 \begin{align}
    \frac{2(d_t(\Pxy^\top\bm{\Psi}_\alpha))^2}{\mathcal{D}(\bm{\Psi}_{\alpha})} \leq \frac{(d_t(\Pxy^\top\bm{\Psi}_\alpha))^2}{\norm{\nabla_{G_\alpha}\bm{\Psi}_\alpha}_{L_\alpha(\Pxy)}^2} \leq \EPR^\mathrm{ex,\;loc}_\alpha(\Pxy)\leq \EPR^\mathrm{ex}_\alpha(\Pxy),  \label{TURhierachy}
\end{align}
is obtained. This implies that when performing the inference by maximizing the quantity on the left-hand side while varying the observable $\bm{\Psi}_\alpha$, one can obtain only the local excess EPR, not the partial excess EPR. However, in situations where the partial excess EPR and the local excess EPR are nearly identical, the partial excess EPR can also be inferred. We will discuss specific examples of such situations later.

\subsubsection{Speed limit}
\label{subsubsec:speed_limit}
We now introduce the thermodynamic speed limit. We consider a solution to the master equation and the corresponding marginal distribution: $\{\Pxy(t)\}_{t\in[0, \tau]}$ and $\{\Pa(t)\}_{t\in[0, \tau]} = \{\Pi_\alpha\Pxy(t)\}_{t\in[0, \tau]}$ for $\alpha\in\{X, Y\}$. We write the initial and final distributions as $\Pxy^{(0)} = \Pxy(0)$, $\Pxy^{(1)} = \Pxy(\tau)$, $\Pa^{(0)} = \Pa(0)$, and $\Pa^{(1)} = \Pa(\tau)$.

First, the triangle inequality yields $\Wsub(\Pa(0), \Pa(\tau)) \leq \int_0^{\tau}v_{\alpha}(\Pa(t)) dt$.
From the Cauchy--Schwarz inequality, we obtain
\begin{align}
    \Wsub(\Pa(0), \Pa(\tau))^2 
    &\leq \left(\int_0^{\tau}v_{\alpha} dt\right)^2\notag\\  &\leq \tau\int_0^{\tau}v_{\alpha}^2dt\notag\\ 
    &\leq\tau\int_0^{\tau}\EPR^\mathrm{ex,\;loc}_\alpha dt \notag\\
    &\leq \tau\int_0^{\tau}\EPRex_\alpha dt, \label{velocity_inequality}
\end{align}
where we used Eq.~\eqref{ineq_localexcess_speed} and Eq.~\eqref{EPRex_inequality}. Here, we introduce the quantities $\tilde{\Sigma}^{\mathrm{ex}}_{\alpha}(\tau) = \int_0^{\tau}\eprex_{\alpha}dt$ and $\Delta_{\alpha}I_{XY}(\tau) = \int_0^{\tau}\iflowex_{\alpha}dt$, which satisfy
$\tilde{\Sigma}^{\mathrm{ex}}_{\alpha}(\tau)- \Delta_{\alpha}I_{XY}(\tau) =\int_0^{\tau}\EPRex_\alpha dt =:\Sigma^\mathrm{ex}_\alpha(\tau)$.
Therefore, Eq.~\eqref{velocity_inequality} is rewritten as
\begin{align}
    \frac{\Wsub(\Pa(0), \Pa(\tau))^2}{\tau} \leq \tilde{\Sigma}^{\mathrm{ex}}_{\alpha}(\tau) - \Delta_{\alpha}I_{XY}(\tau),\label{information_TSL}
\end{align}
which is called the information-thermodynamic speed limit.

This inequality can be understood in three different contexts. 
Firstly,  Eq.~\eqref{information_TSL} corresponds to the thermodynamic speed limit [Eq.~\eqref{TSL}] in the information-thermodynamic scheme for bipartite systems. The main difference between these two inequalities is that the right-hand side of the former comprises both an entropic quantity ($\tilde{\Sigma}^{\mathrm{ex}}_{\alpha}$) and an informational quantity ($\Delta_{\alpha}I_{XY}$). Secondly, Eq.~\eqref{information_TSL} provides a tighter bound than the generalization of the second law of information thermodynamics [Eq.~\eqref{information_second_law_ex}], which can be rewritten as $ \tilde{\Sigma}^{\mathrm{ex}}_{\alpha}(\tau) \geq \Delta_{\alpha}I_{XY}(\tau)$ for a finite time. Thirdly, Eq.~\eqref{information_TSL} corresponds to the information-thermodynamic speed limit, which was originally proposed by Nakazato and Ito for overdamped Langevin systems~\cite{nakazato2021geometrical}.

We briefly characterize the quantity $\Delta_{\alpha}I_{XY}(\tau)$. This quantity $\Delta_{\alpha}I_{XY}(\tau)$ implies the change in mutual information due to subsystem $\alpha$. Specifically, according to Eq.~\eqref{information_flow_derivative_excess}, we have
\begin{align}
    &\Delta_XI_{XY}(\tau) + \Delta_YI_{XY}(\tau) = \int_0^{\tau}d_tI(\hat{X}_t; \hat{Y}_t)dt \notag\\
    &= I(\hat{X}_\tau; \hat{Y}_\tau) - I(\hat{X}_0; \hat{Y}_0). \label{mutual_information_difference}
\end{align}
We note that $\Delta_{\alpha}I_{XY}(\tau)$ does not diverge for stationary systems in the limit $\lim_{\tau\to\infty}|\Delta_{\alpha}I_{XY}(\tau)|<\infty$, even when the detailed balance condition is violated and $\iflow_\alpha(\Pxy^\mathrm{st}) \neq 0$ is satisfied. On the other hand, the integral of $\iflow_\alpha$, defined as $\tilde{\Delta}_\alpha I_{XY}(\tau) \coloneqq\int^\tau_0\iflow_\alpha dt$, formally satisfies the relation equivalent to Eq.~\eqref{mutual_information_difference}: $\tilde{\Delta}_X I_{XY}(\tau) + \tilde{\Delta}_Y I_{XY}(\tau) = I(\hat{X}_\tau; \hat{Y}_\tau) - I(\hat{X}_0; \hat{Y}_0)$. However, $\tilde{\Delta}_\alpha I_{XY}(\tau)$ may diverge in the limit $\lim_{\tau\to\infty}|\tilde{\Delta}_\alpha I_{XY}(\tau)| =\infty$ since the housekeeping contribution of the information flow exists. Therefore,  the quantity $\Delta_\alpha I_{XY}(\tau)$ has a physically sound property in this sense.

Finally, we derive a tighter inequality by defining the thermodynamic length as $\mathcal{L}_\alpha(t)\coloneqq \int_0^{t}v_{\alpha} dt$. From Eq.~\eqref{velocity_inequality}, we obtain
\begin{align}
    \frac{\mathcal{L}^2_\alpha(t)}{t} \leq \tilde{\Sigma}^{\mathrm{ex}}_{\alpha}(t) - \Delta_{\alpha}I_{XY}(t) \label{information_TSL_2}
\end{align}
for $t\in (0, \tau]$.
While $\Wsub(\Pa(0), \Pa(\tau))$ corresponds to the geodesic, $\mathcal{L}_\alpha$ quantifies the length of the path $\{\Pa(t)\}_{t\in[0, \tau]}$. Therefore, unlike the 2-Wasserstein distance for subsystems, this quantity $\mathcal{L}_\alpha(t)$ requires information not only about the initial and final states $\Pa(0)$ and $\Pa(\tau)$, but also about the path $\{\Pa(t)\}_{t\in[0, \tau]}$.

\section{Examples}
\label{sec:example}

\begin{figure}
    \centering
    \includegraphics[width=\linewidth]{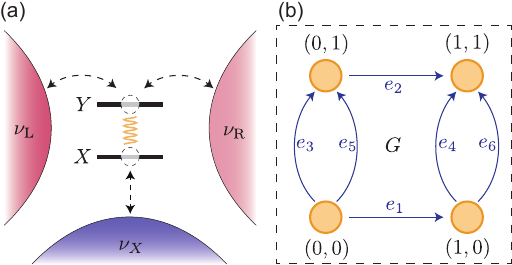}
    \caption{(a) Schematic of the model. The system $X$ exchanges particles with the particle bath that has the chemical potential $\mu_X$ and the inverse temperature $\beta_X$. The system $Y$ exchanges particles with the two particle baths that have different chemical potentials $\mu_{\mathrm{L}}$ and $\mu_{\mathrm{R}}\;(<\mu_{\mathrm{L}})$ and the same inverse temperature $\beta\;(<\beta_X)$. There is also an interaction between $X$ and $Y$. 
    (b) The graph corresponding to the model. The presence or absence of a particle in each system represents the states of $X$ and $Y$, which are written as $x$ and $y$, respectively. All possible states of the system can be described as $(x, y) \in \{(0, 0), (0, 1), (1, 0), (1, 1)\}$ where $1$ represents an occupied state and $0$ an unoccupied one. The nodes are connected by edges $\{e_i\}_{i\in\{1,\ldots, 6\}}$. }
    \label{fig:schematic}
\end{figure}

We illustrate our results using a simple model. The model, shown in Fig.~\ref{fig:schematic}(a), consists of two binary-state systems $X$ and $Y$. The state of $X$ is represented by $x \in\{0, 1\} \eqqcolon\calX$, and the state of $Y$ by $y \in\{0, 1\}\eqqcolon \calY$. 
The state with no particle in $X\;(Y)$ is represented as $x=0\; (y=0)$, while the one with one particle is $x=1\;(y=1)$. Accordingly, the system is described by the four states $(x, y) \in \calX\times\calY = \{(0, 0), (0, 1), (1, 0), (1, 1)\}$. The system $X$ is coupled to a particle bath $\nu_X$ characterized by the inverse temperature $\beta_X$ and the chemical potential $\mu_X$. The system $Y$ exchanges particles with two baths $\nu_{\mathrm{L}}$ and $\nu_{\mathrm{R}}$.   The baths $\nu_{\mathrm{L}}$ and $\nu_{\mathrm{R}}$ are characterized by the inverse temperature $\beta\;(< \beta_X)$ and their respective chemical potentials, $\mu_{\mathrm{L}}$ and $\mu_{\mathrm{R}}$, that satisfy $\mu_{\mathrm{L}} > \mu_{\mathrm{R}}$.
The energy is set to $0$ when neither subsystem is occupied, i.e., when $(x,y) = (0,0)$. It is set to  $E_X$ ($E_Y$) when only one state is occupied, i.e., when $(x,y) = (1,0)$ ($(x,y) = (0,1)$). When both states are occupied, i.e., when $(x, y) = (1, 1)$, an interaction energy $u$ arises, and the total energy of this state is given by $E_X + E_Y + u$.

The graph corresponding to the model is shown in Fig.~\ref{fig:schematic}(b). Here, the set of edges is given by $\edgeset \coloneqq \{e_i\}_{i\in\{1, \ldots, 6\}}$ with
\begin{gather}
    e_1 = (\nu_X, (0, 0)\to (1, 0)),\;e_2 = (\nu_X, (0, 1)\to (1, 1)), \notag\\
    e_3 = (\nu_{\mathrm{L}}, (0, 0)\to (0, 1)),\;e_4 = (\nu_{\mathrm{L}}, (1, 0)\to (1, 1)), \notag\\
    e_5 = (\nu_{\mathrm{R}}, (0, 0)\to (0, 1)),\;e_6 = (\nu_{\mathrm{R}}, (1, 0)\to (1, 1)).
\end{gather}
Thus, the graph $G \coloneqq (\calX\times\calY, \edgeset)$ satisfies the condition for a bipartite system [Eq.~\eqref{bipartite_def1}]. The set of edges is split into $\edgeset_X = \{e_1, e_2\}$ and $\edgeset_Y = \{e_3, e_4, e_5, e_6\}$.
The incidence matrix of $G$ and its marginalization matrices are given as follows:
\begin{align}
    \div &= 
    \begin{pmatrix}
        -1 & 0 & -1 & 0 & -1 & 0\\
        0 & -1 & 1 & 0 & 1 & 0\\
        1 & 0 & 0 & -1 & 0 & -1\\
        0 & 1 & 0 & 1 & 0 & 1
    \end{pmatrix}, \\
    \Pi_X &= 
    \begin{pmatrix}
        1 & 1 & 0 & 0\\
        0 & 0 & 1 & 1
    \end{pmatrix},\;
    \Pi_Y = 
    \begin{pmatrix}
        1 & 0 & 1 & 0\\
        0 & 1 & 0 & 1
    \end{pmatrix}.
\end{align}

For $e_i\in\edgeset_X$ ($i \in \{1,2 \}$), we define the transition rates as $R_{e_1} = \Gamma f_1$, $R_{-e_1} = \Gamma(1-f_1)$, $R_{e_2} = \Gamma f_2$, and $R_{-e_2} = \Gamma(1-f_2)$ with $f_1 = 1/(1 + e^{\beta_X(E_X -\mu_X)})$ and $f_2 = 1/(1 + e^{\beta_X(E_X + u -\mu_X)})$, where $\Gamma$ is a positive constant. 
For $e_i\in\edgeset_Y\;(i\in\{3, 4, 5, 6\})$, we define the transition rates as $R_{e_i} = \Gamma_i f_i$ and $R_{-e_i} = \Gamma_i(1 - f_i)$, where $\Gamma_i$ are positive constants with $f_3 = 1/(1 + e^{\beta(E_Y -\mu_{\mathrm{L}})})$, $f_4 = 1/(1 + e^{\beta(E_Y + u -\mu_{\mathrm{L}})})$, $f_5 = 1/(1 + e^{\beta(E_Y -\mu_{\mathrm{R}})})$, and $f_6 = 1/(1 + e^{\beta(E_Y + u -\mu_{\mathrm{R}})})$. 

For the numerical calculations, we consistently use the following parameters: $\beta_X = 10, \beta = 1$, $\Gamma = 16$, 
$\Gamma_3 = \Gamma_6 = 1.5$, $\Gamma_4 = \Gamma_5 = 0.5$, $E_X = E_Y = 1$, $u = 0.2$, $\mu_{\mathrm{L}} = 1.1$, $\mu_{\mathrm{R}} = 0.9$, and $\mu_X = 0.9$. 

\begin{figure*}
    \centering
    \includegraphics[width=\linewidth]{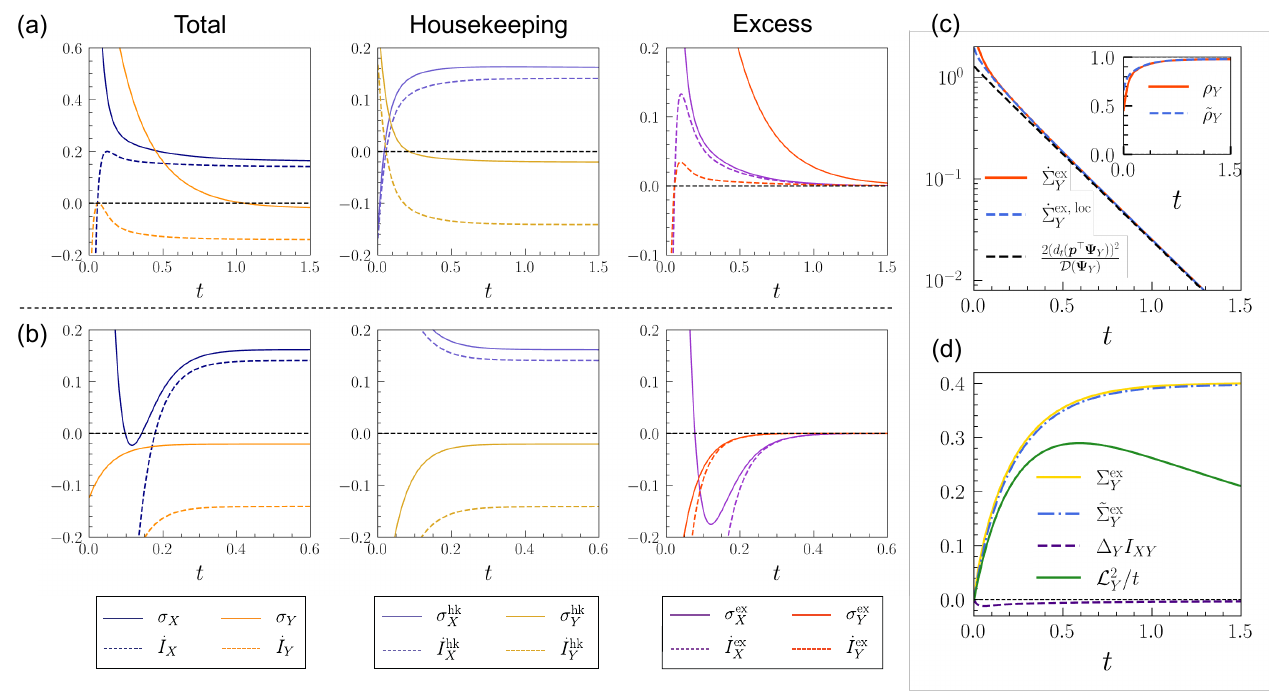}
    \caption{The second law of information thermodynamics [Eq.~\eqref{information_second_law}], the generalized second laws of information thermodynamics [Eqs.~\eqref{information_second_law_hk} and~\eqref{information_second_law_ex}], the thermodynamic uncertainty relation [Eq.~\eqref{TUR_2}] and the information-thermodynamic speed limit [Eq.~\eqref{information_TSL_2}].
    (a) In the left panel, we observe an apparent violation of the second law of thermodynamics $\epr_Y<0$ for $t$ roughly greater than $1.1$. We also see that $\eprhk_Y$ is negative for $t$ roughly greater than $0.2$ in the central panel, while $\eprex_Y$ is always positive in the right panel. 
    (b) In the left panel, $\epr_X$ is negative when $0.1<t<0.15$. Unlike (a), we see that $\eprhk_X$ is positive, while $\eprex_X$ is negative when $0.08<t$ in the right panel. Note that $\sigma_Y$ is always negative, as are $\eprhk_Y$ and $\eprex_Y$.
    %(c) The thermodynamic uncertainty relation$ 2(d_t(\Pxy^\top\bm{\Psi}_Y))^2/(\mathcal{D}(\bm{\Psi}_Y)) \leq \EPRex_Y$.
    %The observable used to obtain the lower bound is defined as $\bm{\Psi}_Y \coloneqq \Pi^\top_Y(0, 1)^\top$. The-upper right panel shows the ratio of the lower bound to the partial excess EPR: $\rho_Y \coloneqq 2(d_t(\Pxy^\top\bm{\Psi}_Y))^2/(\EPRex_Y\mathcal{D}(\bm{\Psi}_Y))$. 
    (c) The thermodynamic uncertainty relations with respect to the partial excess EPR and the local excess EPR [Eqs.~\eqref{TUR_2} and~\eqref{eq:looser_TUR_local}].
    The observable used to obtain the lower bound is defined as $\bm{\Psi}_Y \coloneqq \Pi^\top_Y(0, 1)^\top$. The upper right panel shows the ratios of the lower bound to the partial excess EPR and to the local excess EPR: $\rho_Y \coloneqq 2(d_t(\Pxy^\top\bm{\Psi}_Y))^2/(\EPRex_Y\mathcal{D}(\bm{\Psi}_Y))$ and $\tilde{\rho}_Y \coloneqq 2(d_t(\Pxy^\top\bm{\Psi}_Y))^2/(\EPR^\mathrm{ex,\;loc}_Y\mathcal{D}(\bm{\Psi}_Y))$. 
    }
    (d) The information-thermodynamic speed limit $\mathcal{L}^2_Y/t\leq\int^t_0 \EPRex_Ydt \eqqcolon \Sigma^\mathrm{ex}_Y=\tilde{\Sigma}^{\mathrm{ex}}_{Y} - \Delta_{Y}I_{XY}$. The bound is relatively tight when $t \in [0, 0.6]$.
    \label{fig:numeric_calculation}
\end{figure*}

%In information thermodynamics, apparent violations of the second law of thermodynamics may be caused by information flow. This effect can be discussed separately in terms of two factors: excess information flow and housekeeping information flow. %To this end, we sequentially illustrate the second law of information thermodynamics [Eq.~\eqref{information_second_law}] and the generalized second laws of information thermodynamics [Eqs.~\eqref{information_second_law_hk} and~\eqref{information_second_law_ex}] in Figs.~\ref{fig:numeric_calculation}(a) and~\ref{fig:numeric_calculation}(b). 
Apparent violations of the second law of thermodynamics are represented by the negativity of entropy change rates of subsystems, which is necessarily accompanied by negative information flow. 
In Sec.~\ref{subsec:generalization_second_law}, we discussed that this effect can be separated in terms of two factors: excess and housekeeping information flows.
To numerically confirm this fact, we sequentially illustrate the second law of information thermodynamics [Eq.~\eqref{information_second_law}] and the generalized second laws of information thermodynamics [Eqs.~\eqref{information_second_law_hk} and~\eqref{information_second_law_ex}] in Figs.~\ref{fig:numeric_calculation}(a) and~\ref{fig:numeric_calculation}(b). 

We can observe cases (i) $\sigma^{\rm hk}_X>0$ and $\sigma^{\rm ex}_X>0$, (ii) $\sigma^{\rm hk}_Y<0$ and $\sigma^{\rm ex}_Y>0$ in Fig.~\ref{fig:numeric_calculation}(a). We can also observe cases (iii)
$\sigma^{\rm hk}_X>0$ and $\sigma^{\rm ex}_X<0$, and (iv)
$\sigma^{\rm hk}_Y<0$ and $\sigma^{\rm ex}_Y<0$ in Fig.~\ref{fig:numeric_calculation}(b).

First, let us discuss Fig.~\ref{fig:numeric_calculation}(a), in which the initial distribution is set to be $\Pz(0) = (0.85, 0.05, 0.05, 0.05)^\top$. In the left panel, $\sigma_Y$ is negative for $t$ roughly greater than $1.1$, which can be interpreted as the conventional concept of an apparent violation of the second law of thermodynamics in the subsystem $Y$.
In the central panel, we see that $\eprhk_Y$ turns negative at $t\simeq 0.2$, while in the right panel, $\eprex_Y$ remains positive.
%Therefore, the apparent violation shown in the left panel is caused by the housekeeping information flow from $Y$ to $X$ where the subsystem $X$ is considered to be Maxwell's demon with regard to housekeeping dissipation. 

Therefore, the apparent violation shown in the left panel is caused by the housekeeping information flow from $Y$ to $X$ where the subsystem $X$ is considered to be a housekeeping demon.

For the system $X$, we can also observe that both $\eprhk_X$ and $\eprex_X$ are positive.
We also note that the antisymmetric relation $\iflowhk_X = -\iflowhk_Y$ [Eq.~\eqref{iflowhk_antisymmetry}] is confirmed. 

Second, we refer to Fig.~\ref{fig:numeric_calculation}(b), in which the initial distribution is set to be $\Pz(0) = (0.01, 0.49, 0.49, 0.01)^\top$. 
In this case, the left panel shows that $\sigma_X$ is temporarily negative around $t = 0.1$. 
In the central panel, $\eprhk_X$ remains positive throughout, while in the right panel, $\eprex_X$ temporarily becomes negative and converges towards zero as the system relaxes. 
%Therefore, the apparent violation of the second law of thermodynamics in the subsystem $X$ can be understood as arising from Maxwell's demon in subsystem $Y$ around $t \simeq 0.1$, which is associated with excess dissipation. 

Therefore, the apparent violation of the second law for the subsystem $X$ can be understood as arising from an excess demon (the subsystem $Y$).

Note that $\sigma_Y$ is always negative, and $X$ always acts as Maxwell's demon. Furthermore, since both $\eprhk_Y$ and $\eprex_Y$ are negative, the apparent violation of the second law of thermodynamics arises from the combined effects of excess and housekeeping information flows. 
%Due to the asymmetric relation $\iflowhk_X = -\iflowhk_Y$, neither system can become Maxwell's demon with respect to housekeeping dissipation. However, as shown in this case, both systems can temporarily become Maxwell's demons with respect to excess dissipation.

Due to the antisymmetric relation $\iflowhk_X = -\iflowhk_Y$, the two subsystems never behave as housekeeping demons simultaneously. However, as shown in this case, both systems can temporarily become excess demons.

\begin{comment}
In Fig.~\ref{fig:numeric_calculation}(c), we illustrate the short-time TUR [Eq.~\eqref{TUR_2}] for the subsystem $Y$. The initial distribution is given as $\Pz(0) = (0.85, 0.05, 0.05, 0.05)^{\top}$. We use observable $\bm{\Psi}_Y \coloneqq \Pi^\top_Y(0, 1)^\top = (0, 1, 0, 1)^\top$ to give the lower bound $2(d_t(\Pxy^\top{\bm{\Psi}_Y}))^2/\mathcal{D}(\bm{\Psi}_Y)$. Note that observables that depend only on $y$ give the same lower bound. This is because such observables can be obtained by adding a constant vector or multiplying a scalar to $\bm{\Psi}_Y$, and 
these manipulations do not change the lower bound. As the inset graph, we plot the ratio of the lower bound to the partial excess EPR, $\rho_Y\coloneqq 2(d_t(\Pxy^\top\bm{\Psi}_Y))^2/(\EPRex_Y\mathcal{D}(\bm{\Psi}_Y))$. 
As this lower bound is relatively tight, it can be used to estimate $\EPRex_Y$.
\end{comment}

In Fig.~\ref{fig:numeric_calculation}(c), we illustrate the short-time TURs with respect to the partial excess EPR and the local excess EPR [Eqs.~\eqref{TUR_2} and~\eqref{eq:looser_TUR_local}, respectively]. The initial distribution is given as $\Pz(0) = (0.85, 0.05, 0.05, 0.05)^{\top}$. 
We use an observable $\bm{\Psi}_Y \coloneqq \Pi^\top_Y(0, 1)^\top = (0, 1, 0, 1)^\top \in\im\Pi^\top_Y$ to give the lower bound $2(d_t(\Pxy^\top{\bm{\Psi}_Y}))^2/\mathcal{D}(\bm{\Psi}_Y)$, which is common to both Eqs.~\eqref{TUR_2} and~\eqref{eq:looser_TUR_local}.
Note that observables that depend only on $y$ give the same lower bound~\footnote{This is because such observables can be obtained by adding a constant vector or multiplying a scalar to $\bm{\Psi}_Y$, and 
these manipulations do not change the lower bound.}.
It is also observed that the inequality $\EPRex_Y \geq \EPR^\mathrm{ex, loc}_Y$ [Eq.~\eqref{EPRex_inequality}] holds. 

In the inset of Fig.~\ref{fig:numeric_calculation}(c), we plot the ratios of the lower bound to the partial and local excess EPRs, denoted by
\begin{align}
    \rho_Y \coloneqq \frac{2(d_t(\Pxy^\top\bm{\Psi}_Y))^2}{\EPRex_Y\mathcal{D}(\bm{\Psi}_Y)}, \quad
    \tilde{\rho}_Y \coloneqq \frac{2(d_t(\Pxy^\top\bm{\Psi}_Y))^2}{\EPR^\mathrm{ex,\;loc}_Y\mathcal{D}(\bm{\Psi}_Y)}.
\end{align}
Regarding the ratios, the inequality $\rho_Y \leq \tilde{\rho}_Y \leq 1$ is satisfied because of the hierarchy [Eq.~\eqref{TURhierachy}]. After a long time, both values approach $1$. This demonstrates that it is possible to infer both the local and partial excess EPRs from the observable $\bm{\Psi}_Y$.

In Fig.~\ref{fig:numeric_calculation}(d), we demonstrate the information-thermodynamic speed limit [Eq.~\eqref{information_TSL_2}] for the subsystem $Y$. The initial distribution is given as $\Pz(0) = (0.85, 0.05, 0.05, 0.05)^{\top}$. 
We define the partial excess entropy production of the subsystem $Y$ as $\Sigma^\mathrm{ex}_Y \coloneqq \int^t_0 \EPRex_Ydt$ so that Eq.~\eqref{information_TSL_2} is expressed as $\mathcal{L}^2_Y/t\leq \Sigma^\mathrm{ex}_Y =\tilde{\Sigma}^{\mathrm{ex}}_{Y}(t) - \Delta_{Y}I_{XY}(t)$. %$\mathcal{L}_Y^2/t$ has a peak at $t \simeq 0.6$, and the bound is relatively tight when $t \in [0, 0.6]$. Note that since the contribution of $\Delta_{Y}I_{XY}(t)$ does not change significantly over time, the influence of $\tilde{\Sigma}^{\mathrm{ex}}_{Y}(t)$ largely determines the bound on $\mathcal{L}_Y^2/t$. As $t$ increases, $\Sigma^{\mathrm{ex}}_Y$ also increases and converges to a positive value, while $\mathcal{L}_Y^2/t$ decreases. These behaviors imply that the bound eventually loses its significance.

As $t$ increases, $\Sigma^\mathrm{ex}_Y$ also increases monotonically and converges to a finite positive value. 
In contrast, the lower bound $\mathcal{L}_Y^2/t$ exhibits a peak around $t \simeq 0.6$ and asymptotically approaches zero. 
Therefore, in this case, the speed limit is relatively tight for $t \in [0, 0.6]$. 
We also note that the informational term $\Delta_YI_{XY}$ is significantly smaller than the entropic term $\tilde{\Sigma}^\mathrm{ex}_{Y}$. 
This implies that while $\Delta_Y I_{XY}$ is expected to be of the same order as the change in mutual information, no such constraint is imposed on $\tilde{\Sigma}^\mathrm{ex}_Y$.

\section{Discussion}
\label{sec:discussion}
In this paper, we propose a geometric housekeeping-excess decomposition of information flow for bipartite systems described by Markov jump processes. 
Our decomposition is based on the orthogonal decomposition of a thermodynamic force into its conservative and nonconservative components. The housekeeping information flow, driven by the nonconservative force, satisfies the antisymmetric relation [Eq.~\eqref{iflowhk_antisymmetry}]. Consequently, it does not contribute to the time derivative of mutual information between the two subsystems. 
Conversely, the excess information flow, driven by the conservative force, substantially changes the mutual information (see Eq.~\eqref{information_flow_derivative_excess}).

Our decomposition yields four main consequences. 

The first is the generalization of the second law of information thermodynamics. We derived two inequalities, Eqs.~\eqref{information_second_law_hk} and~\eqref{information_second_law_ex}, by combining the housekeeping/excess entropy change rates with the housekeeping/excess information flows.  
These two generalizations of the second law of information thermodynamics allow us to classify Maxwell's demons into those related to housekeeping dissipation and those related to excess dissipation.

The second consequence is the generalization of cyclic decomposition. By extending partial cycle affinity to non-steady states, we decomposed partial housekeeping EPR into contributions related to each cycle [Eq.~\eqref{partial_EPRhk_cyclic_decomposition}]. 
%In Appendix~\ref{app:proof_cyclic_decomposition_1}, we also demonstrate that this decomposition can be described by cycle affinities and cyclic currents for cycles within the projected graph (i.e., projected cycles).  
Furthermore,  by formally defining the information affinity for non-steady states, we demonstrated that the housekeeping information flow can be expressed solely in terms of contributions from global cycles [Eq.~\eqref{information_flow_cyclic_decomposition}]. 

The third consequence is the short-time TUR [Eq.~\eqref{TUR}]. Although Eq.~\eqref{TUR_2} is weaker than Eq.~\eqref{TUR}, it enables us to obtain a lower bound for partial excess EPR using only quantities that are more easily accessible: the rate of change of the expected value and the diffusivity of an observable. Thus, Eq.~\eqref{TUR_2} may be useful for inferring the partial excess EPR. 

The fourth is the information-thermodynamic speed limit [Eq.~\eqref{information_TSL}], which is based on optimal transport theory. We also proposed a geometric decomposition of partial EPR and discussed its connection to the Wasserstein distance for subsystems.

We discuss the application of optimal transport theory to information thermodynamics. This theory is employed to analyze minimal dissipation of information processing in information thermodynamics~\cite{nakazato2021geometrical,fujimoto2024game,nagase2024thermodynamically,kamijima2024optimal,kamijima2025finite} and to derive the finite-time Landauer principle~\cite{nakazato2021geometrical,van2023thermodynamic}. Because the problem of minimal dissipation depends on what is fixed and how optimization is performed~\cite{nagayama2025infinite,mulder2025bit}, various approaches are possible to formulate the problem of minimal dissipation.
For example, while the $2$-Wasserstein distance is typically employed in information thermodynamics for Langevin systems~\cite{nakazato2021geometrical,kamijima2024optimal}, the 1-Wasserstein distance is mainly used for Markov jump processes~\cite{kamijima2025finite,nagase2024thermodynamically,nagase2025thermodynamic}. Although this use of the 1-Wasserstein distance simplifies the analysis, it has some drawbacks. The 1-Wasserstein distance is not directly related to housekeeping and excess EPRs, and it requires the treatment of activity, which is a kinetic quantity. Our extension of the $2$-Wasserstein distance to the subsystems of Markov jump processes allows us to derive thermodynamic trade-off relations and analyze the optimality of information heat engines, as in the Langevin systems. It also naturally connects to the housekeeping-excess decomposition in information flow. It would be interesting to compare our generalization of the $2$-Wasserstein distance with a variant of the 1-Wasserstein distance that focuses on subsystems~\cite{kamijima2025finite,nagase2024thermodynamically,nagase2025thermodynamic}. 

We note that there may be alternative approaches to decomposing information flow. In this study, we decompose information flow using the geometric decomposition of the EPR, which is based on our definition of the inner product with the edgewise Onsager coefficients. This method is directly related to optimal transport theory and enables us to derive thermodynamic trade-off relations. For example, there are other options for the housekeeping-excess decomposition of the EPR in Markov jump processes: the traditional Hatano--Sasa decomposition~\cite{hatano2001steady,esposito2010three}, the decomposition based on Hessian geometry induced by the Legendre duality between force and current~\cite{kobayashi2022hessian}, and the decomposition based on information geometry of unidirectional fluxes~\cite{kolchinsky2024generalized}. 
Although it is unclear how to decompose information flow using these options, it may also be possible to do so based on these other options. 
Even if the decomposition succeeds, it is unlikely that alternative approaches will possess natural properties, such as Eqs.~\eqref{iflowhk_antisymmetry} and~\eqref{information_flow_derivative_excess}, in our decomposition of information flow. However, since information-geometric decomposition has advantages when decomposing the total EPR into partial EPRs in non-bipartite systems~\cite{ito2020unified}, it may have other desirable properties when an information-geometric decomposition is successful.

Finally, we discuss future perspectives. Information thermodynamics is studied in settings broader than bipartite systems described by Markov jump processes. These settings include non-bipartite systems~\cite{ito2013information,ito2016backward, chetrite2019information, wolpert2020uncertainty}, overdamped Langevin systems~\cite{allahverdyan2009thermodynamic,sagawa2012nonequilibrium, ito2013information,horowitz2014second, ito2015maxwell, nakazato2021geometrical}, underdamped Langevin systems~\cite{ito2011effects,horowitz2014second,kumasaki2025thermodynamic, dechant2025precision}, deterministic chemical reaction networks~\cite{amano2022insights,penocchio2022information}, and open quantum systems~\cite{sagawa2008second, ptaszynski2019thermodynamics, yada2022quantum}. The geometric housekeeping-excess decomposition of the EPR is possible in these systems~\cite{yoshimura2023housekeeping,dechant2022geometric1,dechant2022geometric2,yoshimura2025force}. Thus, the information exchange in these systems could also be decomposed into housekeeping and excess parts. In linear Langevin systems, the housekeeping EPR can also be decomposed into oscillatory modes~\cite{sekizawa2024decomposing}. Therefore, we may be able to decompose the housekeeping information flow into oscillatory modes. This notion is analogous to decomposing the housekeeping information flow with the cycle basis in this study. It would also be interesting to clarify how the housekeeping and the excess information flows behave in concrete systems, such as the macroscopic limit of electric circuits~\cite{freitas2022maxwell,freitas2023information}, information-driven heat engines~\cite{leighton2024information}, and biological processes such as membrane transport~\cite{yoshida2022thermodynamic}.

\begin{acknowledgments}
The authors thank Koya Katayama, Artemy Kolchinsky, and Andreas Dechant for discussions. We also thank Daan Mulder for his helpful comments on the manuscript.
K.Y.\ is supported by the Special Postdoctoral Researchers Program at RIKEN and JSPS KAKENHI Grants No.~24H00834.
S.I.\ is supported by JSPS KAKENHI Grants No.~22H01141, No.~23H00467, and No.~24H00834, JST ERATO Grant No.~JPMJER2302 and UTEC-UTokyo FSI Research Grant Program.
R.N.\ is supported by JSR Fellowship, the University of Tokyo, and JSPS KAKENHI Grants No.~25KJ0931.
\end{acknowledgments}

\appendix

\section{Derivation of the explicit form of information flow}
\label{app:information_flow_potential}
We derive the explicit form of information flow given by Eq.~\eqref{information_flow_potential}. Because of the symmetry between $X$ and $Y$, it is sufficient to consider $\dot{I}_X$. 
Here, we write the probability of the transition from $(x', y')$ to $(x, y)$ occurring within the time interval $[t, t+dt]$ as $P(x, y; t+dt | x', y'; t)$. As $X$ and $Y$ cannot change simultaneously in a bipartite system, $P(x, y; t+dt | x', y'; t) = o(dt)$ holds if $x \neq x'$ and $y \neq y'$, where $dt$ is an infinitesimal time interval and $o(dt)$ is the Landau symbol.

The joint probability $P(x; t+dt, y; t)$, which is needed to compute the mutual information $I(\hat{X}_{t+dt};\hat{Y}_t)$, is given in terms of the transition probability as 
\begin{align}
    &P(x; t+dt, y; t) \notag\\
    &= \sum_{x'\in\calX}\sum_{y'\in\calY} P(x, y'; t+dt| x', y; t)\pxy(x', y; t). 
\end{align}
Because 
\begin{align}
    &\sum_{y'\in\calY} P(x, y'; t+dt| x', y; t)\notag\\
    &=\begin{cases}
        P(x,y;t+dt|x',y;t) + o(dt) & (x\neq x')\\
        \sum_{y'\in\cal Y} P(x,y';t+dt|x,y;t) & (x= x'),
    \end{cases}
\end{align}
it can be further rewritten as 
\begin{align}
    &P(x;t+dt,y;t)\notag\\
    &=\sum_{x'(\neq x)}P(x,y;t+dt|x',y;t)p(x',y;t)\notag\\
    &+ \sum_{y'} P(x,y';t+dt|x,y;t)p(x,y;t) + o(dt). 
\end{align}
In the short-time limit, the transition probability is given by the transition rates as 
\begin{align}
    &P(x,y';t+dt|x',y;t)\notag\\
    &=\begin{cases}
        R(x,y|x',y)\delta_{y, y'}dt + o(dt) & (x\neq x')\\
        R(x,y'|x,y)\delta_{x, x'}dt + o(dt) & (y\neq y')\\
        1-\sum_{(x',y')\in \Delta_{x,y}}R(x',y'|x,y)dt + o(dt) & (\text{otherwise}),
    \end{cases}
\end{align}
where $R(\cdot|\cdot)=\sum_\nu R^{(\nu)}(\cdot|\cdot)$ and we defined 
\begin{align}
    \Delta_{x,y}
    =\{(x',y') \mid (x'=x\,\&\,y'\neq y)\;\text{or}\;(x'\neq x\,\&\,y'=y)\}. 
\end{align}
Therefore, we find 
\begin{align}
    &P(x;t+dt,y;t)\notag\\
    &= \sum_{x'(\neq x)}R(x,y|x',y)p(x',y;t)dt\notag\\
    &+ \sum_{y'(\neq y)}R(x,y'|x,y)p(x,y;t)dt\notag\\
    &+ p(x,y;t)
    - \sum_{(x',y')\in \Delta_{x,y}}R(x',y'|x,y)p(x, y; t)dt
    +o(dt). \\
    &= p(x,y;t) + \sum_{x'(\neq x)}R(x,y|x',y)p(x',y;t)dt\notag\\
    &- \sum_{x'(\neq x)} R(x',y|x,y)p(x, y; t)dt +o(dt). 
\end{align}
For fixed $y$, each choice of $x'(\neq x)$ and $\nu$ corresponds to an element in $\edgeset_X$ as $e=(\nu,(x,y)\to(x',y))$ or $(\nu,(x',y)\to(x,y))$. Thus, we can easily see 
\begin{align}
    &\sum_{x'(\neq x)}\big(R(x,y|x',y)p(x',y;t)-R(x',y|x,y)p(x, y; t)\big)\notag\\
    &=\sum_{e\in\edgeset_X}(\delta_{(x,y),\target(e)}-\delta_{(x,y),\start(e)})J_e(\Pxy(t)). 
\end{align}
Because the coefficient on the right-hand side coincides with the incidence matrix as 
$[\div_{G_X}]_{(x, y), e} = \delta_{(x, y), \target(e)} - \delta_{(x, y), \start(e)}$, we obtain 
\begin{align}
    P(x; t+dt, y; t) = p(x, y; t) + [\div_{G_X}\Current_X]_{(x,y)}dt + o(dt).
    \label{probability_partial_derivative}
\end{align}
Multiplying $\Pi_X$, we can also find 
\begin{align}
    p_X (x; t+dt) = p_X(x;t) + [\Pi_X\div_{G_X}\Current_X]_x dt + o(dt) \label{marginal_probability_partial_derivative}
\end{align}
because $[\Pi_X]_{x', (x,y)} =\delta_{x',x}$ and $\sum_y P(x;t+dt,y;t)=p_X(x,t+dt)$. 

We derive Eq.~\eqref{information_flow_potential} from Eqs.~\eqref{probability_partial_derivative} and \eqref{marginal_probability_partial_derivative}. 
Recall that the mutual information $I(\hat{X}_{t+dt};\hat{Y}_t)$ is defined as 
\begin{align}
    &I(\hat{X}_{t+dt};\hat{Y}_t)\notag\\
    &=D_{\mathrm{KL}}(\bm{P}(t+dt,t)\|\Px\otimes\Py(t+dt,t)) \notag\\
    &=\sum_{x,y}
    P(x;t+dt,y;t)\ln\frac{P(x;t+dt,y;t)}{p_X(x;t+dt)p_Y(y;t)}. 
\end{align}
Through straightforward computation, those relations lead to 
\begin{align}
        &\ln\left(\frac{P(x; t+dt, y; t)}{p_X(x;t+dt)p_Y(y;t)}\right) 
        = \ln\left(\frac{\pxy(x, y; t)}{p_X(x;t)p_Y(y;t)}\right) \nonumber
        \\&\;\;+ \left(\frac{[\div_{G_X}\Current_X]_{(x,y)}}{\pxy(x, y; t)} - \frac{[\Pi_X\div_{G_X}\Current_X]_x}{p_X(x;t)}\right)dt + o(dt). 
\end{align}
We also find 
\begin{align}
        & \sum_{x,y} p(x,y;t) \left(\frac{[\div_{G_X}\Current_X]_{(x,y)}}{\pxy(x, y; t)} - \frac{[\Pi_X\div_{G_X}\Current_X]_x}{p_X(x;t)}\right) \nonumber\\
        &= \sum_{x,y}[\div_{G_X}\Current_X]_{(x,y)} - \sum_x [\Pi_X\div_{G_X}\Current_X]_x =0. 
\end{align}
Therefore, the mutual information $I(\hat{X}_{t+dt}; \hat{Y}_t)$ reads 
\begin{align}
    &I(\hat{X}_{t+dt}; \hat{Y}_t)\notag\\
    &= \sum_{x, y}P(x; t+dt, y;t)\ln\left(\frac{p(x, y; t)}{p_X(x; t+dt)p_Y(y; t)}\right)\notag\\
    &+o(dt) \notag\\
    &= I(\hat{X}_t; \hat{Y}_t) + \sum_{x, y}dt [\div_{G_X}\Current_X]_{(x,y)}\ln\left(\frac{p(x, y; t)}{p_X(x;t)p_Y(y;t)}\right) 
    \notag\\
    &+o(dt).
\end{align}
Combining the above equality with Eq.~\eqref{information_potential_component} leads to
\begin{align}
    &\frac{I(\hat{X}_{t + dt}; \hat{Y}_t) -I(\hat{X}_t; \hat{Y}_t)}{dt} \notag\\
    &= (\div_{G_X}\Current_X(\Pxy(t)))^{\top}\potentialit(\Pxy(t)) + o(dt) \notag\\
    &= \innerprod{\Current_X(\Pxy(t))}{\nabla_{G_X}\potentialit(\Pxy(t))}+ o(dt),
\end{align}
and taking the limit $dt \to +0$ yields
\begin{align}
    \iflow_X(\Pxy(t)) = \innerprod{\Current_{X}(\Pxy(t))}{\nabla_{G_X}\potentialit(\Pxy(t))}.
\end{align}

\section{Graph projection}
\label{app:graph_projection}
\begin{figure}
    \centering
    \includegraphics[width=\linewidth]{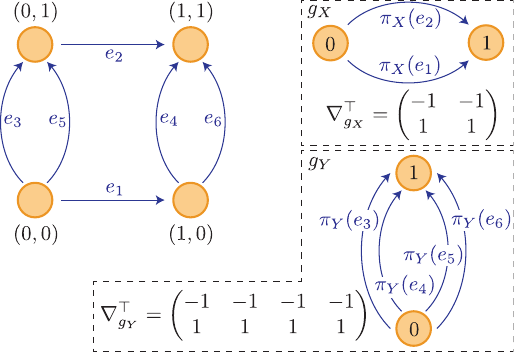}
    \caption{An example of the graph projection method using the graph $G$ shown in Fig.~\ref{fig:decomposed_graphs}. We can confirm that the incidence matrices of $g_X$ and $g_Y$ are given by $\div_{g_X} = \Pi_X\div_{G_X}$ and $\div_{g_Y} = \Pi_Y\div_{G_Y}$, respectively.}
    \label{fig:graph_projection}
\end{figure}
Here, we explain the graph projection method for a graph $G=(\calX\times\calY,\edgeset_X\cup\edgeset_Y)$ (see also Fig.~\ref{fig:graph_projection} for an example). 
A projected graph $g_\alpha=(\mathcal{A}, \varepsilon_\alpha)$ is defined for $(\alpha, \mathcal{A}) \in\{(X, \calX), (Y, \calY)\}$ by a set of edges $\varepsilon_\alpha$ constructed from $\edgeset_\alpha$ as follows.
For an edge $e = (\nu, (x, y)\to(x', y)) \in \edgeset^\mathrm{all}_X$, the one-to-one map $\pi_X$ is defined as $\pi_X(e) = ((\nu, y), x\to x')$. Here, to ensure that the map is one-to-one, we include $y = \starty(e) = \targety(e)$ in the label of $\pi_X(e)$ in addition to $\nu$. Similarly, for an edge $e = (\nu, (x, y)\to(x, y')) \in \edgeset^\mathrm{all}_Y$, the one-to-one map $\pi_Y$ is defined as $\pi_Y(e) = ((\nu, x), y\to y')$. 
Using these maps $\pi_X$ and $\pi_Y$, the sets of edges $\varepsilon_X$ and $\varepsilon_Y$ are defined as $\varepsilon_X=\{\pi_X(e) \mid e\in\edgeset_X \}$ and $\varepsilon_Y=\{\pi_Y(e) \mid e\in\edgeset_Y \}$, respectively.

We can confirm that the incidence matrix of the projected graph $g_{\alpha}$, written as $\div_{g_\alpha}$, is obtained by marginalizing the incidence matrix of $G_\alpha$ for $\alpha \in \{ X, Y\}$.
First note that since every $e'\in\varepsilon_\alpha$ has a unique counterpart $e$ in $\edgeset_\alpha$ (i.e., $\pi_\alpha^{-1}(e')$), a matrix whose rows/columns correspond to edges in $\varepsilon_\alpha$ can be labeled by $e\in\edgeset_\alpha$. 
Indicating the element of $\div_{g_\alpha}$ representing the pair $(x,\pi_X(e))\in \mathcal{A}\times\varepsilon_\alpha$ by $[\div_{g_X}]_{x, e}$, we can show 
\begin{align}
    \Pi_{\alpha}\div_{G_{\alpha}} = \div_{g_{\alpha}}.\label{projected_divergence}
\end{align}
This is because, when $\alpha=X$, $[\Pi_X \div_{G_X}]_{x, e} = \delta_{x, \targetx(e)} - \delta_{x, \startx(e)} = [\div_{g_X}]_{x, e}$ holds for any node $x\in\mathcal{X}$ and edge $e\in\edgeset_X$. Similarly, we also obtain $[\Pi_Y \div_{G_Y}]_{y, e} = [\div_{g_Y}]_{y, e}$ for any node $y\in\mathcal{Y}$ and edge $e\in\edgeset_Y$. Using Eq.~\eqref{projected_divergence}, we can interpret Eq.~\eqref{partial_continuity_eq} as a continuity equation on the projected graph $g_\alpha$:
\begin{align}
    d_t\Pa = \div_{g_{\alpha}}\Current_{\alpha}(\Pz). \label{modified_partial_continuity_eq}
\end{align}

\label{app:proof_cyclic_decomposition_1}

\section{Proof of Eq.~\eqref{partial_EPRhk_cyclic_decomposition}}
We prove the cyclic decomposition of partial housekeeping EPR [Eq.~\eqref{partial_EPRhk_cyclic_decomposition}].

To this end, we first introduce a matrix notation for cycles in bipartite systems.
Let us recall the vector $\cyclebasis^\mu$ and decompose it vertically as $\cyclebasis^\mu = \cyclebasis^\mu_X\oplus\cyclebasis^\mu_Y$.
By explicitly specifying the cases in which the cycle is local, we can write the vector cycle $\cyclebasis^\mu$ as follows:
\begin{align}
    \cyclebasis^\mu = 
    \begin{cases}
        \cyclebasis^\mu_X\oplus\bm{0} & (\mu\in \mathcal{M}^X),\\
        \bm{0}\oplus\cyclebasis^\mu_Y & (\mu\in \mathcal{M}^Y),\\
        \cyclebasis^\mu_X\oplus\cyclebasis^\mu_Y & (\mu\in \mathcal{M}^{XY}).
    \end{cases} \label{cyclebasis_vectors_correspondence}
\end{align}
It is worth mentioning that the equality
\begin{gather}
    \div_{G_{\alpha}}\cyclebasis^\mu_\alpha = \bm{0}\quad  (\mu\in\mathcal{M}^\alpha) \label{calculation_rule_2}
\end{gather}
holds for any $\mu\in\mathcal{M}^\alpha$ with $\alpha \in \{X,Y \}$. This equality follows from $\div\cyclebasis^\mu = \bm{0}$ and $\div = (\div_{G_X}\,\div_{G_Y})$ [Eq.~\eqref{divergence_decomposition}].

Then, let us derive Eq.~\eqref{partial_EPRhk_cyclic_decomposition}. 
Focusing on the expansion of $\Currenthk$ [Eq.~\eqref{housekeeping_cyclic_decomp}], the decomposition $\Currenthk = \Currenthk_X\oplus\Currenthk_Y$, and Eq.~\eqref{cyclebasis_vectors_correspondence}, we obtain the following expansion:
\begin{align}
    \Currenthk_\alpha(\Pxy) = \sum_{\mu\in\mathcal{M}^\alpha\cup\mathcal{M}^{XY}}\calJ^\mathrm{hk}(\Pxy; C^\mu)\cyclebasis^\mu_\alpha. \label{eq:expansion_partial} %expansion of the partial housekeeping current
\end{align}
On the other hand, we can also rewrite the partial cycle affinity as
\begin{align}
    \calF^\mathrm{hk}_\alpha(\Pxy; C^\mu) &= \sum_{e\in\edgeset_\alpha}[\cyclebasis^\mu]_e[\Forcehk (\Pxy)]_e \notag\\
    &= \sum_{e\in\edgeset_\alpha}[\cyclebasis^\mu_\alpha]_e[\Forcehk_\alpha(\Pxy)]_e \notag\\
    &= \innerprod{\cyclebasis^\mu_\alpha}{\Forcehk_\alpha(\Pxy)}.
\end{align}
Therefore, we finally obtain the desired equality [Eq.~\eqref{partial_EPRhk_cyclic_decomposition}]:
\begin{align}
    \EPRhk_\alpha(\Pz) &= \innerprod{\Currenthk_{\alpha}(\Pxy)}{\Forcehk_{\alpha}(\Pxy)} \notag\\
    &= \bigg\langle\sum_{\mu\in\mathcal{M}^\alpha\cup\mathcal{M}^{XY}}\calJ^\mathrm{hk}(\Pz; C^\mu)\cyclebasis^\mu_\alpha,\;\Forcehk_\alpha(\Pz)\bigg\rangle \notag\\
    &= \sum_{\mu\in\mathcal{M}^\alpha\cup\mathcal{M}^{XY}}\calJ^\mathrm{hk}(\Pz; C^\mu)\innerprod{\cyclebasis^\mu_\alpha}{\Forcehk_\alpha(\Pz)} \notag\\
    &= \sum_{\mu\in\mathcal{M}^\alpha\cup\mathcal{M}^{XY}}\calJ^\mathrm{hk}(\Pz; C^\mu)\calF^\mathrm{hk}_\alpha(\Pz; C^\mu).
\end{align}

\section{Information affinity for a local cycle and proof of Eq.~\eqref{information_flow_cyclic_decomposition}}
\label{app:proof_cyclic_decomposition_2}
%First, we discuss the information affinity for a local cycle. Then, we rewrite the information affinity as 

We derive the cyclic decomposition of the housekeeping information flow [Eq.~\eqref{information_flow_cyclic_decomposition}].
Let us first show $\calF^\mathrm{info}_\alpha(\Pz; C^\mu) = 0$ for $\mu\in\mathcal{M}^X\cup\mathcal{M}^Y$ by rewriting the information affinity as follows:

\begin{align}
\calF^\mathrm{info}_\alpha(\Pz; C^\mu) &= \sum_{e\in\edgeset_\alpha}[\cyclebasis^\mu]_e([\potentialit(\Pz)]_{\target(e)} - [\potentialit(\Pz)]_{\start(e)}) \nonumber \\
&= \sum_{e\in\edgeset_\alpha}[\cyclebasis^\mu_\alpha]_e \sum_{(x, y)\in \mathcal{X} \times \mathcal{Y} }[\nabla_{G_\alpha}]_{e, (x,y)}[\potentialit(\Pz)]_{(x,y)}\nonumber\\
&= \innerprod{\cyclebasis^\mu_\alpha}{\nabla_{G_\alpha}\potentialit(\Pz)}.
\label{infoaffinity}
\end{align}
Because $\cyclebasis^\mu_\alpha = \bm{0}$ for $\mu\in\mathcal{M}^{\alphac}$
 and $\div_{G_\alpha}\cyclebasis^\mu_\alpha = \bm{0}$ for $\mu\in\mathcal{M}^\alpha$ [Eq.~\eqref{calculation_rule_2}], we find that $\calF^\mathrm{info}_\alpha(\Pz; C^\mu) = 0$ for $\mu\in\mathcal{M}^X\cup\mathcal{M}^Y$ based on this expression.

We next prove Eq.~\eqref{information_flow_cyclic_decomposition}. From Eq.~\eqref{eq:expansion_partial} and Eq.~\eqref{infoaffinity}, we obtain
\begin{align}
    \iflowhk_\alpha(\Pz) &=\innerprod{\Currenthk_{\alpha}(\Pz)}{\nabla_{G_{\alpha}}\potentialit(\Pz)} \nonumber\\
    &= \bigg\langle\sum_{\mu\in\mathcal{M}^{\alpha} \cup \mathcal{M}^{XY}}\calJ^\mathrm{hk}(\Pz; C^\mu)\cyclebasis^\mu_{\alpha},\;\nabla_{G_\alpha}\potentialit(\Pz)\bigg\rangle \notag\\
    &= \sum_{\mu\in\mathcal{M}^{\alpha} \cup \mathcal{M}^{XY}}\calJ^\mathrm{hk}(\Pz; C^\mu)\innerprod{\cyclebasis^\mu_\alpha}{\nabla_{G_\alpha}\potentialit(\Pz)} \notag\\
    &= \sum_{\mu\in\mathcal{M}^{XY}}\calJ^\mathrm{hk}(\Pz; C^\mu)\calF^\mathrm{info}_\alpha(\Pz; C^\mu),
\end{align}
where we used $\calF^\mathrm{info}_\alpha(\Pz; C^\mu) = 0$ for $\mu\in \mathcal{M}^{\alpha}$.

\section{Proof of Eq.~\eqref{diffusivity_inequality}}
\label{app:diffusivity_inequality}
We prove Eq.~\eqref{diffusivity_inequality}. First, we derive an explicit expression for the diffusivity of a time-independent observable $\bm{\Phi}$. Then we apply this expression to $\bm{\Psi}_\alpha$ and show Eq.~\eqref{diffusivity_inequality}.

Here, let us calculate the diffusivity of $\bm{\Phi}$, which is given by the short-time limit of the variance of $\delta\bm{\Phi}$. The variance is defined as 
\begin{align}
    \mathrm{Var}[\delta\bm{\Phi}] \coloneqq \mathbb{E}[(\delta\bm{\Phi})^2] - (\mathbb{E}[\delta\bm{\Phi}])^2, \label{variance}
\end{align}
where $\mathbb{E}[\cdot]$ denotes an ensemble average. To calculate the ensemble averages, it is necessary to consider the joint probability of $z(t) = z$ and $z(t+\Delta t) = z'$, denoted by $\mathbb{P}(z'; t+\Delta t, z; t)$. Also, we write $P(z'; t+\Delta t|z; t)$ for the conditional probability of $z(t+\Delta t) = z'$ given $z(t) = z$. Since $P(z'; t+\Delta t|z; t) = \delta_{z, z'}+ \sum_{\nu}R^{(\nu)}(z'|z)\Delta t +o(\Delta t)$ for sufficiently small $\Delta t$, we have
\begin{align}
    &\mathbb{P}(z'; t+\Delta t, z; t) \notag\\
    =& P(z'; t+\Delta t|z; t)p(z; t)\notag\\
    =& \delta_{z, z'}p(z; t) + \sum_{\nu}R^{(\nu)}(z'|z)p(z; t)\Delta t +o(\Delta t) \notag\\
    =& \delta_{z, z'}[p(z; t)+ \sum_{\nu}R^{(\nu)}(z|z)p(z; t)\Delta t ] \nonumber \\ 
    &+ \Delta t\sum_{e\in\edgeset^\mathrm{all}}J^+_e(\Pz(t))\delta_{z', \target(e)}\delta_{z, \start(e)} + o(\Delta t).
\end{align}
To obtain the last equality, we used the fact that for $z'\neq z$, $R^{(\nu)}(z'|z)>0 \iff (\nu, z\to z')\in\edgeset^\mathrm{all}$. Therefore, the ensemble average of $\delta\bm{\Phi}$ is asymptotically given as
\begin{align}
    &\mathbb{E}[\delta\bm{\Phi}] = \sum_{(z, z')\in\mathcal{Z}\times\mathcal{Z}}([\bm{\Phi}]_{z'} - [\bm{\Phi}]_{z})\mathbb{P}(z'; t+\Delta t, z; t) \notag\\
    \begin{split}
        &= \Delta t\sum_{(z, z')\in\mathcal{Z}\times\mathcal{Z}}\sum_{e\in\edgeset^\mathrm{all}}([\bm{\Phi}]_{z'} - [\bm{\Phi}]_{z})J^+_e(\Pz(t))\delta_{z', \target(e)}\delta_{z, \start(e)}\\
        &\quad\quad\quad\quad\quad\quad\quad\quad\quad\quad\quad\quad\quad\quad\quad\quad\quad\quad\quad\quad
        +o(\Delta t)
    \end{split}
    \notag\\
    &= \Delta t\sum_{e\in\edgeset^\mathrm{all}}J^+_e(\Pz(t))([\bm{\Phi}]_{\target(e)} - [\bm{\Phi}]_{\start(e)}) + o(\Delta t).
\end{align}
In the same way, we have
\begin{align}
    &\mathbb{E}[(\delta\bm{\Phi})^2] = \sum_{(z, z')\in\mathcal{Z}\times\mathcal{Z}}([\bm{\Phi}]_{z'} - [\bm{\Phi}]_{z})^2\mathbb{P}(z'; t+\Delta t, z; t) \notag\\
    &= \Delta t\sum_{e\in\edgeset^\mathrm{all}}J^+_e([\bm{\Phi}]_{\target(e)} - [\bm{\Phi}]_{\start(e)})^2 + o(\Delta t) \notag\\
    &= \Delta t\sum_{e\in\edgeset}\sum_{e'\in\{e, -e\}}J^+_{e'}([\bm{\Phi}]_{\target(e')} - [\bm{\Phi}]_{\start(e')})^2 + o(\Delta t) \notag\\
    &= \Delta t\sum_{e\in\edgeset}(J^+_e + J^-_e)([\bm{\Phi}]_{\target(e)} - [\bm{\Phi}]_{\start(e)})^2 + o(\Delta t).
\end{align}
Here, we used $J^+_{-e} = J^-_e$ and $\target(-e) = \start(e)$ to obtain the last equality. Combining the above calculations with Eq.~\eqref{variance} yields 
\begin{align}
    &\mathrm{Var}[\delta\bm{\Phi}]\notag\\
    &= \Delta t\sum_{e\in\edgeset}(J^+_e + J^-_e)([\bm{\Phi}]_{\target(e)} - [\bm{\Phi}]_{\start(e)})^2 + o(\Delta t),
\end{align}
which leads to the following formula:
\begin{align}
    \mathcal{D}(\bm{\Phi}) &= \lim_{\Delta t\to +0}\frac{\mathrm{Var}[\delta\bm{\Phi}]}{\Delta t} \notag\\
    &= \sum_{e\in\edgeset}(J^+_e + J^-_e)([\bm{\Phi}]_{\target(e)} - [\bm{\Phi}]_{\start(e)})^2. \label{diffusivity_formula}
\end{align}

We now prove Eq.~\eqref{diffusivity_inequality} by applying Eq.~\eqref{diffusivity_formula} to $\bm{\Psi}_\alpha = \Pi^\top_\alpha\bm{\psi}_\alpha$. To this end, we focus on the fact that for $e\in\edgeset_{\alphac}$, $[\bm{\Psi}_\alpha]_{\target(e)} - [\bm{\Psi}_\alpha]_{\start(e)} = 0$. It can be confirmed by $[\bm{\Psi}_\alpha]_{\target(e)} = [\bm{\psi}_\alpha]_{\targeta(e)}$, $[\bm{\Psi}_\alpha]_{\start(e)} = [\bm{\psi}_\alpha]_{\starta(e)}$, and $\targeta(e) = \starta(e)$ for $e\in\edgeset_{\alphac}$. Therefore, we can rewrite $\mathcal{D}(\bm{\Psi}_\alpha)$ as follows:
\begin{align}
    \mathcal{D}(\bm{\Psi}_\alpha) &= \sum_{e\in\edgeset}(J^+_e + J^-_e)([\bm{\Psi}_\alpha]_{\target(e)} - [\bm{\Psi}_\alpha]_{\start(e)})^2 \notag\\
    &= \sum_{e\in\edgeset_\alpha}(J^+_e + J^-_e)([\bm{\Psi}_\alpha]_{\target(e)} - [\bm{\Psi}_\alpha]_{\start(e)})^2 \notag\\
    &= \sum_{e\in\edgeset_\alpha}(J^+_e + J^-_e)[\nabla_{G_\alpha}\bm{\Psi}_\alpha]^2_e.
\end{align}
Here, we used $[\nabla_{G_\alpha}\bm{\Psi}_\alpha]_e = [\bm{\Psi}_\alpha]_{\target(e)} - [\bm{\Psi}_\alpha]_{\start(e)}$ to obtain the last equality. Finally, by using the inequality $J^+_e + J^-_e\geq 2(J^+_e - J^-_e)/(\ln J^+_e - \ln J^-_e) = 2l_e$, we arrive at
\begin{align}
    &2\norm{\nabla_{G_\alpha}\bm{\Psi}_\alpha}^2_{L_\alpha} \notag\\
    &= 2\sum_{(e, e')\in\edgeset_\alpha\times\edgeset_\alpha}[\nabla_{G_\alpha}\bm{\Psi}_\alpha]_e[L_\alpha]_{e, e'}[\nabla_{G_\alpha}\bm{\Psi}_\alpha]_{e'} \notag\\
    &= 2\sum_{(e, e')\in\edgeset_\alpha\times\edgeset_\alpha}[\nabla_{G_\alpha}\bm{\Psi}_\alpha]_el_e\delta_{e, e'}[\nabla_{G_\alpha}\bm{\Psi}_\alpha]_{e'} \notag\\
    &= \sum_{e\in\edgeset_\alpha}2l_e[\nabla_{G_\alpha}\bm{\Psi}_\alpha]^2_{e} \notag\\
    &\leq \sum_{e\in\edgeset_\alpha}(J^+_e + J^-_e)[\nabla_{G_\alpha}\bm{\Psi}_\alpha]^2_e = \mathcal{D}(\bm{\Psi}_\alpha).
\end{align}

\section{The conservativeness of the minimizer}
\label{app:conservativeness}
We prove the formula~\eqref{subsystem_Wasserstein_distance_potential}, which implies the conservativeness of the minimizer $\force^*_{\alpha}(t) =-\nabla_{g_{\alpha}}\bm{\varphi}^*_\alpha(t)$. First, we consider the following functional associated with a triple $(\bm{p}\in\bbR^{|\calX\times\calY|}, \force_\alpha\in\bbR^{|\edgeset_{\alpha}|}, \bm{\varphi}_\alpha\in\bbR^{|\mathcal{A}|})$ with $(\alpha, \mathcal{A})\in\{(X, \calX), (Y, \calY)\}$:
\begin{align}
    &\mathbb{I}[\bm{p}, \force_\alpha, \bm{\varphi}_\alpha] =\notag\\
    \begin{split}
        &\int_0^{\tau}\biggl(\frac{1}{2}\norm{\force_\alpha(t)}^2_{L_\alpha(\bm{p}(t))}\\
        &\quad\quad-(\bm{\varphi}_\alpha(t))^\top [d_t\Pi_{\alpha}\bm{p}(t) - \div_{g_{\alpha}}L_\alpha(\bm{p}(t))\force_\alpha(t)]\biggr)dt \label{functional}
    \end{split}.
\end{align}
The Lagrange multiplier method corresponding to Eq.~\eqref{subsystem_Wasserstein_distance} under the constraints Eqs.~\eqref{cond1_projection} and~\eqref{cond2_dynamics_conservation} provides the optimization problem:
\begin{align}
    \frac{1}{2\tau}\Wsub(\Pa^{(0)}, \Pa^{(1)})^2 = \inf_{\bm{p}, \force_\alpha}\max_{\bm{\varphi}_\alpha}\mathbb{I}[\bm{p}, \force_\alpha, \bm{\varphi}_\alpha], \label{Lagrange_multiplier}
\end{align}
subject to $\Pi_{\alpha}\bm{p}(0) = \Pa^{(0)}$ and $\Pi_{\alpha}\bm{p}(\tau) = \Pa^{(1)}$.  
The functional derivative of $\mathbb{I}$ with respect to $[\force_\alpha (t)]_e=:f_e (t)$ is given by
\begin{align}
    \frac{\delta\mathbb{I}[\bm{p}, \force_\alpha, \bm{\varphi}_\alpha]}{\delta f_e (t)} = l_e(\bm{p}(t))f_e(t) + l_e(\bm{p}(t))[\nabla_{g_{\alpha}}\bm{\varphi}_\alpha(t)]_e,
\end{align}
and the minimizer $(\bm{p}^*, \force^*_\alpha, \bm{\varphi}^*_\alpha)$ satisfies
\begin{align}
    \frac{\delta\mathbb{I}[\Pxy^*, \force^*_\alpha, \bm{\varphi}^*_\alpha]}{\delta f_e(t)} = 0
\end{align}
for all $e\in\edgeset_\alpha$ and $t \in [0, \tau]$.
Therefore, we find that the optimal force $\force^*_\alpha$ is given by
\begin{align}
    \force^*_\alpha (t)= -\nabla_{g_{\alpha}}\bm{\varphi}^*_\alpha (t).\label{minimizer_force}
\end{align}

\section{Proof of the symmetry }
\label{app:symmetry_subsystem_Wasserstein_distance}
We prove the symmetry $\Wsub(\Pa^{(0)}, \Pa^{(1)}) = \Wsub(\Pa^{(1)}, \Pa^{(0)})$ under the assumption that a minimizer  of Eq.~\eqref{subsystem_Wasserstein_distance} exists.
Let $\{(\Pxy^*(t), \force^*_\alpha(t))\}_{t\in[0, \tau]}$ be the minimizer of Eq.~\eqref{subsystem_Wasserstein_distance}. Then, the time-reversal pair $\{(\Pxy^\dag(t), \force_\alpha^\dag(t))\}_{t\in[0, \tau]} \coloneqq \{(\Pxy^*(\tau - t), -\force_\alpha^*(\tau - t))\}_{t\in[0, \tau]}$ satisfies conditions Eqs.~\eqref{cond1_projection},~\eqref{cond2_dynamics_conservation}, and~\eqref{cond3_initial_final_distributions} for the optimization problem [Eq.~\eqref{subsystem_Wasserstein_distance}], where $\Pxy^{(0)}$ and $\Pxy^{(1)}$ are exchanged. Therefore, the following inequality
\begin{align}
\left(\tau\int_0^\tau\norm{\force^\dag_{\alpha}(t)}_{L_\alpha(\Pxy^\dag(t))}^2dt\right)^{1/2} \geq \Wsub(\Pa^{(1)}, \Pa^{(0)}),
\label{maybeoptimal}
\end{align}
holds. Combining 
\begin{align}
    \Wsub(\Pa^{(0)}, \Pa^{(1)}) &= \left(\tau\int_0^\tau\norm{\force^*_{\alpha}(t)}_{L_\alpha(\Pxy^*(t))}^2dt\right)^{1/2} \notag\\
    &= \left(\tau\int_0^\tau\norm{\force^\dag_{\alpha}(t)}_{L_\alpha(\Pxy^\dag(t))}^2dt\right)^{1/2},
\end{align}
with the inequality Eq.~\eqref{maybeoptimal} yields $\Wsub(\Pa^{(0)}, \Pa^{(1)}) \geq \Wsub(\Pa^{(1)}, \Pa^{(0)})$.
The reverse inequality $\Wsub(\Pa^{(0)}, \Pa^{(1)}) \leq \Wsub(\Pa^{(1)}, \Pa^{(0)})$ can also be obtained in the same way, thereby proving the symmetry $\Wsub(\Pa^{(0)}, \Pa^{(1)}) = \Wsub(\Pa^{(1)}, \Pa^{(0)})$.

\section{Proof of the triangle inequality}
\label{app:triangle_inequality}
Here, we prove the triangle inequality of the $2$-Wasserstein distance for the subsystem under the assumption that a minimizer exists. Since a geodesic travels at a constant speed, i.e.,  $\norm{\force_{\alpha} (t) }_{L_\alpha(\Pxy(t))} = {\rm const.}$ (see also Appendix~\ref{app:geodesic}), the definition of the $2$-Wasserstein distance for the subsystem can be rewritten as follows:
\begin{align}
    &\Wsub(\Pa^{(0)}, \Pa^{(1)}) \nonumber\\ &= \min_{(\Pxy(t), \force_{\alpha}(t))_{0 \leq t \tau}}\left(\tau\int_0^{\tau}\norm{\force_{\alpha}(t)}_{L_\alpha(\Pxy(t))}^2dt\right)^{\frac{1}{2}} \nonumber \\
    &= \min_{(\Pxy(t), \force_{\alpha}(t))_{0 \leq t \tau}}\int_0^{\tau}\norm{\force_{\alpha} (t) }_{L_\alpha(\Pxy(t))}dt, \label{subsystem_Wasserstein_distance_alternative_def}
\end{align}
under the constraints [Eqs.~\eqref{cond1_projection}, \eqref{cond2_dynamics_conservation}, and \eqref{cond3_initial_final_distributions}].
This definition has the same minimizer as the one in  Eq.~\eqref{subsystem_Wasserstein_distance}. For simplicity, we set $\tau = 1$. Let $(\Pxy^{(10)}, \force_{\alpha}^{(10)})$ and $(\Pxy^{(21)}, \force_{\alpha}^{(21)})$ be the minimizers that give $\Wsub(\Pa^{(0)}, \Pa^{(1)})$ and $\Wsub(\Pa^{(1)}, \Pa^{(2)})$, respectively. Then, we define the pair $(\Pxy', \force'_{\alpha})$ as 
\begin{gather}
    \Pxy'(t) = 
    \begin{cases}
        \Pxy^{(10)}(2t) & \text{if $t\in [0, 1/2]$},\\
        \Pxy^{(21)}(2t-1) & \text{if $t\in (1/2, 1]$},
    \end{cases}\\
    \force'_{\alpha}(t) = 
    \begin{cases}
        2\force_{\alpha}^{(10)}(2t) & \text{if $t\in [0, 1/2]$},\\
        2\force_{\alpha}^{(21)}(2t-1) & \text{if $t\in (1/2, 1]$}.
    \end{cases}
\end{gather}
Note that the pair $(\Pxy', \force'_{\alpha})$ satisfies the constraints, Eq.~ \eqref{cond2_dynamics_conservation}, $\Pi_{\alpha}\Pxy'(0) =\Pa^{(0)}$ and $\Pi_{\alpha}\Pxy'(1) =\Pa^{(2)}$. Therefore, we can derive the triangle inequality as follows:
\begin{align}
    &\Wsub(\Pa^{(0)}, \Pa^{(2)})\leq \int_0^1\norm{\force'_\alpha(t)}_{L_\alpha(\Pxy'(t))}dt \notag\\
    \begin{split}
        &=2\int_0^{1/2}\norm{\force^{(10)}_\alpha(2t)}_{L_\alpha(\Pxy^{(10)}(2t))}dt\\
        &\quad+ 2\int_{1/2}^{1}\norm{\force^{(21)}_\alpha(2t-1)}_{L_\alpha(\Pxy^{(21)}(2t-1))}dt
    \end{split} \notag\\
    &= \Wsub(\Pa^{(0)}, \Pa^{(1)}) + \Wsub(\Pa^{(1)}, \Pa^{(2)}),
\end{align}
where we used  $2\int_0^{1/2}\norm{\force^{(10)}_\alpha(2t)}_{L_\alpha(\Pxy^{(10)}(2t))}dt=\Wsub(\Pa^{(0)}, \Pa^{(1)})$ and $2\int_{1/2}^{1}\norm{\force^{(21)}_\alpha(2t-1)}_{L_\alpha(\Pxy^{(21)}(2t-1))}dt=\Wsub(\Pa^{(1)}, \Pa^{(2)})$.

\section{Proof of the constant speed property of a geodesic}
\label{app:geodesic}
We show here that a geodesic has a constant speed under the assumption that a minimizer exists. Firstly, using the Cauchy--Schwarz inequality, we obtain
\begin{align}
    \tau\int_0^\tau \norm{\force_{\alpha (t)}}^2_{L_\alpha(\Pxy (t))}dt\geq \left(\int_0^\tau \norm{\force_{\alpha}(t)}_{L_\alpha(\Pxy (t) )}dt\right)^2,\label{forward_inequality}
\end{align}
for any pair $(\Pxy, \force_{\alpha})$. Secondly, as we will show later, the minimizer $(\Pxy^*, \force^*_{\alpha} = -\nabla_{g_{\alpha}}\bm{\varphi}^*_{\alpha})$ of Eq.~\eqref{subsystem_Wasserstein_distance} satisfies the following inverse inequality:
\begin{align}
    \tau\int_0^\tau \norm{\force^*_{\alpha }(t)}^2_{L_\alpha(\Pxy^*(t))}dt \leq \left(\int_0^\tau \norm{\force^*_{\alpha }(t)}_{L_\alpha(\Pxy^*(t))}dt\right)^2. \label{inverse_inequality}
\end{align}
Combining Eqs.~\eqref{forward_inequality} and~\eqref{inverse_inequality} yields
\begin{align}
      \left(\int_0^\tau \norm{\force^*_{\alpha }(t)}_{L_\alpha(\Pxy^*(t))}dt\right)^2 &=\tau\int_0^\tau \norm{\force^*_{\alpha }(t)}^2_{L_\alpha(\Pxy^*(t))}dt \nonumber\\
      &= \Wsub(\Pa^{(0)}, \Pa^{(1)})^2.
\end{align}
Since the equality in the Cauchy--Schwarz inequality holds if and only if $\norm{\force^*_{\alpha }(t)}^2_{L_\alpha(\Pxy(t))} = \mathrm{const.}$, we conclude that the geodesic has constant speed and satisfies
\begin{align}
     \norm{\force^*_{\alpha }(t)}_{L_\alpha(\Pxy^*(t))} 
     = \frac{\Wsub(\Pa^{(0)}, \Pa^{(1)})}{\tau}.
\end{align}

To complete the proof, we next show Eq.~\eqref{inverse_inequality}. Let $(\Pxy', \force'_{\alpha})$ be a pair satisfying the constraints [Eqs.~\eqref{cond1_projection}, \eqref{cond2_dynamics_conservation}, and \eqref{cond3_initial_final_distributions}]. We define a functional of the pair $(\Pxy', \force'_{\alpha})$ parametrized by $\eta > 0$ as
\begin{align}
    \tilde{s}_{\eta}(t) = \int_0^{t}\left(\eta + \norm{\force'_\alpha(t')}^2_{L_\alpha(\Pxy'(t'))}\right)^{1/2}dt'.
\end{align}
Since $d_t\tilde{s}_\eta(t) > 0$, the inverse function $\tilde{t}_\eta \coloneqq \tilde{s}_\eta^{-1}$ is well defined. Using $\tilde{t}_\eta$, we introduce the following quantities,
\begin{gather}
    \tilde{\bm{p}}(s) = \Pxy'(\tilde{t}_\eta(s)),~\tilde{\bm{p}}_{\alpha}(s) = \Pi_{\alpha}\tilde{\bm{p}}(s),\\
    \tilde{\force}_{\alpha}(s) = (d_s\tilde{t}_\eta(s))\force'_{\alpha}(\tilde{t}_\eta(s)).
\end{gather}
The pair $(\tilde{\bm{p}}, \tilde{\force}_{\alpha})$ also satisfies the constraints [Eqs.~\eqref{cond1_projection},~\eqref{cond2_dynamics_conservation}, and~\eqref{cond3_initial_final_distributions}] on the time interval $s \in [0, \tilde{s}_{\eta}(\tau)]$:
\begin{align}
    \tilde{\bm{p}}_\alpha(0) = \Pi_{\alpha}\Pxy'(0) = \Pxy_{\alpha}^{(0)},\;\tilde{\bm{p}}_\alpha(\tilde{s}_\eta(\tau)) = \Pi_{\alpha}\Pxy'(\tau) = \Pxy_{\alpha}^{(1)},
\end{align}
and
\begin{align}
    d_s\tilde{\bm{p}}_{\alpha}(s) &= \left.(d_s\tilde{t}_\eta(s))d_t\Pa'(t)\right|_{t = \tilde{t}_{\eta}(s)} \notag\\
    &= (d_s\tilde{t}_\eta(s))\left.[\div_{g_{\alpha}}L_\alpha(\Pxy'(t))\force'_{\alpha}(t)]\right|_{t = \tilde{t}_\eta(s)} \notag\\
    &= \div_{g_{\alpha}}L_\alpha(\tilde{\bm{p}}(s))\tilde{\force}_{\alpha}(s).
\end{align}
Therefore, we obtain the following inequality:
\begin{align}
    &\Wsub(\Pa^{(0)}, \Pa^{(1)})^2
    \leq \tilde{s}_{\eta}(\tau)\int_0^{\tilde
    {s}_{\eta}(\tau)}\norm{\tilde{\force}_{\alpha}(s)}^2_{L_\alpha(\tilde{\bm{p}}(s))}ds \notag\\
    &= \tilde{s}_{\eta}(\tau)\int_0^{\tau}(\left.d_s\tilde{t}_{\eta}(s)\right|_{s=\tilde{s}_{\eta}(t)})^2\norm{\force'_{\alpha}(t)}^2_{L_\alpha(\Pxy'(t))}d_t\tilde{s}_{\eta}(t)dt \notag\\
    &= \tilde{s}_{\eta}(\tau)\int_0^{\tau}\frac{\norm{\force'_{\alpha}(t)}^2_{L_\alpha(\Pxy'(t))}}{\left(\eta + \norm{\force'_{\alpha}(t)}^2_{L_\alpha(\Pxy'(t))}\right)^{1/2}}dt \leq \tilde{s}_{\eta}(\tau)^2,
\end{align}
where we performed the variable transformation $t = \tilde{t}_\eta(s)$ and used the following relation,
\begin{align}
    \left.d_s\tilde{t}_{\eta}(s)\right|_{s=\tilde{s}_{\eta}(t)} 
    &= \left(d_t\tilde{s}_{\eta}(t)\right)^{-1} \notag\\
    &= \left(\eta + \norm{\force'_\alpha(t)}^2_{L_\alpha(\Pxy'(t))}\right)^{-1/2}.
\end{align}
Taking the limit $\eta \to 0$ yields
\begin{align}
    \Wsub(\Pa^{(0)}, \Pa^{(1)}) \leq \int_0^{\tau} \norm{\force'_\alpha(t)}_{L_\alpha(\Pxy'(t))}dt,
\end{align}
which is an equivalent restatement of Eq.~\eqref{inverse_inequality}.

\section{Proof of Eq.~\eqref{EPRex_inequality}}
\label{app:EPRex_inequality}
We prove Eq.~\eqref{EPRex_inequality} by deriving the following formula:
\begin{align}
    \EPR^\mathrm{ex,\;loc}_\alpha = \inf_{\tilde{\force}_\alpha}\norm{\Force_\alpha-\tilde{\force}_\alpha}^2_{L_\alpha}, \label{local_EPRex_formula}
\end{align}
such that $\tilde{\force}_\alpha \in\ker \div_{g_\alpha}L_\alpha$. This formula can be shown by the following transformation:
\begin{align}
    \norm{\Force_\alpha-\tilde{\force}_\alpha}^2_{L_\alpha}
    &= \norm{\force'^*_\alpha + (\Force_\alpha - \force'^*_\alpha - \tilde{\force}_\alpha)}^2_{L_\alpha} \notag\\
    &= \norm{\force'^*_\alpha}^2_{L_\alpha} + \norm{\Force_\alpha - \force'^*_\alpha - \tilde{\force}_\alpha}^2_{L_\alpha} \notag\\
    &\geq \norm{\force'^*_\alpha}^2_{L_\alpha},
\end{align}
where we used the orthogonality $\innerprod{\force'^*_\alpha}{\Force_\alpha - \force'^*_\alpha}_{L_\alpha} = 0$ and $\innerprod{\force'^*_\alpha}{\tilde{\force}_\alpha}_{L_\alpha} = -(\bm{\psi}'^*_\alpha)^\top\div_{g_\alpha}L_\alpha\tilde{\force}_\alpha = 0$ to obtain the second equality. Since the second term in the second line vanishes when $\tilde{\force}_\alpha = \Force_\alpha - \force'^*_\alpha \in\ker\div_{g_\alpha}L_\alpha$, we can conclude that the right-hand side of Eq.~\eqref{local_EPRex_formula} yields $\EPR^\mathrm{ex,\;loc}_\alpha$.

Note that $\Force_\alpha - \Force^*_\alpha\in\ker\div_{g_\alpha}L_\alpha$ because $\div_{g_\alpha}L_\alpha(\Force_\alpha - \Force^*_\alpha) = \Pi_\alpha\div L(\Force - \Force^*) = \bm{0}$. From Eq.~\eqref{local_EPRex_formula}, we thus obtain the desired inequality:
\begin{align}
    \EPRex_\alpha &= \norm{\Force_\alpha -(\Force_\alpha - \Force^*_\alpha)}^2_{L_\alpha} \notag \\ 
    &\geq \inf_{\tilde{\force}_\alpha\in\ker\div_{g_\alpha}L_\alpha}\norm{\Force_\alpha-\tilde{\force}_\alpha}^2_{L_\alpha}
    = \EPR^\mathrm{ex,\;loc}_\alpha.
\end{align}

\end{document}